\documentclass[twocolumn]{aastex62}

\received{January 1, 2019}
\revised{?, 2019}
\accepted{\today}
\submitjournal{ApJ}

\shorttitle{SEDs of AGN dust I: synthetic spectra}
\shortauthors{Gonz\'alez-Mart\'in et al.}

\begin{document}

\title{Exploring the mid-infrared SEDs of six AGN dusty torus models I: synthetic spectra }

\correspondingauthor{Omaira Gonz\'alez-Mart\'in, faculty}
\email{o.gonzalez@irya.unam.mx}

\author{Omaira Gonz\'alez-Mart\'in}
\affil{Instituto de Radioastronom\'ia y Astrof\'isica (IRyA-UNAM), 3-72 (Xangari), 8701, Morelia, Mexico}
\author{Josefa Masegosa}
\affil{Instituto de Astrof\'isica de Andaluc\'ia, CSIC, Glorieta de la Astronom\'ia s/n E-18008, Granada, Spain}
\author{Ismael Garc\'ia-Bernete}
\affil{Instituto de F\'isica de Cantabria (CSIC-UC), Avenida de los Castros, 39005 Santander, Spain}
\author{Cristina Ramos Almeida}
\affil{Instituto de Astrof\'isica de Canarias (IAC), C/V\'ia L\'actea, s/n, E-38205 La Laguna, Spain}
\affil{Departamento de Astrof\'isica, Universidad de La Laguna (ULL), E-38205 La Laguna, Spain}
\author{Jos\'e Miguel Rodr\'iguez-Espinosa}
\affil{Instituto de Astrof\'isica de Canarias (IAC), C/V\'ia L\'actea, s/n, E-38205 La Laguna, Spain}
\affil{Departamento de Astrof\'isica, Universidad de La Laguna (ULL), E-38205 La Laguna, Spain}
\author{Isabel M\'arquez}
\affil{Instituto de Astrof\'isica de Andaluc\'ia,  CSIC, Glorieta de la Astronom\'ia s/n E-18008, Granada, Spain}
\author{Donaji Esparza-Arredondo}
\affil{Instituto de Radioastronom\'ia y Astrof\'isica (IRyA-UNAM), 3-72 (Xangari), 8701, Morelia, Mexico}
\author{Natalia Osorio-Clavijo}
\affil{Instituto de Radioastronom\'ia y Astrof\'isica (IRyA-UNAM), 3-72 (Xangari), 8701, Morelia, Mexico}
\author{Mariela Mart\'inez-Paredes}
\affil{Instituto de Radioastronom\'ia y Astrof\'isica (IRyA-UNAM), 3-72 (Xangari), 8701, Morelia, Mexico}
\affil{Korea Astronomy and Space Science Institute 776, Daedeokdae-ro, Yuseong-gu, Daejeon, Republic of Korea (34055)}
\author{C\'esar Victoria-Ceballos}
\affil{Instituto de Radioastronom\'ia y Astrof\'isica (IRyA-UNAM), 3-72 (Xangari), 8701, Morelia, Mexico}
\author{Alice Pasetto}
\affil{Instituto de Radioastronom\'ia y Astrof\'isica (IRyA-UNAM), 3-72 (Xangari), 8701, Morelia, Mexico}
\author{Deborah Dultzin}
\affil{Instituto de Astronom\'ia (IA-UNAM), Apartado Postal 70-264, 04510, Mexico DF, Mexico}



\begin{abstract}

At distances from the active galaxy nucleus (AGN) where the ambient temperature falls below $\rm{\sim}$1500-1800 K, dust is able to survive. It is thus possible to have a large dusty structure present which surrounds the AGN. This is the first of two papers aiming at comparing six dusty torus models with available SEDs, namely \citet{Fritz06}, \citet{Nenkova08B}, \citet{Hoenig10B}, \citet{Siebenmorgen15}, \citet{Stalevski16}, and \citet{Hoenig17}. In this first paper we use synthetic spectra to explore the discrimination between these models and under which circumstances they allow to restrict the torus parameters, while our second paper analyzes the best model to describe the mid-infrared spectroscopic data. We have produced synthetic spectra from current instruments: GTC/CanariCam and \emph{Spitzer}/IRS and future \emph{JWST}/MIRI and \emph{JWST}/NIRSpec instruments. We find that for a reasonable brightness ($\rm{F_{12\mu m}>100 mJy}$) we can actually distinguish among models except for the two pair of parent models. We show that these models can be distinguished based on the continuum slopes and the strength of the silicate features. Moreover, their parameters can be constrained within 15\% of error, irrespective of the instrument used, for all the models but \citet{Hoenig17}. However, the parameter estimates are ruined when more than 50\% of circumnuclear contributors are included. Therefore, future high spatial resolution spectra as those expected from \emph{JWST} will provide enough coverage and spatial resolution to tackle this topic. 

\end{abstract}


\section{Introduction} \label{sec:intro}

\citet{Barvainis87} was the first to propose that the excess of emission observed at $\rm{\sim}$3\,$\rm{\mu m}$ in active galactic nuclei (AGN), compared to the disk emission at optical/UV frequencies, was probably due to multiple temperature dust components. Since then, many attempts have been made to solve the radiative transfer equations to produce the spectral energy distribution (SED) due to dust\footnote{The dust in AGN was historically thought to be arranged as a torus-like geometry. However, other options as disks or winds have been proposed lately. We refer to these models as dust models hereafter, avoiding the use of a particular geometry.}. Pioneers works on the subject were \citet{Krolik88}, \citet{Pier92}, \citet{Granato94}, \citet{Stenholm94}, \citet{Efstathiou95}, and \citet{Nenkova02}.

Initially, for the sake of simplicity, most authors used smooth dust distributions with different radial and vertical density profiles \citep[e.g.][]{Pier92,Granato94,Efstathiou95, vanBemmel03,Schartmann05,Fritz06}. However, before the development of the first geometrical torus models, it was known that dust was probably organized into clouds because it would be very difficult for the dust to survive in the region otherwise. The dust should be moving with random velocities of hundreds of km/s to reproduce the geometrical thickness required to obscure a substantial solid angle around the central source.  If the dust was distributed homogeneously the temperature would be of the order of $\rm{\sim 10^6}$\,K, preventing the dust to survive \citep[e.g.][]{Krolik88,Tacconi94}. Several radiative transfer models were developed to account for 2D or 3D clumpy dust distributions \citep{Nenkova08A,Nenkova08B,Hoenig10A,Hoenig10B}. A mix of smooth plus clumpy distributions have also been proposed \citep{Stalevski12,Siebenmorgen15}. More recently, motivated by the discovery of a polar dust contribution to the mid-infrared SEDs, a more complex scenario has been proposed to explain the infrared nuclear emission of Seyfert galaxies \citep[e.g.][]{Hoenig13,Asmus16}. \citet{Hoenig17} produced a model that includes a compact, geometrically thin disk in the equatorial region of the AGN, and an extended, elongated polar structure, which is co-spatial with the outflow region of the AGN on larger scales \citep[see also the recently published semi-empirical SED libraries for quasars by][]{Lyu18}. They claim that this model reproduces the $\rm{3-5\,\mu m}$ bump observed in many type 1 AGN that other models could not reproduce. Using infrared interferometry of 23 Seyfert galaxies (for which 7 sources have three or more well spaced directions in the $(u,v)$ plane, the point-like source contributes less than 70\%, and the uncertainties are below 10\%) this mid-infrared polar emission has been detected so far in six sources; five by \citet[][]{Lopez-Gonzaga16} and an additional one reported by \citet{Leftley18}. We give a short overview of the six models used in this paper in Section\,\ref{sec:models} \citep[see also][for further details of different AGN dust models]{Ramos-Almeida17}.

Apart from the obvious differences on the torus geometry and smoothness/clumpyness, these models also invoke different grain sizes and chemical compositions. \citet{Fritz06} considered typical silicate and graphite grain sizes of $\rm{a_{G}=0.005-0.25\,\mu m}$ and  $\rm{a_{G}=0.025-0.25\,\mu m}$, respectively. \citet{Nenkova08A} assumed that dust grains are spherical, with size distribution from \citet{Mathis77} and a standard ISM composition of 53\% silicates and 47\% graphites. \citet{Hoenig10B} included the standard ISM composition, although they also used ISM large grains (sizes between 0.1-1\,$\rm{\mu m}$), dominated by intermediate to larger graphite grains. 

Minor effort has been put into comparing the various models or to disentangle which of them better reproduces the data. Most comparisons are confronting parent models developed by the same group \citep{Schartmann08,Garcia-Gonzalez17}. Although some papers have tried to qualitatively compare different models \citep[e.g.][]{vanBemmel03}. \citet{Feltre12} is the only work comparing the smooth model by \citet{Fritz06} and the clumpy model by \citet{Nenkova08B} using matched parameters. They found that the behavior of the silicate features and spectral slopes below $\rm{\sim 7\, \mu m}$ are different.

The question we would like to answer in this paper is if the data allow to distinguish which of the proposed models better describes AGN infrared SEDs. For this purpose we created a set of synthetic spectra using six of these well known models (see Section \ref{sec:models}) using different instruments. A comparison between models and real AGN mid-infrared spectroscopic data is performed by Gonzalez-Martin et al. 2019B (hereafter Paper II). The paper is organized as follows. Section \ref{sec:models} gives a brief summary of the dusty models used along this paper. Section \ref{sec:modelprecedure} describes how to convert multi-parameter models to the spectral fitting tool environment XSPEC \citep{Arnaud96}, how to create synthetic spectra using the most relevant current and future mid-infrared instrumentation, and how to fit these synthetic spectra. The main results are included in Section \ref{sec:results}. These results are discussed in Section \ref{sec:discussion} and the paper ends with a report of the main findings in Section \ref{sec:summary}. 

\begin{table*}
\footnotesize
\begin{center}
\begin{tabular}{ l c c c c l}
\hline \hline
Model    &   Dust             &    Dust             & N. &  wv. range ($\rm{\mu m}$)  &   Parameters  \\ 
         &   distribution     &    composition      & SEDs  &  \& N. bins  &           \\ (1) &        (2)            &      (3)  & (4)   &     (5)     &   (6)          \\
         \hline
\citet{Fritz06} &   Smooth    &  Silicate \&    &  24,000   &    0.001-1,000       &    $\rm{i = [0,10,20,30,40,50,60,70,80,90] }$       \\
$\rm{[Fritz06]}$		&	torus	&	Graphite 	&		&	178	&	$\rm{\sigma=[20,40,60]}$ 	\\
		&		&		&		&		&	$\rm{\Gamma=[0,2,4,6]}$	\\
		&		&		&		&		&	$\rm{\beta=[-1,-0.75,,-0.5,-0.25,0]}$ 	\\
		&		&		&		&		&	$\rm{Y= [10,30,60,100,150]}$	\\
		&		&		&		&		&	$\rm{\tau_{9.7\mu m}=[0.1,0.3,0.6,1,2,3,6,10]}$	\\
\citet{Nenkova08B} &    Clumpy  &  Standard ISM  & 1,247,400     &    0.001-1,000       &    $\rm{i = [0,10,20,30,40,50,60,70,80,90]}$     \\
$\rm{[Nenkova08]}$		&	torus	&		&		&	119	&	$\rm{N_{0}=[1,3,5,7,9,11,13,15]}$ 	\\
		&		&		&		&		&	$\rm{\sigma =}$ [15,25,35,45,55,65,70]	\\
		&		&		&		&		&	$\rm{Y= [5,10,20,30,...,80,90,100,150]}$	\\
		&		&		&		&		&	$\rm{ q=[0.0,0.5,1.0,1.5,2.0,2.5,3.0]}$ 	\\
		&		&		&		&		&	$\rm{\tau_{v}= [10,20,40,60,80,120,160,200,300]}$	\\
\citet{Hoenig10B} &    Clumpy  &  Standard ISM      &  1,680    &   0.01-36,000  &     $\rm{i = [0,15,30,45,60,75,90]}$    \\
$\rm{[Hoenig10]}$               &  torus     &  ISM large     &     &   105   &   $\rm{N_{0}=[2.5,5.0,7.5,10.0]}$       \\
                &       &  Gr-dominated      &      &      &   $\rm{\theta= [5,30,45,60]}$     \\
		&		&		&		&		&	$\rm{a=[-2.0,-1.5,-1.0,-0.5,0]}$ 	\\
		&		&		&		&		&	$\rm{\tau_{cl}= [30,50,80]}$	\\
		&		&		&		&		&	($\rm{Y=150}$)	\\
\citet{Siebenmorgen15} &    Smooth \&    &  Silicate \&    &   3,600    &   0.0005-500       &    $\rm{i = [19, 33, 43, 52, 60, 67, 73, 80, 86]}$     \\
$\rm{[Sieben15]}$        &    clumpy    &  Amorphous carbon    &      &     473     &  $\rm{R_{in}=[3,5.1,7.7,10,15.5]}$  \\
    &	torus or/\&	&		&		&		&	$\rm{\eta=[1.5, 7.7, 38.5, 77.7]}$	\\
	&	outflow	&		&		&		&	$\rm{\tau_{cl}=[0, 4.5, 13.5, 45]}$ \\
	&		&		&		&		&	$\rm{\tau_{disk}=[ 0, 30, 100, 300, 1000]}$	\\
 	&		&		&		&		&	($\rm{R_{out} = 170R_{in}}$) \\
\citet{Stalevski16} &   Smooth  \&  &  Silicate \&    &  19,200   &    0.001-1,000       &    $\rm{i = [0,10,20,30,40,50,60,70,80,90] }$       \\
$\rm{[Stalev16]}$		& Clumpy	& Graphite		&		&	132	&	$\rm{\sigma=[10,20,30,40,50,60,70,80]}$ 	\\
		&	torus		&		&		&	& 	$\rm{p=[0,0.5,1.0,1.5]}$	\\
		&		&		&		&		&	$\rm{q=[0,0.5,1.0,1.5]}$ 	\\
		&		&		&		&		&	$\rm{Y= [10,20,30]}$	\\
		&		&		&		&		&	$\rm{\tau_{9.7\mu m}=[3,5,7,11]}$	\\
		&		&		&		&		&	($\rm{R_{in}=0.5}$ pc)	\\		
\citet{Hoenig17} &    Clumpy   &  Standard ISM    &   132,300   &   0.01-36,000       &    $\rm{i = [0,15,30,45,60,75,90]}$     \\
$\rm{[Hoenig17]}$	&	torus \&	&	ISM large 	&		&	105	&	$\rm{N_{0}=[5,7,10]}$ 	\\
	&	outflow	&		&		&		&	$\rm{a=[-3.0,-2.5,-2.0,-1.5,-1.0,-0.5]}$	\\
	&		&		&		&		&	$\rm{\theta}$ = [30,45] 	\\
	&		&		&		&		&	$\rm{\sigma=[7,10,15]}$ \\
	&		&		&		&		&	$\rm{a_{w}=[-2.5,-2.0,-1.5,-1.0,-0.5]}$	\\
	&		&		&		&		&	$\rm{h = [0.1, 0.2, 0.3, 0.4, 0.5]}$  \\
	&		&		&		&		& $\rm{f_{wd}= [0.15,0.3,0.45,0.6,0.75]}$	\\
	&		&		&		&		&	($\rm{Y=500(large)/450(ISM))}$	\\
	&		&		&		&		&	($\rm{R_{cl} = 0.035 \times R}$)	\\
	&		&		&		&		&	($\rm{\tau_{v}=50}$)	\\
\hline \hline
\end{tabular}
\caption{Summary of the dusty torus models used in this work. We show the name as included in XSPEC for each model below the citation. It includes the dust distribution (i.e. morphology of the dusty distribution, Col. 2), dust chemical composition (Col. 3), number of SEDs produced (Col. 4), wavelength coverage and number of bins within the wavelength range (Col. 5), and parameters involved including values to produce the SEDs for each parameter (Col. 6). Fixed parameters are quoted within brackets. Note that the viewing angle is measured in all the cases from the pole to the equator of the system except for [Fritz06], which is measured in the opposite direction. Note that we restricted the Clumpy model \citep{Nenkova08B} within XSPEC using $\rm{N_{0}=}$[1,3,5,7,9,11,13,15] and $\rm{\sigma=}$[15,25,35,45,55,65,70] (see text).}
\label{tab:models}
\end{center}
\end{table*}

\section{SED libraries of AGN dust emission}\label{sec:models}

This paper is devoted to compare as many dust SED libraries as possible. Table \ref{tab:models} shows the compilation of radiative codes with publicly available SEDs. All of them show slightly different dust distributions (Col. 2) and compositions (Col. 3). Furthermore, they map a different range and/or set of parameters (see Col. 6). There are other available radiative calculations to model the dusty structure of AGN but there is no available SED library to test them. Furthermore, note that there are semi-empirical SED libraries that are also out of the scope of this paper \citep[e.g.][]{Lyu18}. Below we describe the SED libraries used. We quote the abbreviation used for each model along this paper within brackets.

$\bullet$ {\bf Smooth torus model by \citet{Fritz06} [Fritz06]} \citep[see also][]{Feltre12}. Radiative transfer code used to produce the SED of a simple toroidal geometry consisting in a flared disc that can be represented as two concentric spheres, delimiting respectively the inner and the outer torus radius, having the polar cones removed. They considered typical silicate and graphite grain sizes of $\rm{a_{G}=0.005-0.25\,\mu m}$ and  $\rm{a_{G}=0.025-0.25\,\mu m}$, respectively. This model uses different sublimation radii for graphite and silicate grains. The free parameters of this model are the viewing angle toward the toroidal structure, $i$, the opening angle, $\rm{\sigma}$, the exponent of the logarithmic azimuthal density distribution, $\rm{\Gamma}$, the exponent of the logarithmic radial profile of the density distribution, $\rm{\beta}$, the outer radius of the torus, $\rm{R_{max}}$ compared to the inner one (expressed as $\rm{Y = R_{max}/R_{min}}$), and the edge-on optical depth at 9.7\,$\rm{\mu m}$, $\rm{\tau_{9.7\,\mu m}}$. The size of the torus is defined by the outer radius, $\rm{R_{max}}$, and the opening angle, $\rm{\sigma}$. The inner radius is defined by the sublimation temperature of dust grains under the influence of the strong nuclear radiation field ($\rm{T_{sub}=1500\,K}$). Its incident radiation spectrum is defined by a combination of power-laws as those described by \citet{Granato94}. The SED library contains 24,000 SEDs in the $\rm{0.001-1,000\,\mu m}$ wavelength range.

$\bullet$ {\bf Clumpy torus model by \citet{Nenkova08B} [Nenkova08]\footnote{We have used the most updated version of the clumpy toroidal model included at www.clumpy.org by the time of the submission of this paper.}} \citep[see also][]{Nenkova08A}. They developed a formalism that accounts for the concentration of dust in clumps or clouds, referred to as clumpy, to describe the nature of the AGN torus. \citet{Nenkova08A} assumed that dust grains are spherical, with size distribution from \citet{Mathis77} and a standard Galactic mix of 53\% silicates and 47\% graphites (i.e. standard ISM). The model assumes a toroidal distribution that depends on the viewing angle toward the torus, $i$, the inner number of clouds, $\rm{N_{0}}$, the half opening angle of the torus, $\rm{\sigma}$, the outer radius of the torus, $\rm{R_{out}}$ (scaled to the inner radius, i.e. $\rm{Y=R_{out}/R_{in}}$), the slope of the radial density distribution, $\rm{q}$, and the optical depth (i.e. at 0.55$\rm{\mu m}$) of the individual clouds, $\rm{\tau_{v}}$. The inner radius is fixed to the dust sublimation radius ($\rm{T_{dust}=1500\,K}$). They produced a library including 1,247,400 SEDs in the 0.001-1,000\,$\rm{\mu m}$ wavelength range.

$\bullet$ {\bf Clumpy torus model by \citet{Hoenig10B} [Hoenig10]} \citep[see also][]{Hoenig06,Hoenig10A}. Radiative transfer model of three-dimensional clumpy dust tori using optically thick dust clouds and a low torus volume filling factor. They included an improved handling of the diffuse radiation field in the torus, which is approximated by a statistical approach and the possibility of different dust compositions and grain sizes. They synthesized three compositions: the standard ISM (47\% graphites and 53\% silicates), ISM large grains (grains between 0.1 and 1\,$\rm{\mu m}$ in size), and Gr-dominated (dominated by intermediate to larger graphite grains; 70\% graphites and 30\% silicates). The parameters of this library of SEDs are: the viewing angle, $i$, the number of clouds along an equatorial line-of-sight, $\rm{N_0}$, the half-opening angle of the distribution of clouds, $\rm{\theta}$, the radial dust-cloud distribution power law index, $a$, and the opacity of the clouds, $\rm{\tau_{cl}}$. The outer torus radius, $\rm{R_{max}}$, is fixed to the inner radius as $\rm{R_{max}=150\,R_{min}}$, where $\rm{R_{min}}$ is set to the dust sublimation radius. This library includes 1,680 SEDs in the 0.01-36,000\,$\rm{\mu m}$ wavelength range.
 
$\bullet$ {\bf Two phase model (clumpy + smooth) torus model by \citet{Siebenmorgen15} [Sieben15]}. They assumed that the dust near the AGN is distributed in a torus-like geometry, which can be described as a clumpy medium or a homogeneous disk, or a combination of the two. They include an isothermal disk that is embedded in a clumpy medium, including pc sized dust clouds passively heated. Moreover, this model is based on the idea that a significant part of the mid-infrared emission appears to come from the ionization cones \citep{Braatz93,Hoenig13}. Therefore, in this model dust exists also in the polar region. The dust particles considered are fluffy and have higher sub-millimeter emissivities than grains in the diffuse ISM. The SED library depends on the viewing angle, $\rm{i}$, the inner radius, $\rm{R_{in}}$, the volume filling factor of the clouds, $\rm{\eta}$, the optical depth (i.e. at 0.55$\rm{\mu m}$) of the individual clouds, $\rm{\tau_{cl}}$, and the optical depth of the disk mid-plane, $\rm{\tau_{disk}}$. The outer radius up to where dust exists and their distribution function are kept to be constant as $\rm{R_{out} = 170\,R_{in}}$. They produced a library including 3,600 SEDs in the 0.0005-500\,$\rm{\mu m}$ wavelength range.  
 
$\bullet$ {\bf Two phase (clumpy + smooth) torus model by \citet{Stalevski16} [Stalev16]} \citep[see also][]{Stalevski12}. They model the dust in a torus geometry with a two-phase medium, consisting in a large number of high-density clumps embedded in a smooth dusty component of low density. This model is claimed to reproduce the near-infrared excess around 5$\rm{\mu m}$, producing at the same time attenuated silicate features \citep{Stalevski12}. One of the improvements of the current set of models described by \citet{Stalevski16} compared to the previous modelling \citep[i.e.][]{Stalevski12} is the inclusion of anisotropic emission. This implies that the inner radius is not constant but shows the same dependency with the polar angle shown by the anisotropy. Dust is also distributed along the radial and polar directions according to a power law distribution \citep[based on the same prescription given by][]{Fritz06}. The dust chemical composition is set to a mixture of silicate and graphite grains. The parameters of the model are the viewing angle toward the observer, $i$, the ratio between the outer and the inner radius of the torus, $\rm{Y= R_{out}/R_{in}}$, the half opening angle of the torus, $\rm{\sigma}$, the indices that sets dust density gradient with the radial $p$ and polar $q$ distribution of dust, and the 9.7$\rm{\mu m}$ average edge-on optical depth, $\rm{\tau_{9.7\mu m}}$. The fraction of total dust mass in clumps compared to the total dust mass is set to $\rm{M_{cl}=0.97}$ and an inner radius of the torus is set to $\rm{R_{in} = 0.5\,pc}$. This SED library contains 19,200 SEDs in the 0.001-1,000\,$\rm{\mu m}$ wavelength range.

$\bullet$ {\bf Clumpy disk and outflow model by \citet{Hoenig17} [Hoenig17]}. This model is built upon the suggestion that the dusty gas around the AGN consists on an inflowing disk and an outflowing wind. In practice, this model consists in clumpy disk-like models \citep[following that described by][]{Hoenig10B} plus a polar outflow. Common parameters for disk and wind are the viewing angle, $i$, and the number of clouds in the equatorial plane, $\rm{N_{0}}$. The disk is also governed by the exponent of the radial distribution of clouds, $a$, and the the optical depth of individual clouds, $\rm{\tau_{cl}}$, which is fixed to $\rm{\tau_{cl}=50}$. The outflow is modeled as a hollow cone and characterized by three parameters: the index of the dust cloud distribution power law along the wind, $\rm{a_w}$, the half-opening angle of the wind, $\rm{\theta}$, and the angular width of the hollow wind cone, $\rm{\sigma}$. Finally, a wind-to-disk ratio, $\rm{f_{wd}}$, defines the ratio between the number of clouds along the cone and $\rm{N_{0}}$. The library includes 132,300 SEDs in the 0.01-36,000\,$\rm{\mu m}$ wavelength range.

A caveat to take into account when comparing all these models is that they are not produced to fully cover similar dusty structures. Indeed, this is impossible when producing SEDs with completely different structural components. The exception is the parameter space from [Fritz06] and [Nenkova08] because the former libraries were produced to match similar physical parameters of the torus \citep[see][]{Feltre12}. However, we rely on the assumption that the parameter space for all the models is set by the authors to better match observational evidence of the AGN dusty structure. In that sense, the comparison shown in this paper is still valid. We defer the study of the adequacy of the parameter space by comparison to an AGN sample with \emph{Spitzer}/IRS spectroscopic data to Paper II.

\section{Multi-parameter model fitting procedure}\label{sec:modelprecedure}

We devoted this paper to test if the models can be self-constrained and if they can be differentiated in their spectral shape among each other. For that purpose we converted the SED libraries to multi-parametric models within the spectral fitting tool XSPEC (Section \ref{sec:XspecModel}) and produced a set of synthetic spectra (Section \ref{sec:Synthdata}) to test these models. 

\subsection{Model conversion into XSPEC format}\label{sec:XspecModel}

XSPEC is a command-driven, interactive, spectral-fitting program within the HEASOFT\footnote{https://heasarc.gsfc.nasa.gov} software. XSPEC has been used to analyze X-ray data as those provided by \emph{ROSAT}, \emph{ASCA}, \emph{Chandra}, \emph{XMM}-Newton, \emph{Suzaku}, \emph{NuSTAR}, or \emph{Hitomi}. XSPEC allows users to fit data with models constructed from individual components. XSPEC already incorporates a large number of models, but new models can be uploaded using the {\sc atable} task (see below).

Here we provide a brief summary of XSPEC capabilities. XSPEC integrates two main statistics analyses required to test the spectral fittings which are required by our analysis: (1) finding the parameters for a given model that provide the best fit to the data and then estimating uncertainties on these parameters; and (2) testing whether the model and its best-fit parameters actually match the data, i.e. determining the goodness-of-fit.

XSPEC uses several statistical methods associated to different likelihoods. We will use the $\rm{\chi^2}$ statistics for Gaussian data distribution. To assess for the goodness-of-fit XSPEC performs a test to reject the null hypothesis that the observed data are drawn from the model (by including the $\rm{\chi^2}$/dof and the null hypothesis probability). The parameter confidence regions are found by surfaces of constant $\rm{\chi^2}$ statistics from the best-fit value ({\sc error} task). Finally, it also allows to calculate fluxes and luminosities at a given wavelength. 

Thus, XSPEC provides a wide range of tools to perform spectral fittings to the data, being able to perform parallel processes in order to speed them up. To use these capabilities we need to convert the SEDs libraries to XSPEC format to upload our models within XSPEC as additive tables. The basic concept of a table model is that the file contains an N-dimensional grid of model spectra with each point on the grid calculated for particular values of the N parameters in the model. XSPEC will interpolate on the grid to get the spectrum for the parameter values required at that point in the fit. 

We have created an additive table for each of the models used in this paper. To do so, we created a one-parameter table (in fits format) associated to all the SEDs using the {\sc flx2tab} task within HEASOFT. We then wrote a python routine to change the headers associating each SED to a set of parameters. Each model has a number of free parameters, including those reported in Table \ref{tab:models}, redshift, and normalization. Note that in the case of [Nenkova08], we were not able to obtain a XSPEC model using the entire SED library due to the unpractical size of the final model (over 100\,GB). Instead, we slightly restricted the number of clouds and the angular width to the torus to the ranges $\rm{N_0=}$[1,3,5,7,9,11,13,15] and $\rm{\sigma=}$[15,25,35,45,55,65,70], respectively, to recover a more transferable model ($\sim$6\,GB)\footnote{Note that the restricted number of SEDs used for [Nenkova08] is $\rm{\sim}$333,000.}. 

\begin{figure*}[!ht]
\begin{center}
\includegraphics[width=2.0\columnwidth]{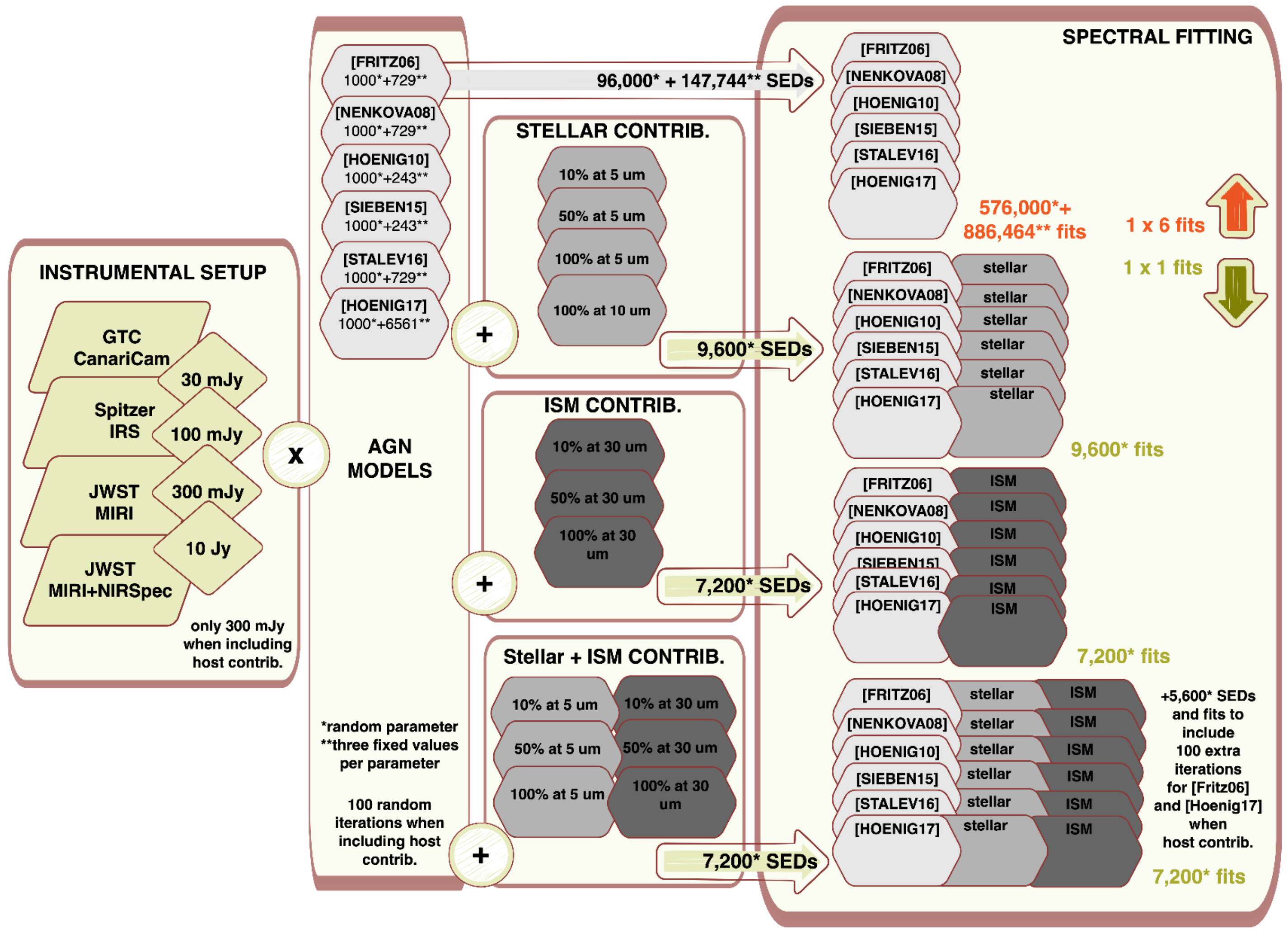}
\caption{Flowchart of the multi-parameter model fitting procedure. On the left hand the instrumental set-up, in the middle panels the models tested (AGN model and circumnuclear contributors), and on the left panel the spectral fits performed. We used orange and green to remark that we fit the synthetic spectra to all the models when using only AGN dust models while we fit the synthetic spectra to its own model when including circumnuclear contributors. Furthermore, we mark with * and ** those synthetic spectra produced with random parameters and with 3 fixed values along each parameter space (see text).}
\label{fig:flowchart_syntheticspectra}
\end{center}
\end{figure*}

\subsection{Synthetic spectra in XSPEC format}\label{sec:Synthdata}

We create synthetic spectra to test the models. We simulated mid-infrared GTC/CanariCam, mid-infrared low-resolution \emph{Spitzer}/IRS, near-infrared \emph{JWST}/NIRSpec, and mid-infrared \emph{JWST}/MIRI spectra to confront them against models. We do not attempt to test models against photometric data because the large variety of filters and their combinations add complexity to the analysis \citep[see e.g.][]{Ramos-Almeida14}. XSPEC is able to simulate spectra with the {\sc fakeit} task using the instrument response\footnote{The instrument response describes the efficiency per unit wavelength.} and a model. The {\sc flx2xsp} task, within HEASOFT, reads a text file containing one or more spectra and their errors and writes out a standard XSPEC pulse height amplitude (PHA\footnote{Engineering unit describing the integrated charge per pixel from an event recorded in a detector.}) and response files. 

We convert into XSPEC format a real spectrum in order to get the response file to produce the synthetic spectra. Note that this response file does not depend on the actual spectrum of the object used but on the instrumental set up. The information on the spectrum is recorded in a different file while the response file contains only the sensitivity/performance of the instrument used. We used the N-band ($\rm{\sim}$7-13\,$\rm{\mu m}$) and Q-band ($\rm{\sim}$17-23\,$\rm{\mu m}$) T-ReCS/Gemini spectra for NGC\,1068 downloaded from the Gemini archive\footnote{https://archive.gemini.edu} and reduced as a point-like source with the RedCan pipeline \citep{Gonzalez-Martin13}. Note that this is representative of the expected wavelength coverage and sensitivity of GTC/CanariCam, because CanariCam and T-ReCS are twin instruments. T-ReCS is not longer available for the community while CanariCam is currently available. GTC/CanariCam mid-infrared simulated spectra are also equivalent to other ground-based facilities (e.g. VLT/VISIR or Michelle/Gemini) due to their similarities on the wavelength range and sensitivities. 

As for the  \emph{Spitzer}/IRS response, we obtained the \emph{Spitzer}/IRS low-resolution spectra for NGC\,1052 using the Combined Atlas of Sources with \emph{Spitzer} IRS Spectra (CASSIS\footnote{http://cassis.sirtf.com}), which provides fully calibrated 1D spectra for both low-resolution (LR) and high-resolution (HR) \emph{Spitzer} instruments.   

The \emph{JWST}/MIRI and \emph{JWST}/NIRSpec spectra were simulated using the \emph{JWST} exposure time calculator (ETC\footnote{https://jwst-docs.stsci.edu}) tool. This tool provides simulated \emph{JWST} data for any uploaded spectrum. We created MIRI integral field units (IFU) data cubes for the 12 wavelength ranges (three channels and four wavelength ranges per channel) and NIRSpec IFU data cubes using G140H, G235H, and G395H filters. We uploaded one of the SEDs produced by [Nenkova08] to produce the simulated data cube for NGC\,1068 using a point-like source as the proposed model distribution. In order to get flux-calibrated spectra, we also produced the same data cubes for a flat spectrum (i.e. with a known constant flux across the wavelength range) to produce a wavelength dependent count-to-flux conversion. We created the simulated spectra by multiplying the simulated [Nenkova08] SED by this count-to-flux conversion. 

An schematic view of the procedure to produce and fit synthetic spectra can be seen in Fig.\,\ref{fig:flowchart_syntheticspectra}. Our synthetic spectra include several instrumental setups as a combination of four instruments (GTC/CanariCam, \emph{Spitzer}/IRS, \emph{JWST}/MIRI, and \emph{JWST}/(MIRI + NIRSpec) and four sensitivities ($\rm{F(12\mu m)}$ = 30\,mJy, 100\,mJy, 300\,mJy, and 10\,Jy). We use the six AGN dust models alone and a combination of AGN dust models and circumnuclear contributors (stellar and/or ISM contributors). The stellar contribution is set to 10\%, 50\%, and 100\% of the AGN at 5\,$\rm{\mu m}$, and 100\% of the AGN at 10\,$\rm{\mu m}$. The ISM contribution is set to 10\%, 50\%, and 100\% of the AGN at 30\,$\rm{\mu m}$. We also explore a combination of 10\%, 50\%, and 100\% of stellar and ISM contribution at 5 and 30\,$\rm{\mu m}$, respectively. Note that only the $\rm{F(12\mu m) = 300\,mJy}$ was used when the circumnuclear contributors were added to the synthetic spectra. We produce 96,000 SEDs using random realization of the parameters for the synthetic spectra. This includes 1,000 SEDs per model and instrumental setup. For the only-AGN synthetic spectra we also produce 147,744 SEDs using three fixed values per model, including 243, 729, and 6,561 SEDs for the 5, 6, and 8 parameter models. These three values are chosen as 2/5, 3/5, and 5/5 of the parameter space. Moreover, we generate 29,600 SEDs using random realizations of the parameters when AGN dust models are combined with circumnuclear contributors. This includes 100 SEDs per model and 100 additional SEDs for [Fritz06] and [Hoenig17] to explore if the limited number of SEDs affects our results. Overall we produced 273,344 synthetic SEDs and $\rm{\sim}$1,5 million spectral fits. This took over 3 months of computational time in three 64-cores, one 32-cores, and two 8-cores dedicated servers\footnote{Servers: HyperCat (IAA-CSIC), IRyAGN1, IRyAGN2, Galaxias, Posgrado04, and Arambolas (IRyA-UNAM).}.

\begin{figure*}[!ht]
\begin{flushleft}
\includegraphics[width=0.75\columnwidth]{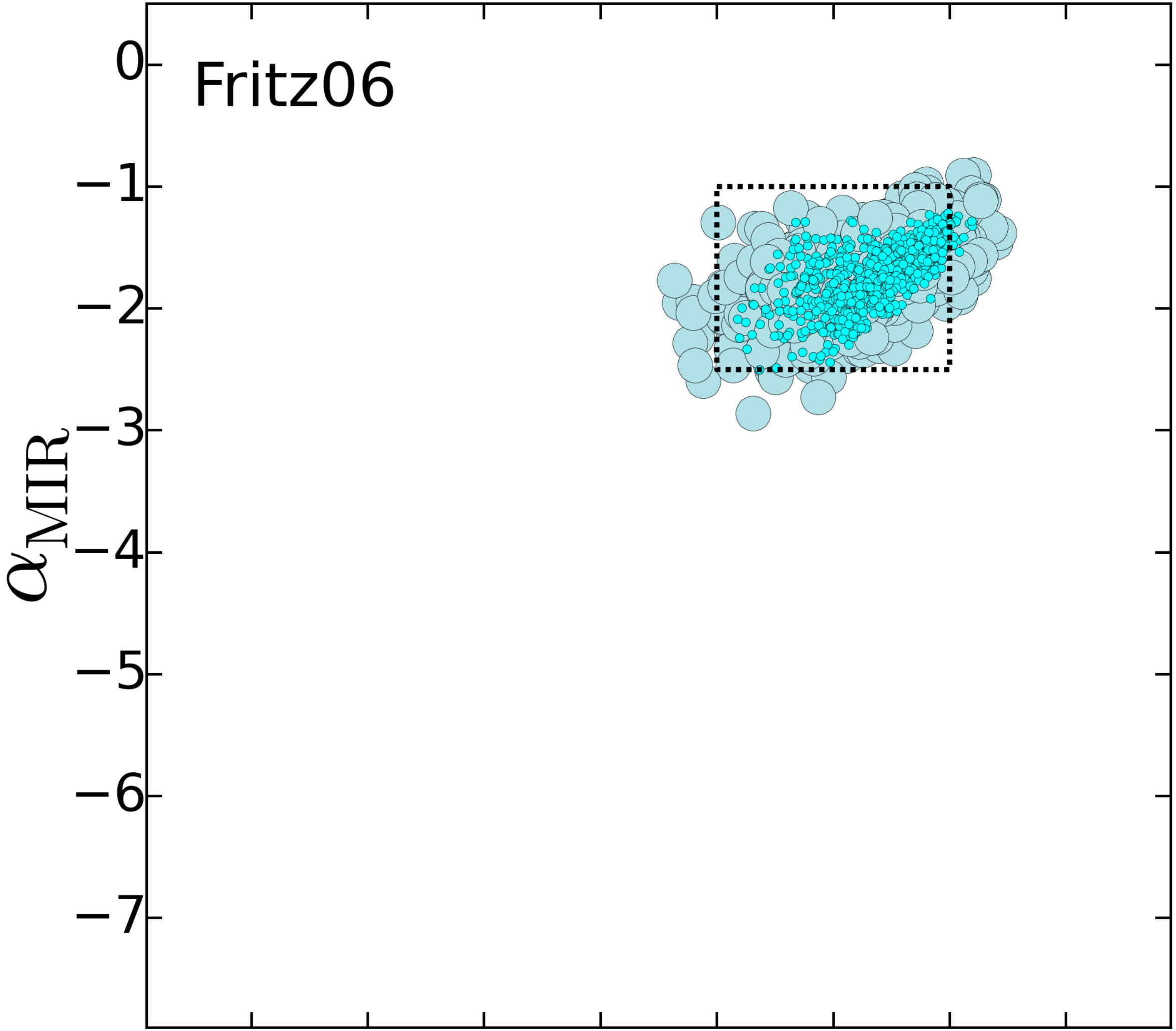}
\includegraphics[width=0.66\columnwidth]{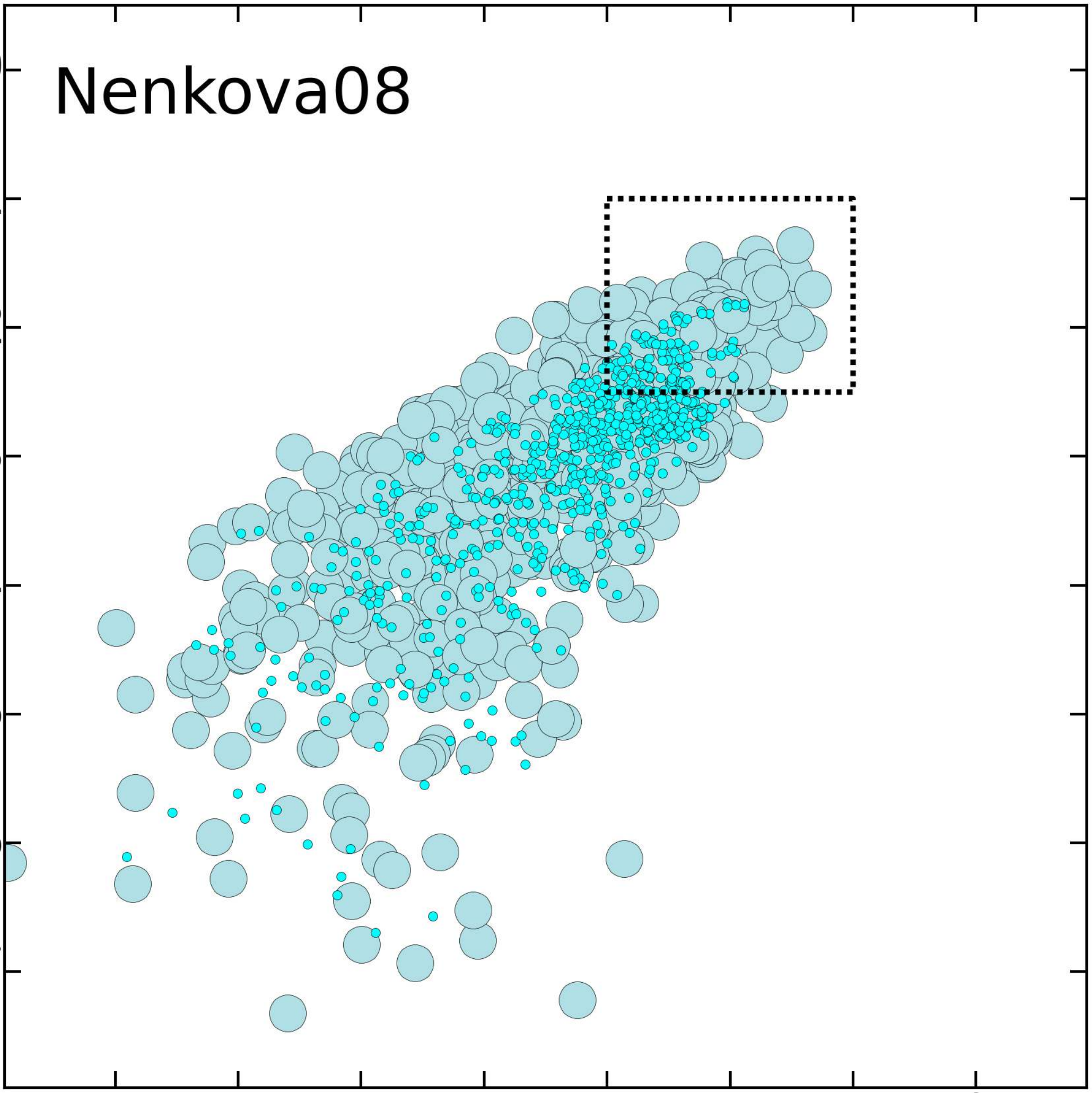}
\includegraphics[width=0.66\columnwidth]{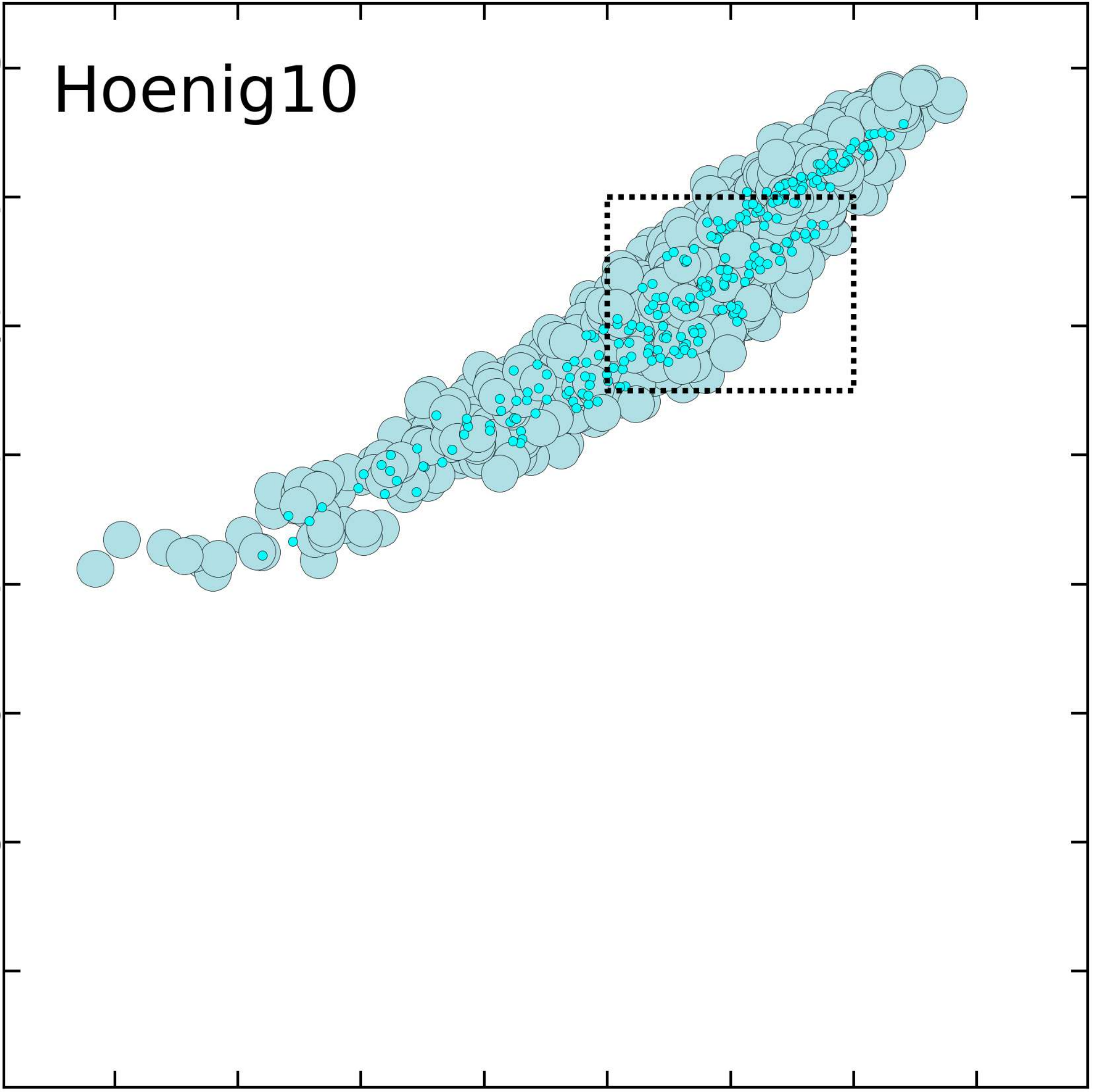}
\includegraphics[width=0.75\columnwidth]{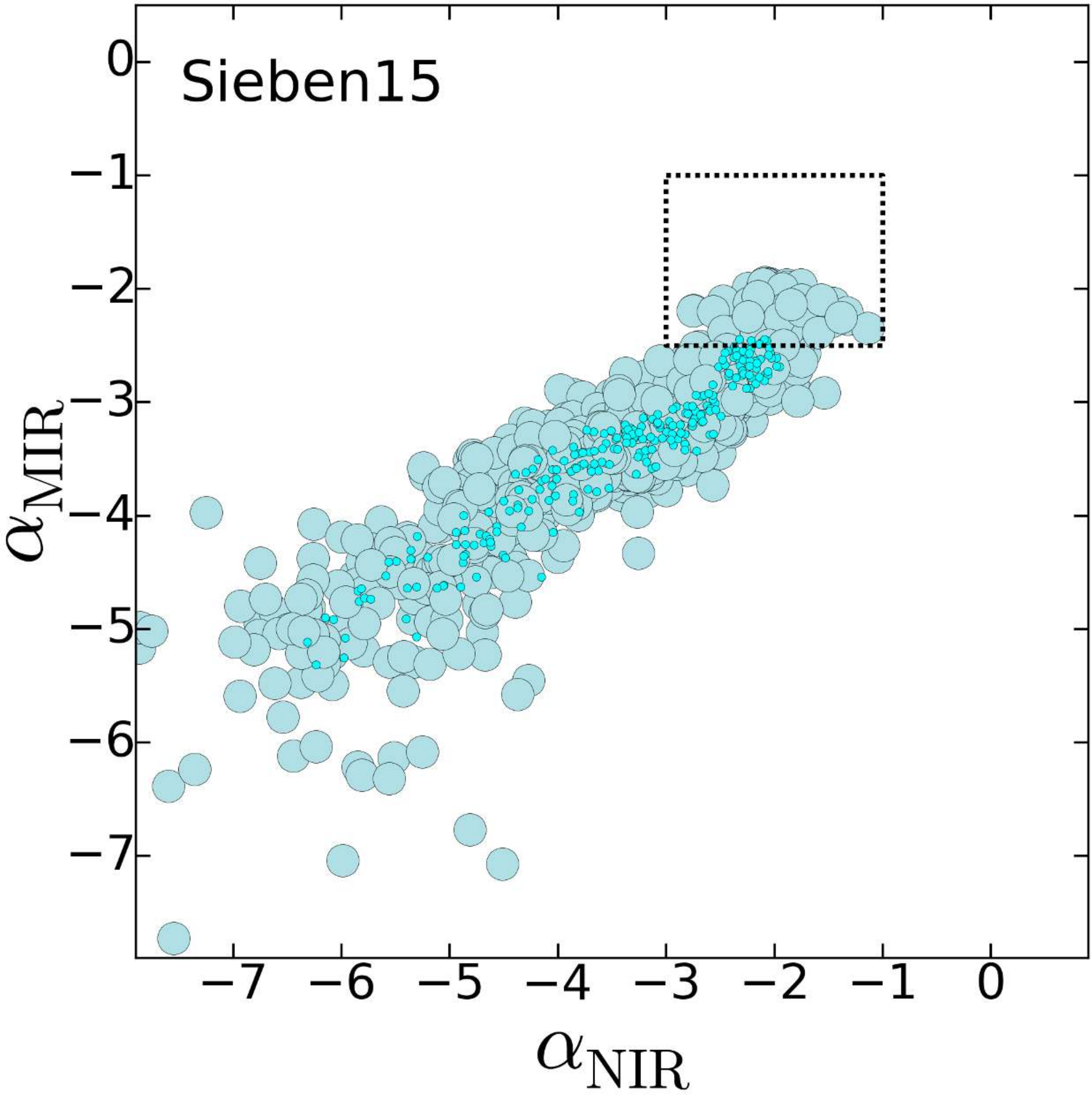}
\includegraphics[width=0.66\columnwidth]{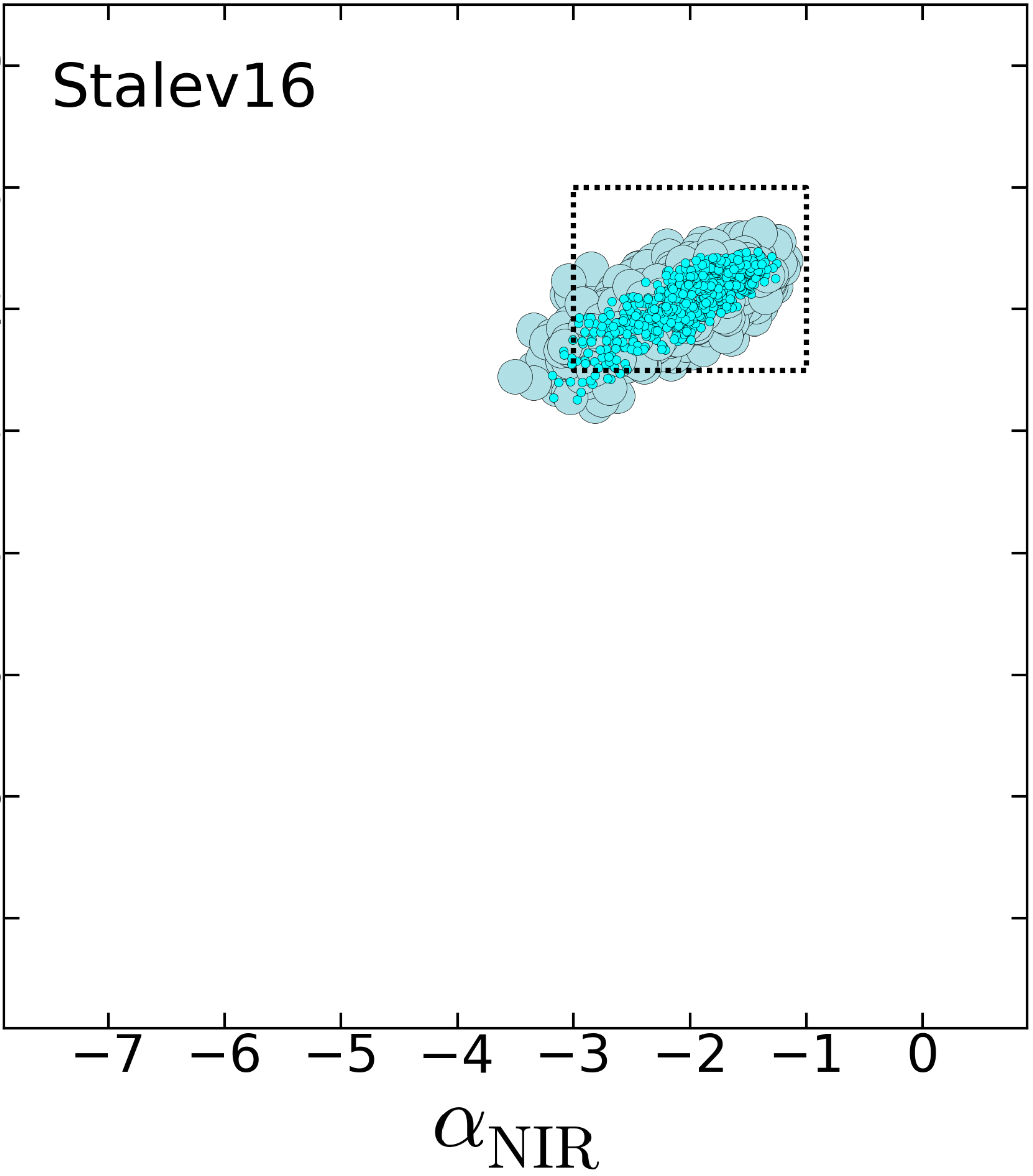}
\includegraphics[width=0.66\columnwidth]{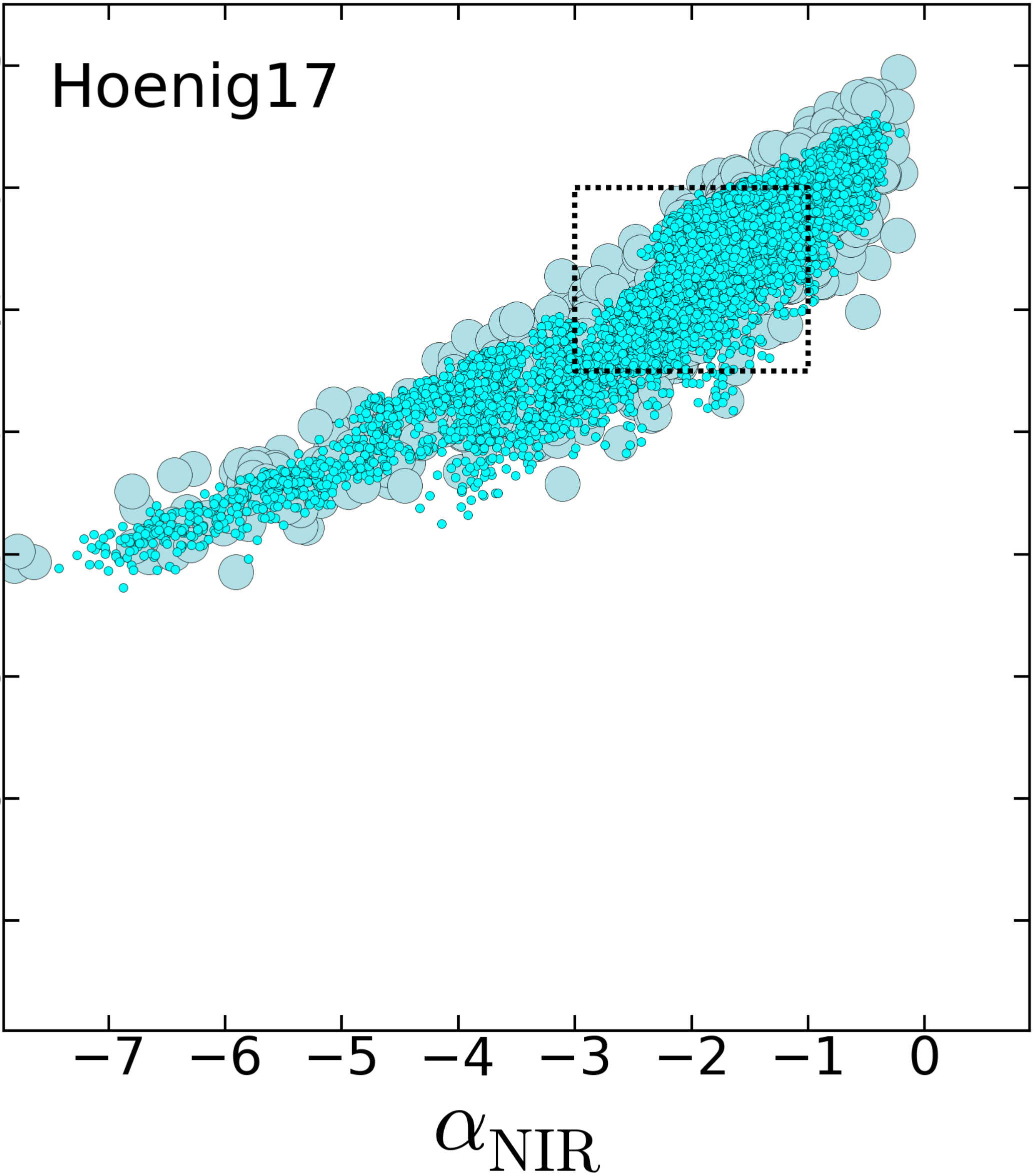}
\end{flushleft}
\begin{center}
\caption{Spectral slope computed as the flux ratio between 14 and 7.5$\rm{\mu m}$ ($\alpha_{MIR}$) versus the spectral slope computed as the flux ratio between the 7.5 and 5.5$\rm{\mu m}$ ($\alpha_{NIR}$). Synthetic spectral results using 1,000 random parameters are shown with turquoise circles and fixed step values at 2/5,  3/5, and 4/5 of each parameter space are shown with turquoise dots. We highlight with a dotted box the area where most of the SEDs fall for [Fritz06] in all the panels for comparison purposes.}
\label{fig:genfit1}
\end{center}
\end{figure*}

\begin{figure*}[!ht]
\begin{flushleft}
\includegraphics[width=0.75\columnwidth]{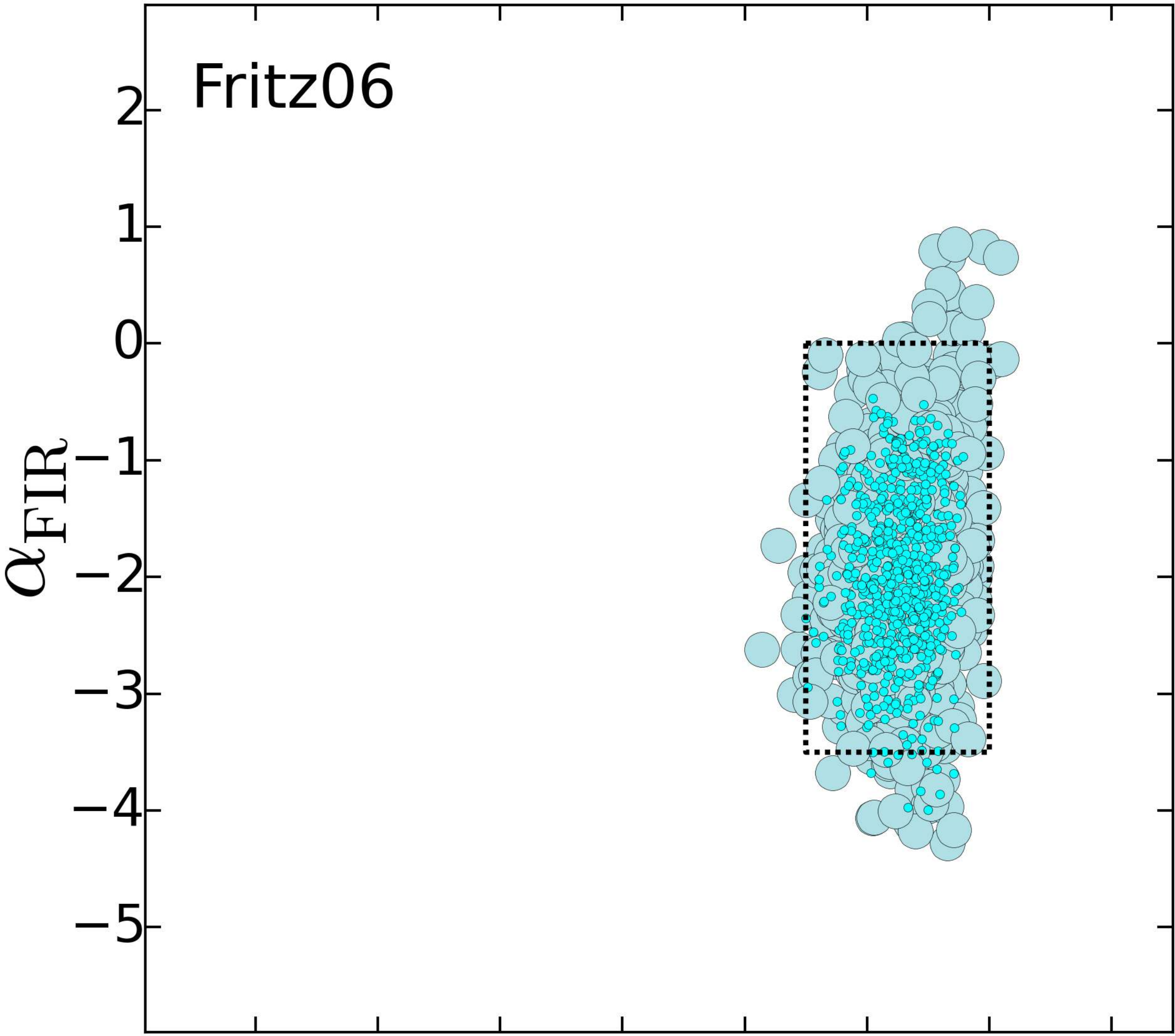}
\includegraphics[width=0.66\columnwidth]{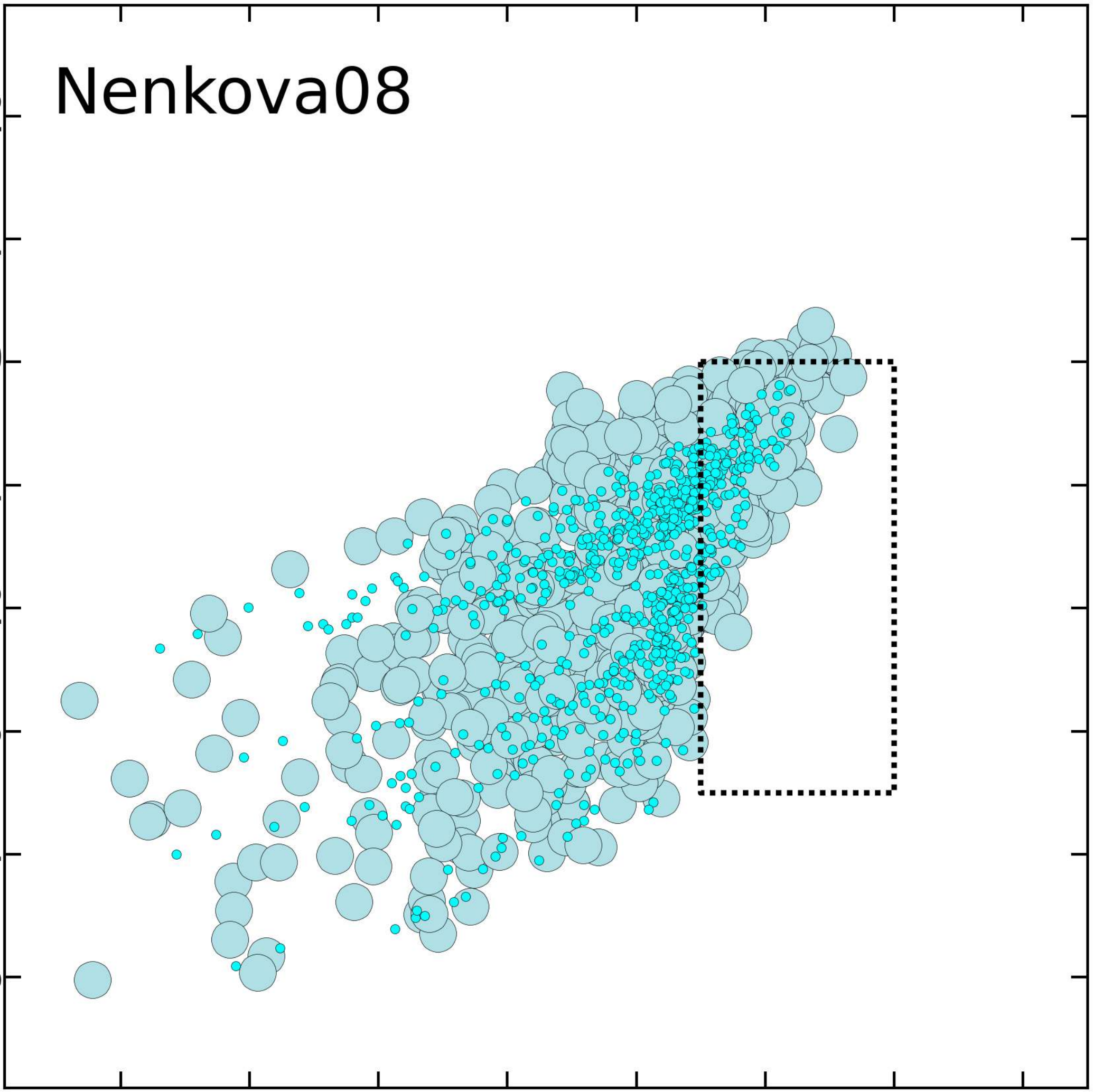}
\includegraphics[width=0.66\columnwidth]{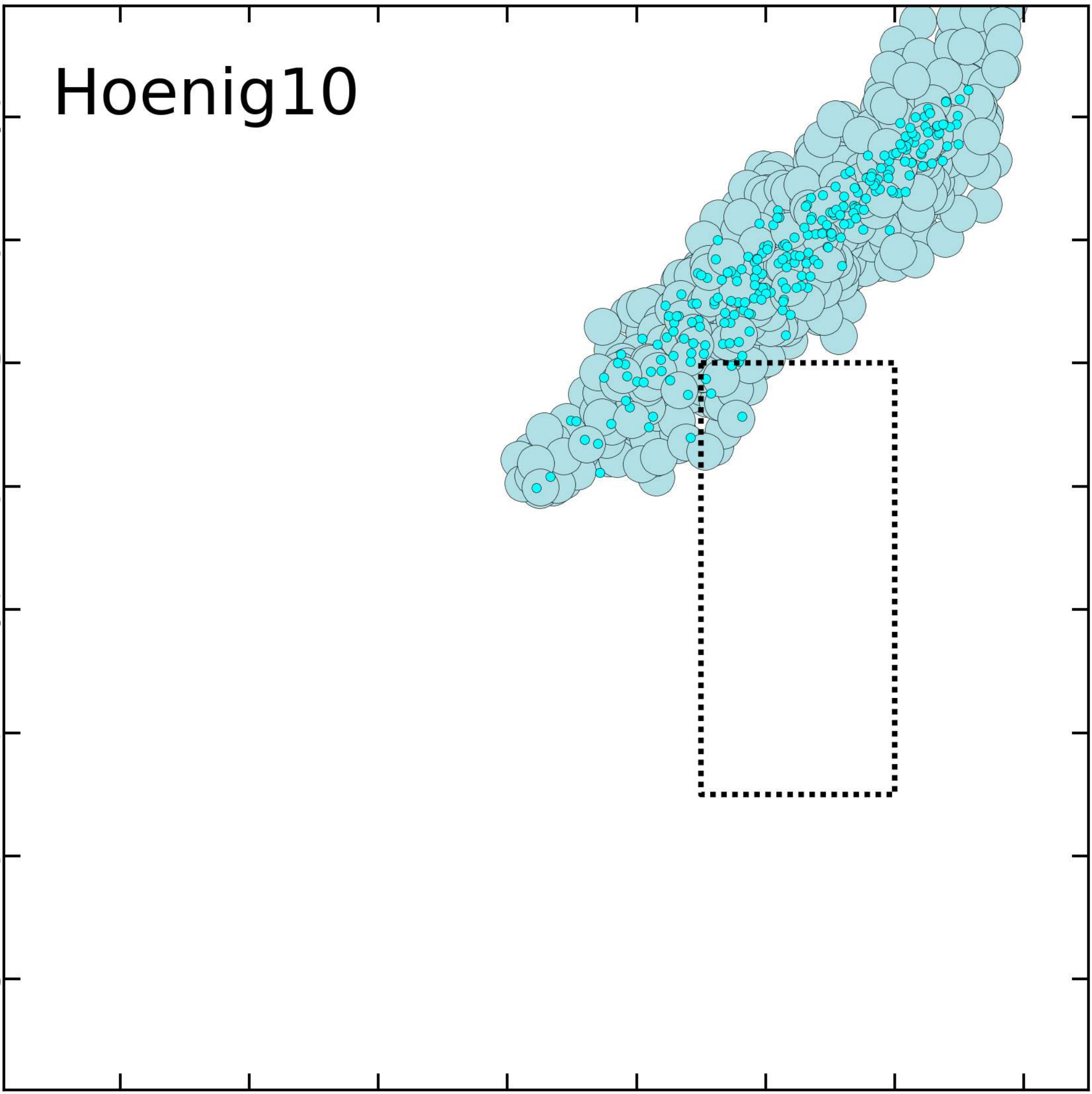}
\includegraphics[width=0.75\columnwidth]{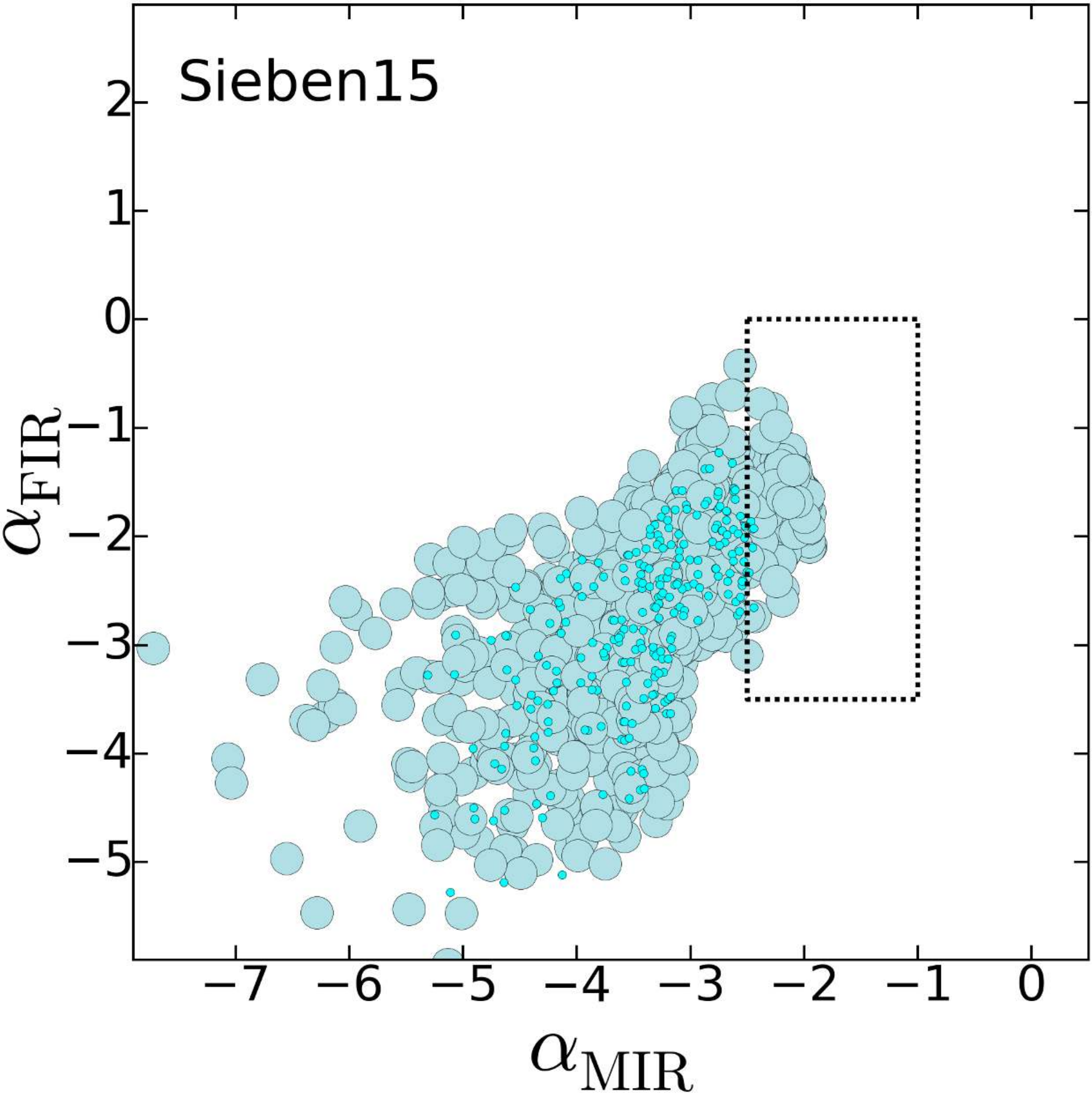}
\includegraphics[width=0.66\columnwidth]{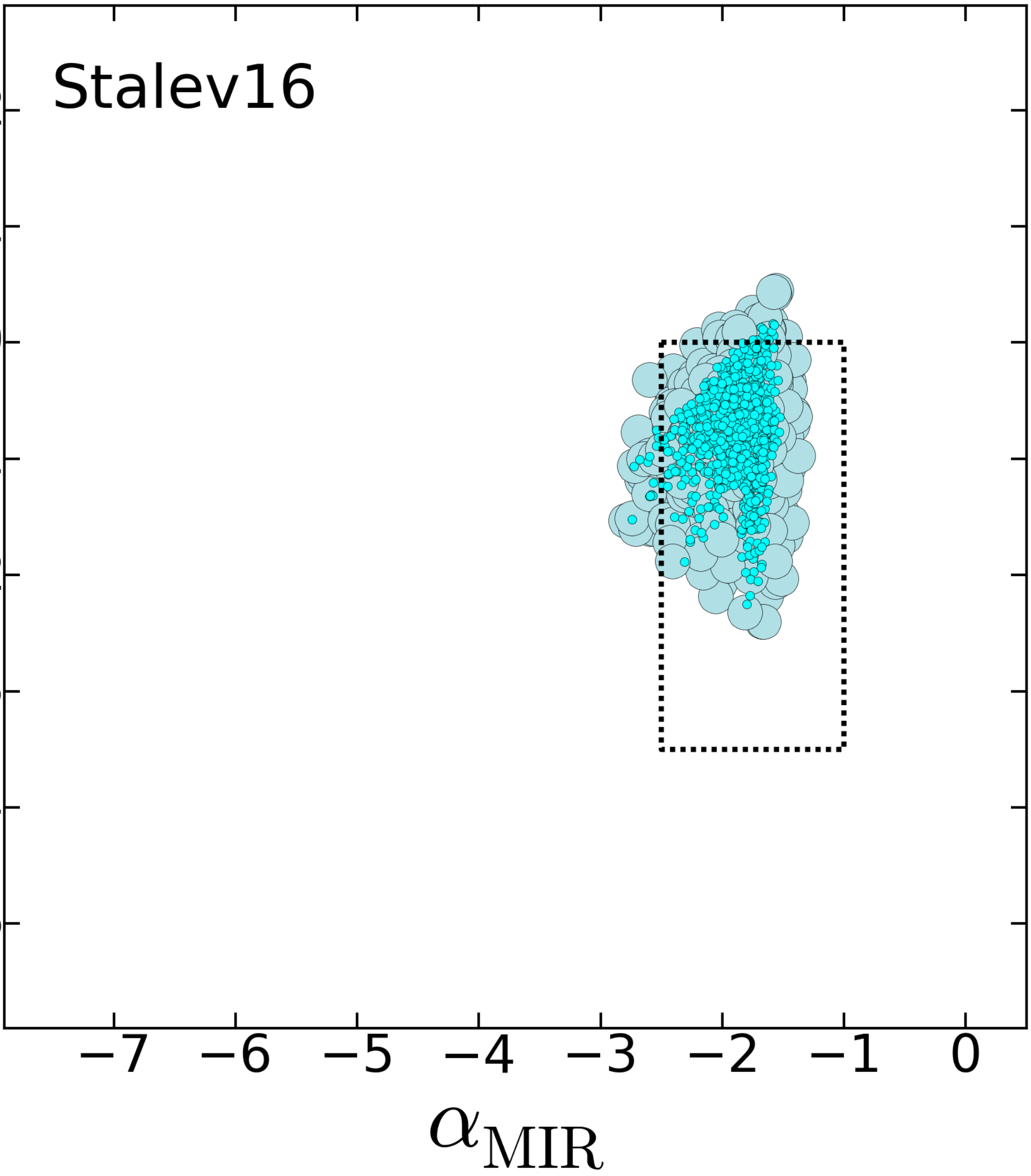}
\includegraphics[width=0.66\columnwidth]{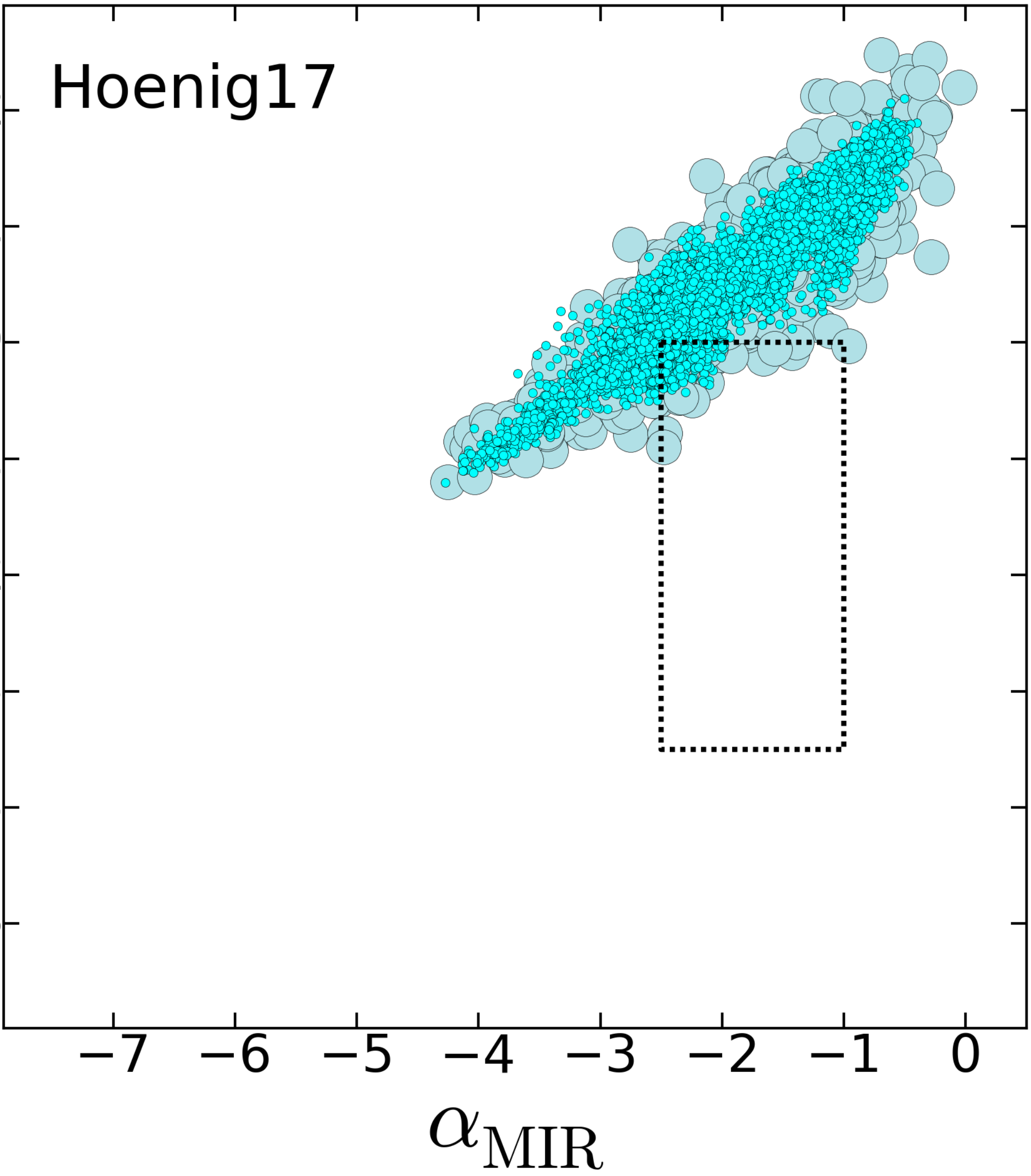}
\end{flushleft}
\begin{center}
\caption{Spectral slope computed as the flux ratio between 30 and 25$\rm{\mu m}$ ($\alpha_{FIR}$) versus the spectral slope computed as the flux ratio between the 14 and 7.5$\rm{\mu m}$ ($\alpha_{MIR}$). Symbols as in Fig.\,\ref{fig:genfit1}.  We highlight with a dotted box the area where most of the SEDs fall for [Fritz06] in all the panels for comparison purposes.}
\label{fig:genfit2}
\end{center}
\end{figure*}

\begin{figure*}[!ht]
\begin{flushleft}
\includegraphics[width=0.78\columnwidth]{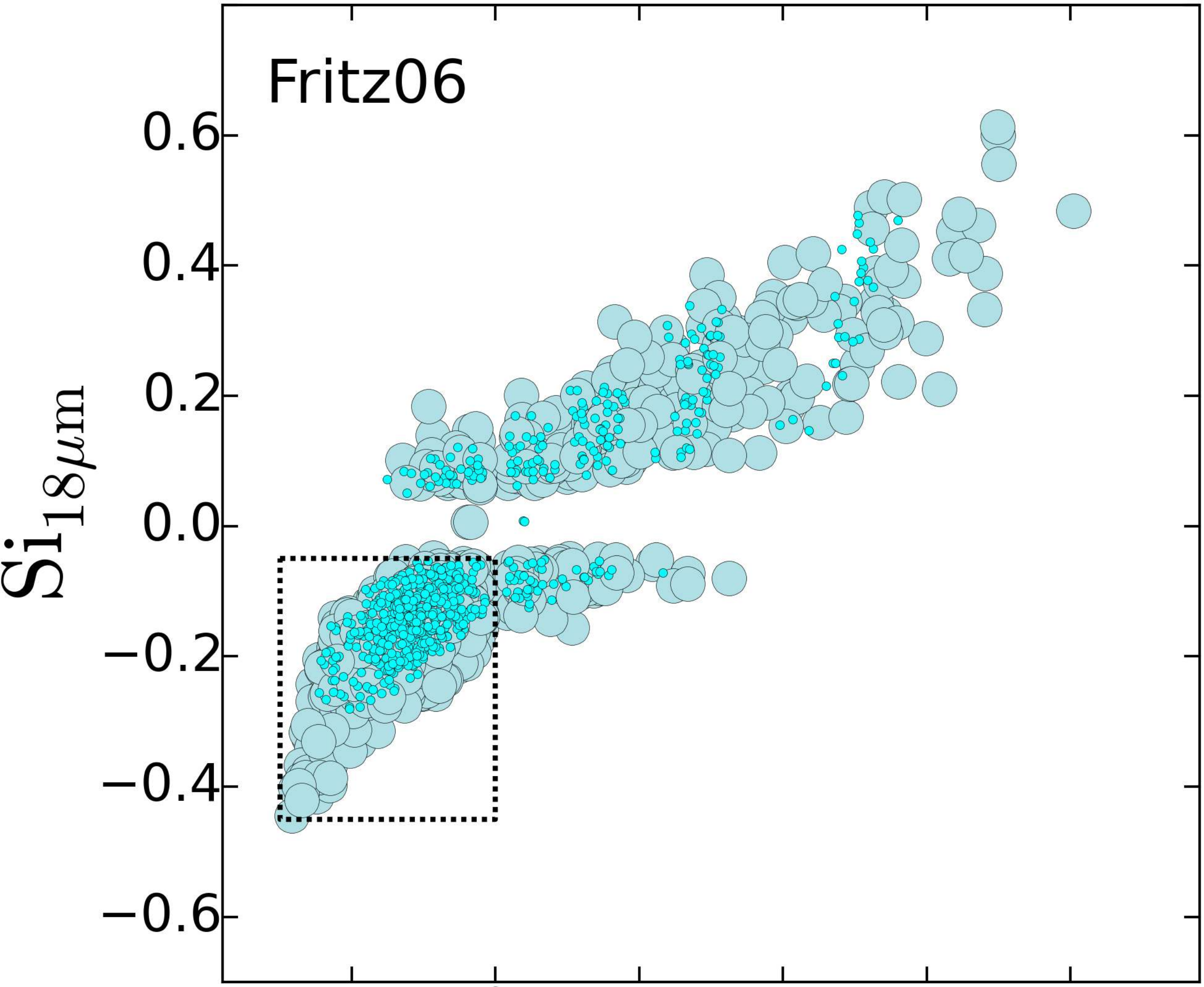}
\includegraphics[width=0.64\columnwidth]{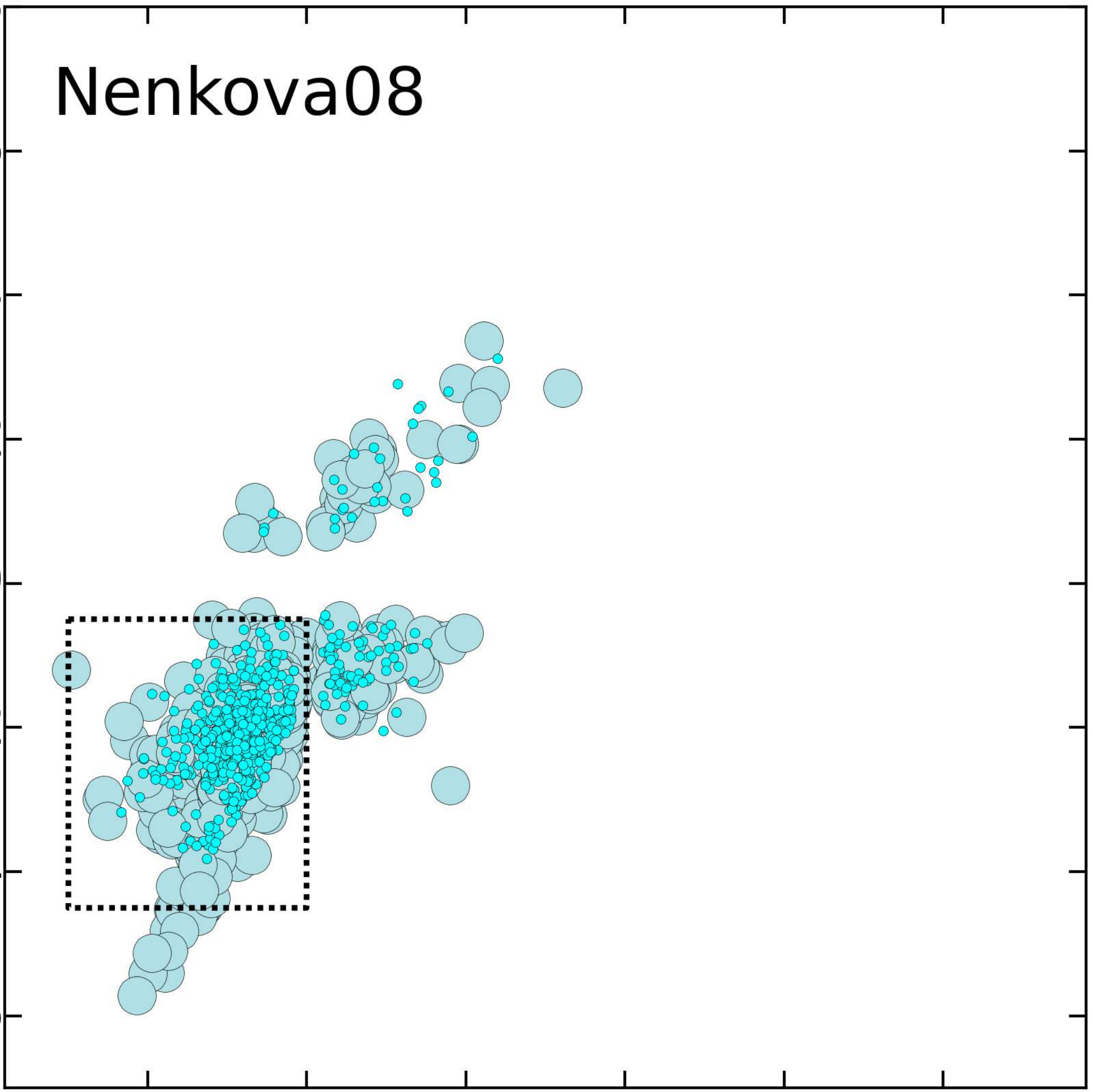}
\includegraphics[width=0.64\columnwidth]{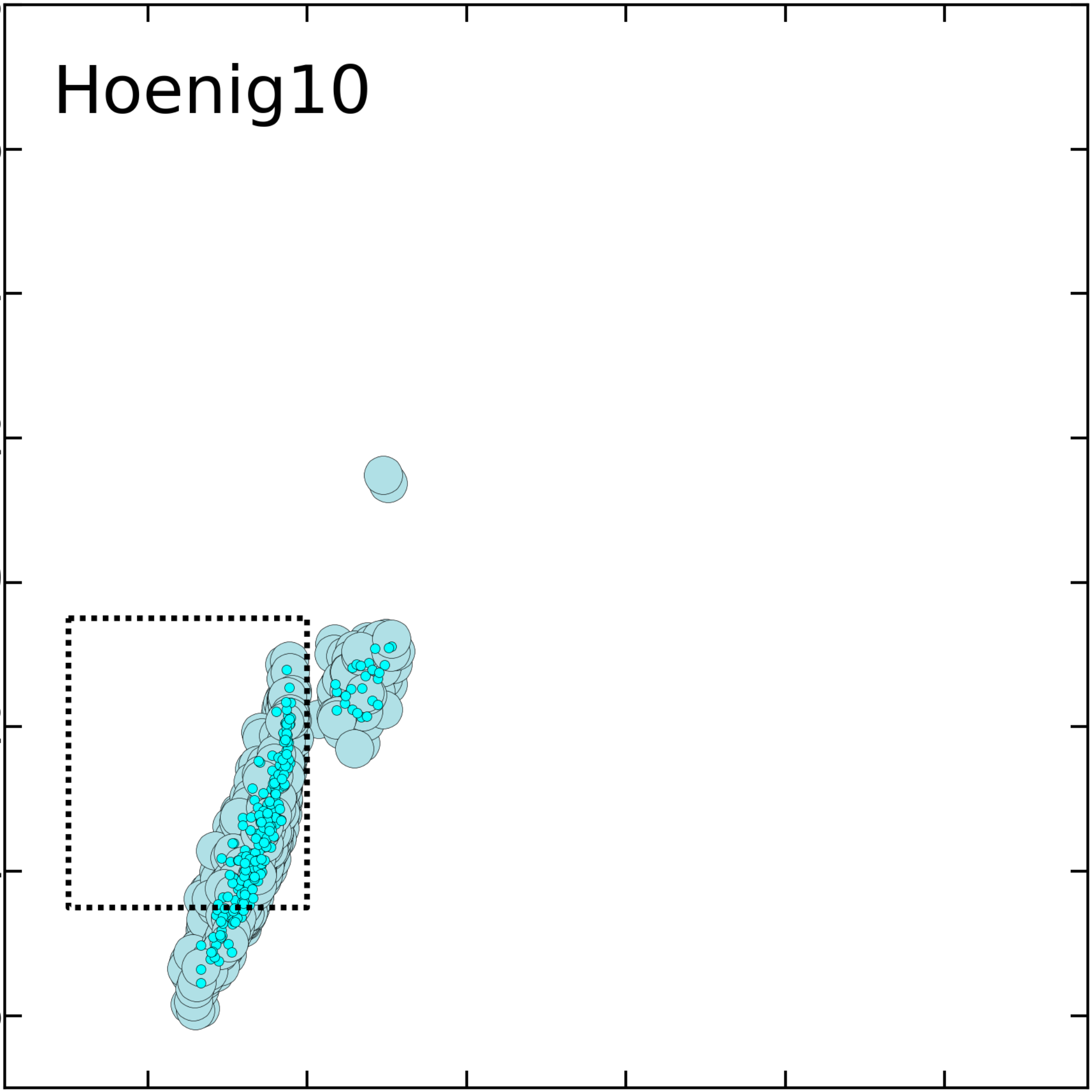}
\includegraphics[width=0.78\columnwidth]{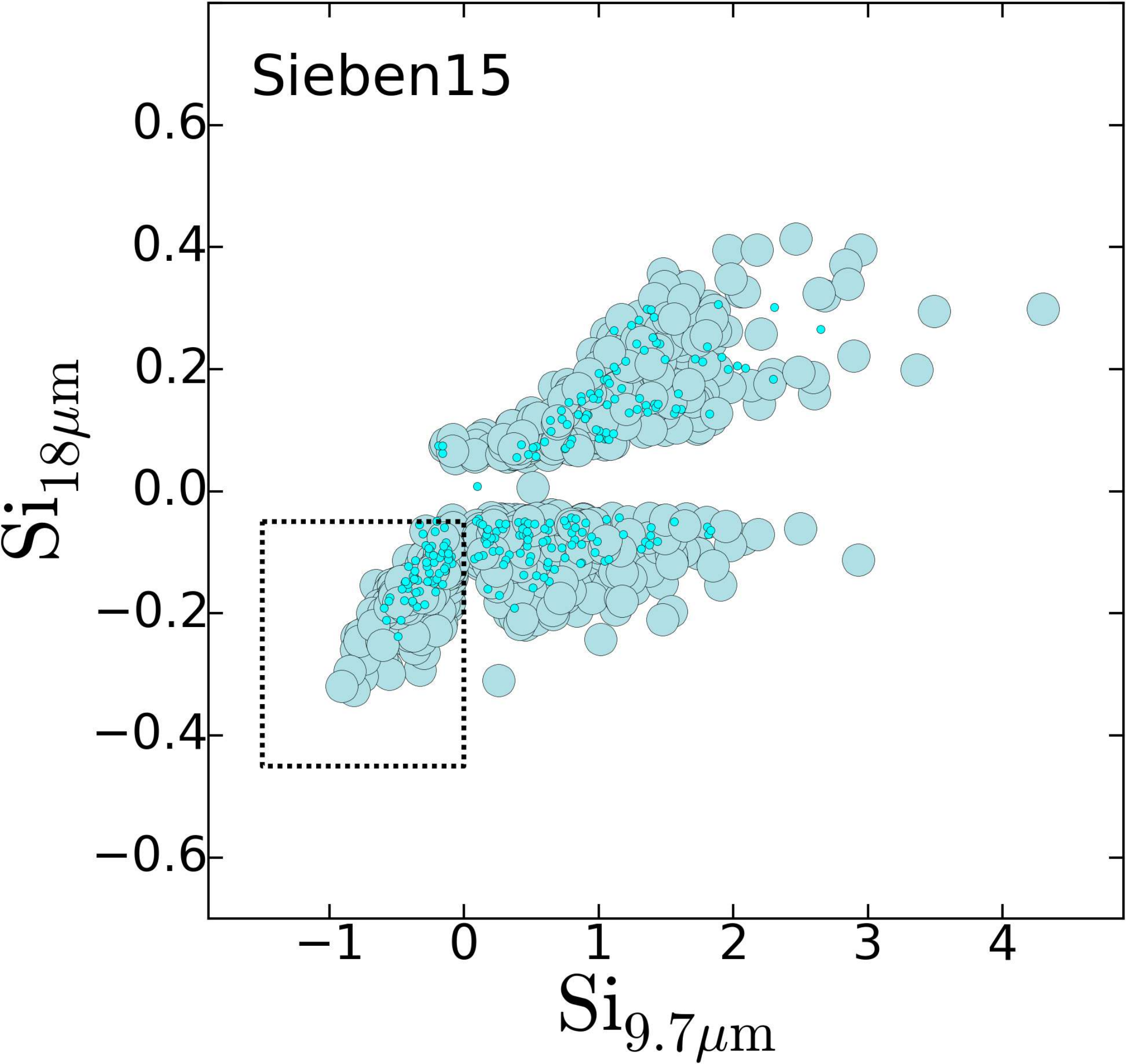}
\includegraphics[width=0.64\columnwidth]{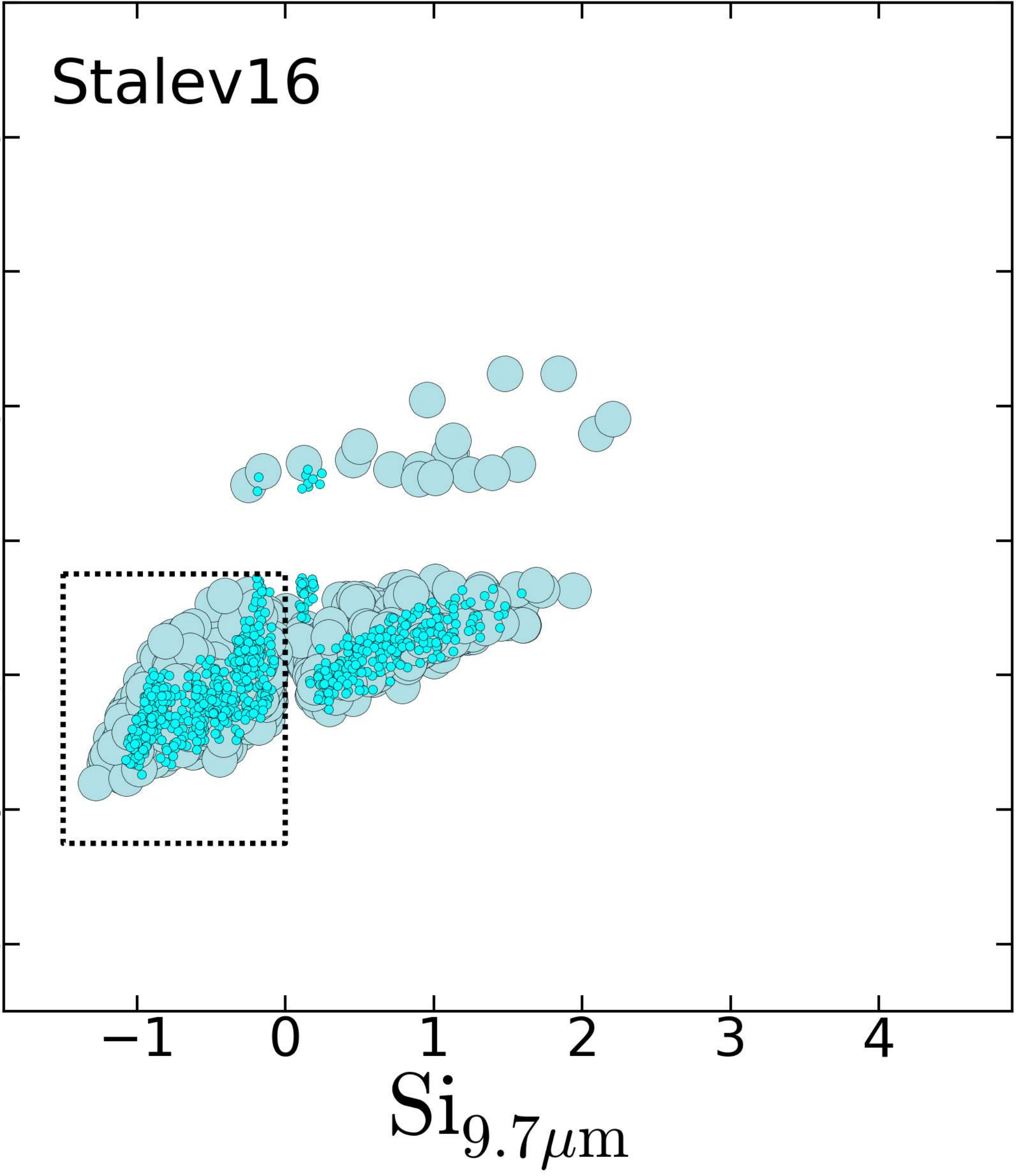}
\includegraphics[width=0.64\columnwidth]{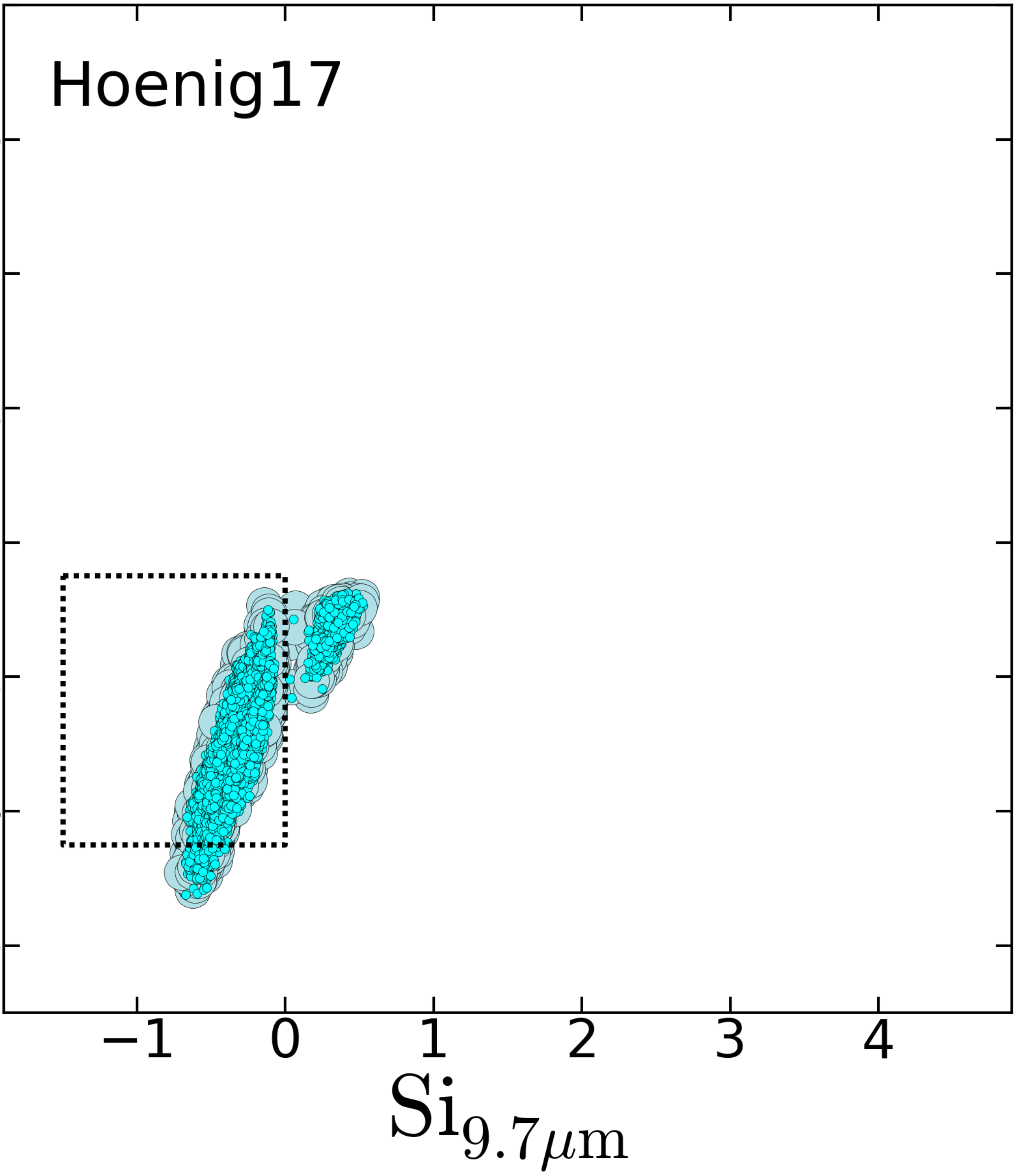}
\end{flushleft}
\begin{center}
\caption{18\,$\rm{\mu m}$ Silicate feature strength versus 9.7\,$\rm{\mu m}$ Silicate feature strength. Symbols as in Fig.\,\ref{fig:genfit1}.  We highlight with a dotted box the area where most of the SEDs with emission features fall for [Fritz06] in all the panels for comparison purposes.}
\label{fig:genfit3}
\end{center}
\end{figure*}

\section{Results} \label{sec:results}

We use the synthetic spectra to study the spectral shapes produced by these models (Section\,\ref{sec:spectralshape}), to test the accuracy of the parameter determination per model (Section\,\ref{sec:Modelconstrain}), if these models can be distinguished among them (Section\,\ref{sec:ModelvsModel}), and how the recovery of the parameters is affected by unresolved circumnuclear contributors to the spectra (Section\,\ref{sec:CircumModelconstrain}).

\begin{figure*}[!ht]
\begin{flushright}
\includegraphics[width=0.7\columnwidth]{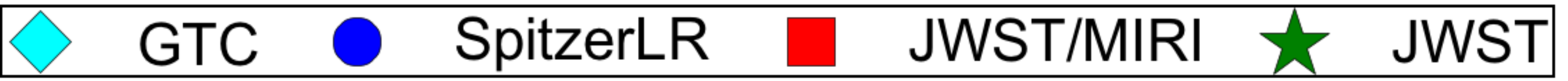}~~~~~~~
\end{flushright}
\begin{center}
\includegraphics[width=2.0\columnwidth]{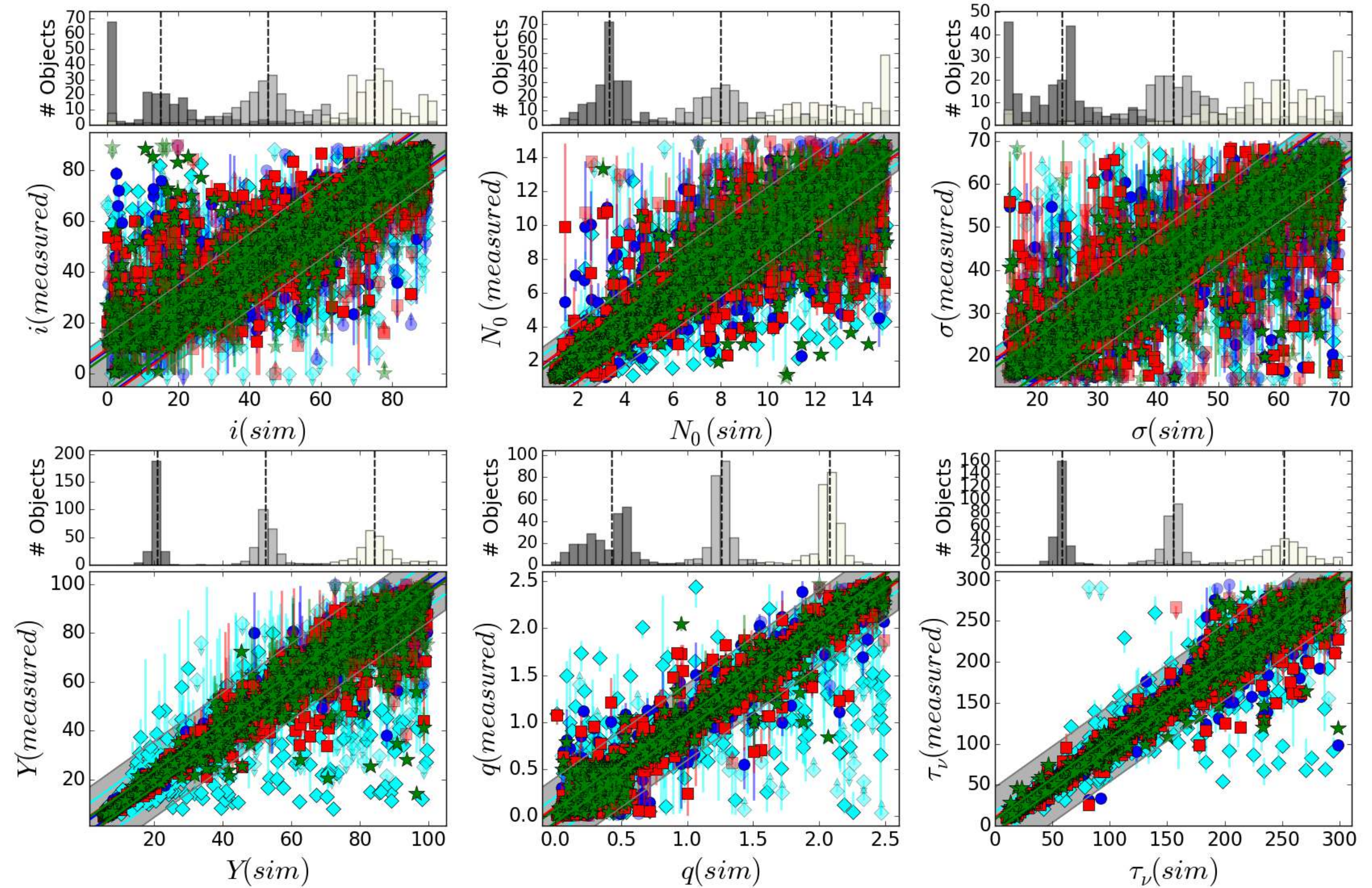}
\caption{ Histograms of the measured values when using fixed parameters for the simulated spectra (top panels) and measured versus simulated value when using random numbers for the parameters (bottom panels). Each two of these panels show the viewing angle, $i$ (top-left): equatorial number of clouds, $N_{0}$ (top-center), half angular width of the torus, $\sigma$ (top-right), ratio between the outer and inner radius of the torus, $Y$ (bottom-left), radial steepness of the cloud distribution, $q$ (bottom-center), and opacity of the clouds, $\tau_{v}$ (bottom-right) for [Nenkova08] model. The data are simulated with a S/N$\rm{\sim 40-60}$, corresponding to a source with a flux of $\rm{f(12\mu m)=300\, mJy}$. The results for the GTC/CanariCam, \emph{Spitzer}/IRS, \emph{JWST}/MIRI, and \emph{JWST}/(MIRI+NIRSpec) spectra are shown using cyan diamonds, blue circles, red squares, and green stars, respectively. The gray area shows the confidence error within 15\%. Cyan long-dashed, blue dot-dashed, red short-dashed, and green dotted lines show the average error using  GTC/CanariCam, \emph{Spitzer}/IRS, \emph{JWST}/MIRI, and \emph{JWST}/(MIRI+NIRSpec) spectra, respectively. 
}
\label{fig:Par2Par}
\end{center}
\end{figure*}

\begin{figure*}[!ht]
\begin{flushleft}
\includegraphics[width=0.69\columnwidth]{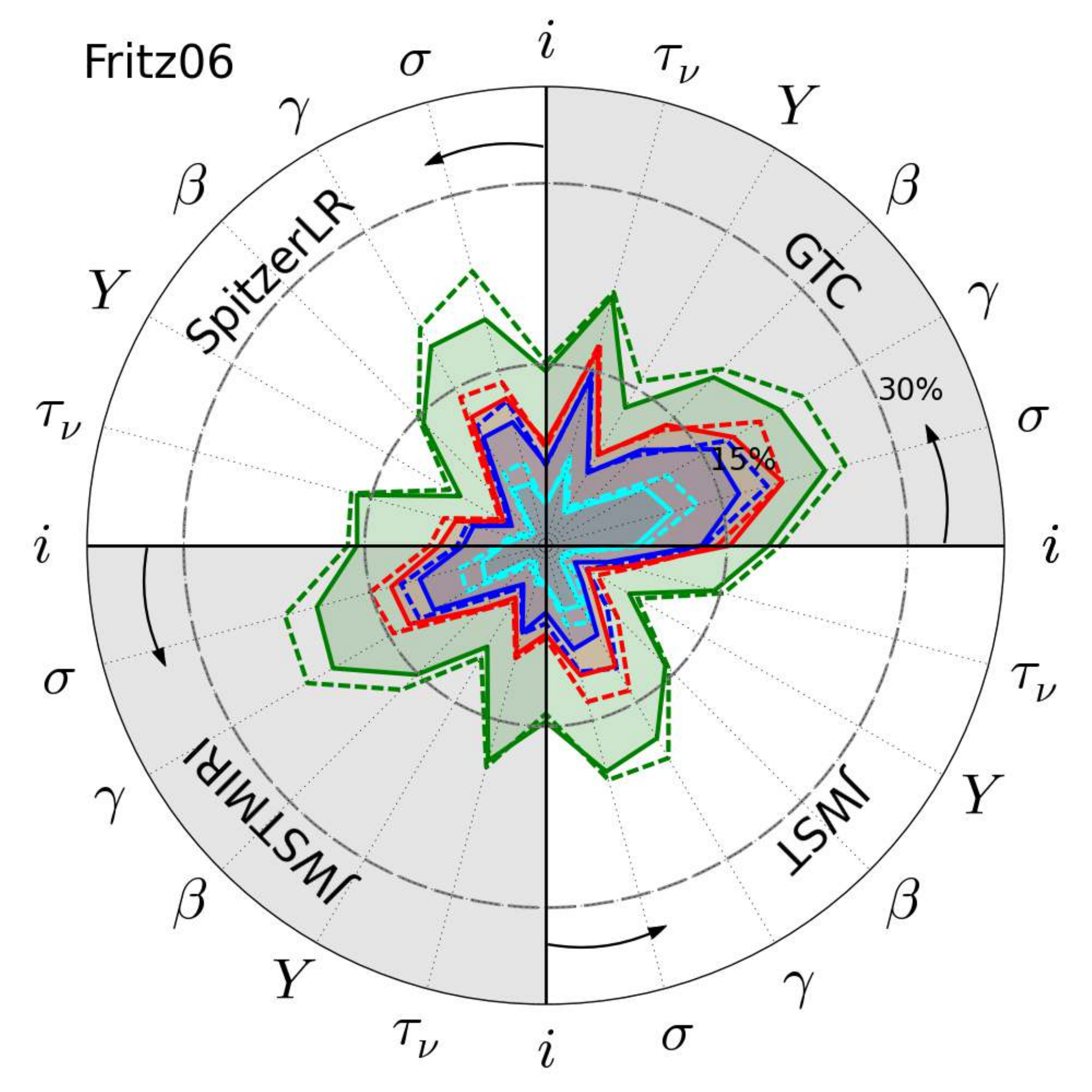}
\includegraphics[width=0.69\columnwidth]{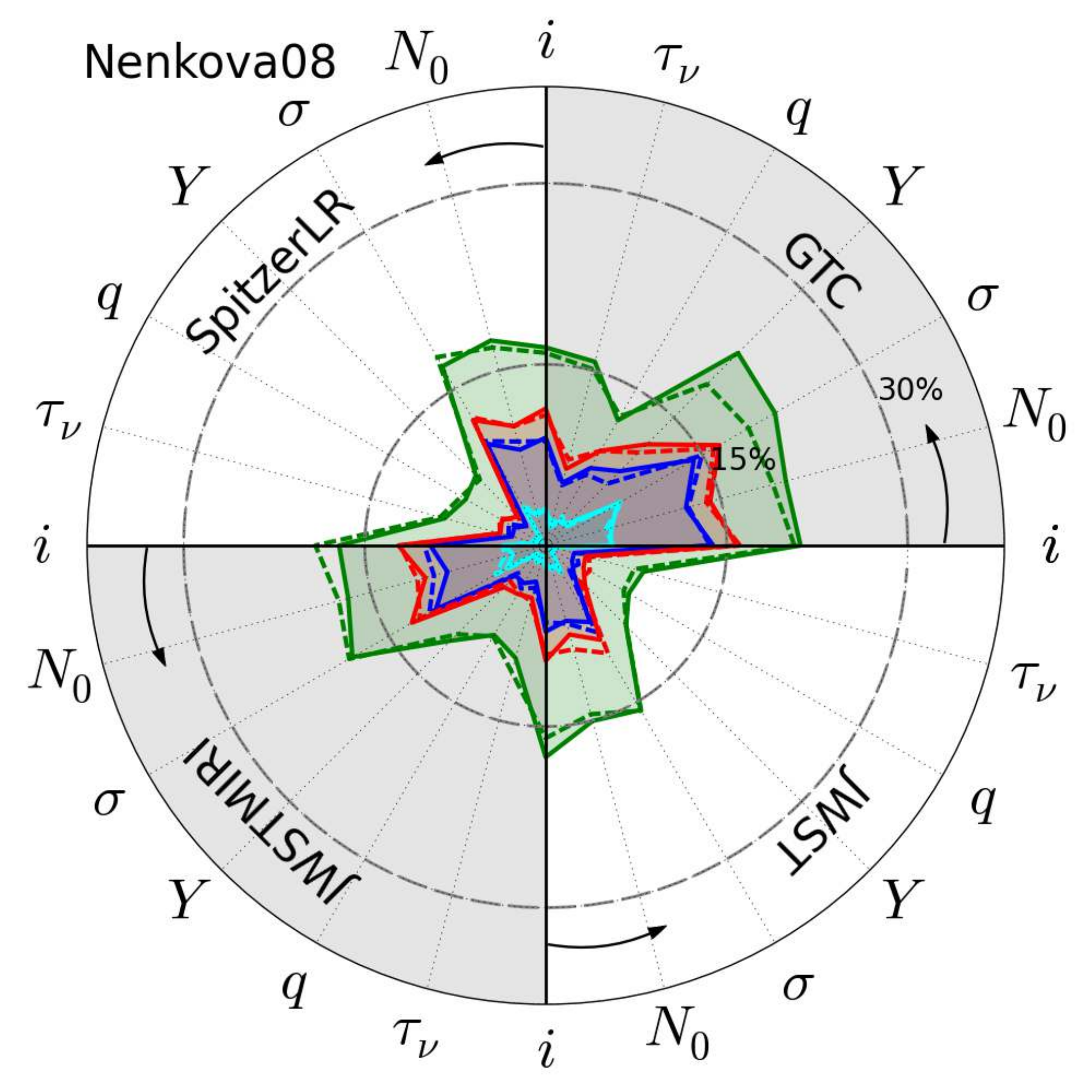}
\includegraphics[width=0.69\columnwidth]{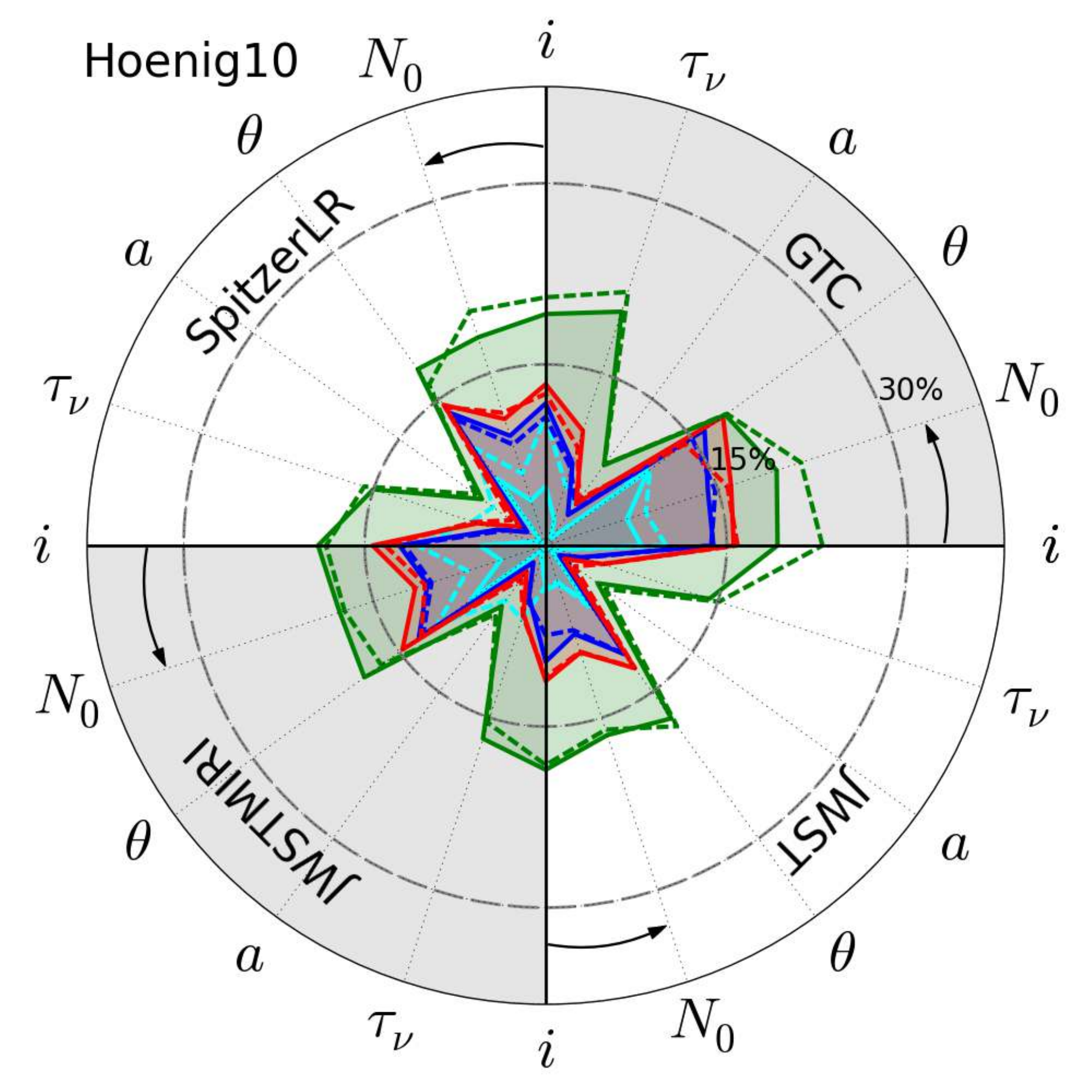}
\includegraphics[width=0.69\columnwidth]{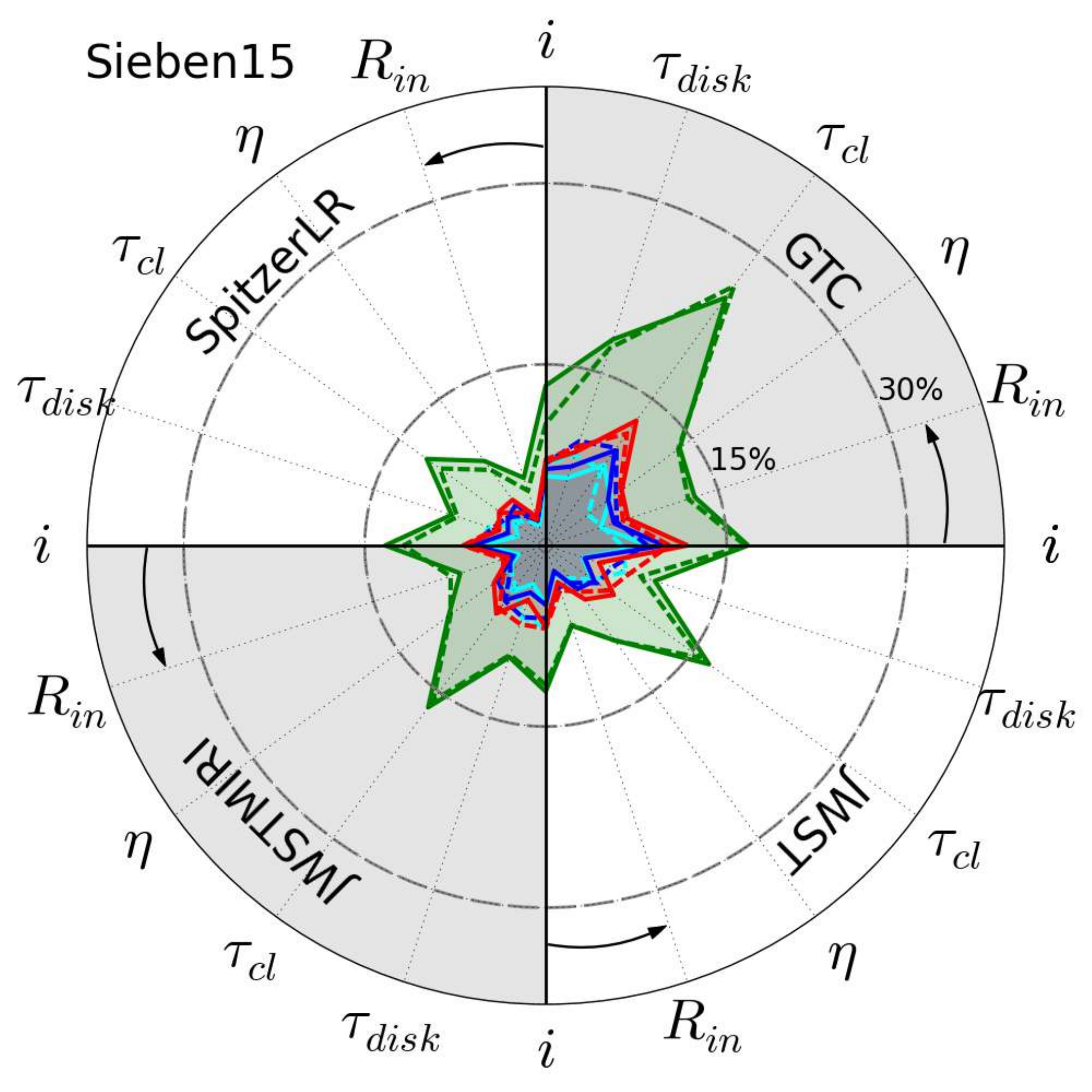}
\includegraphics[width=0.69\columnwidth]{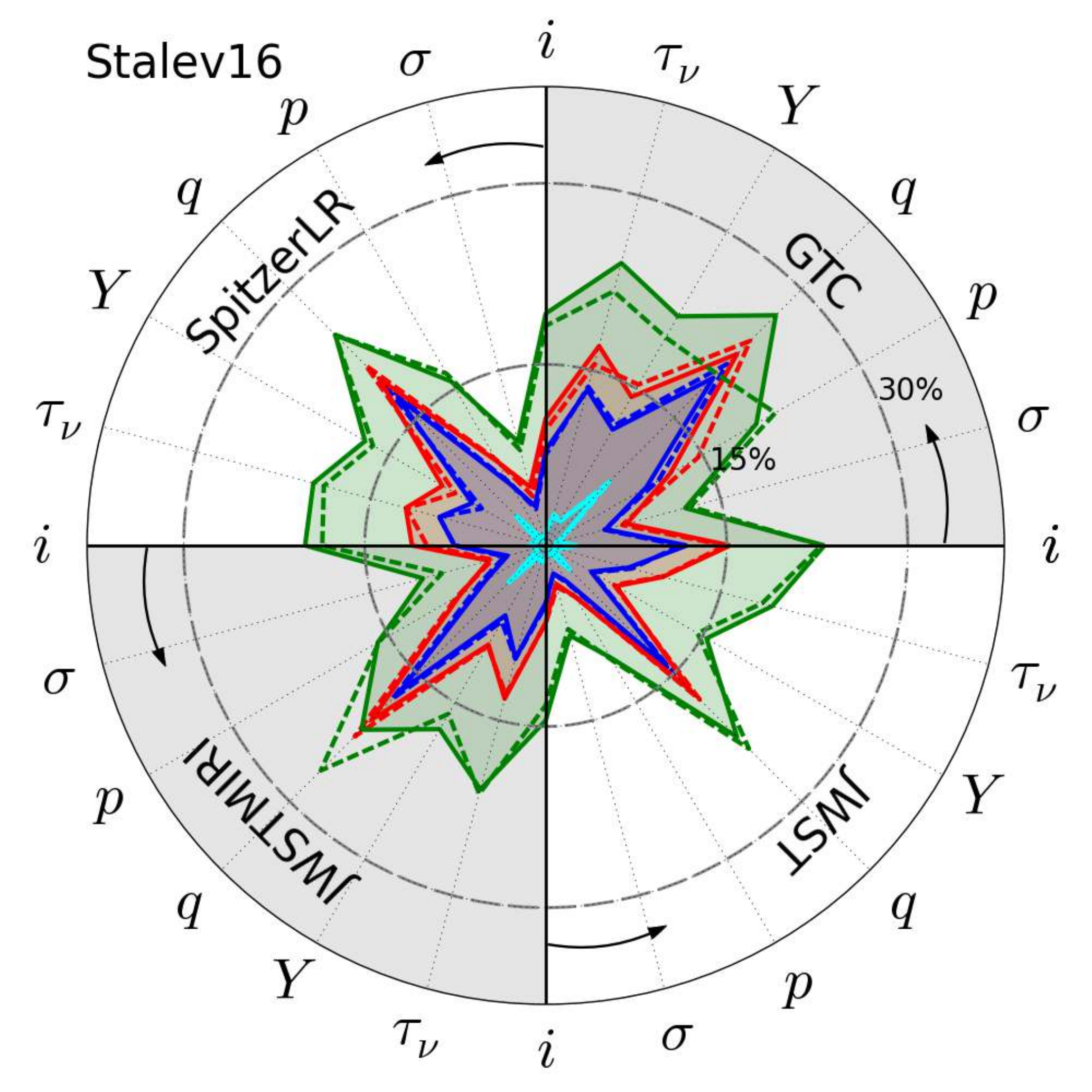}
\includegraphics[width=0.69\columnwidth]{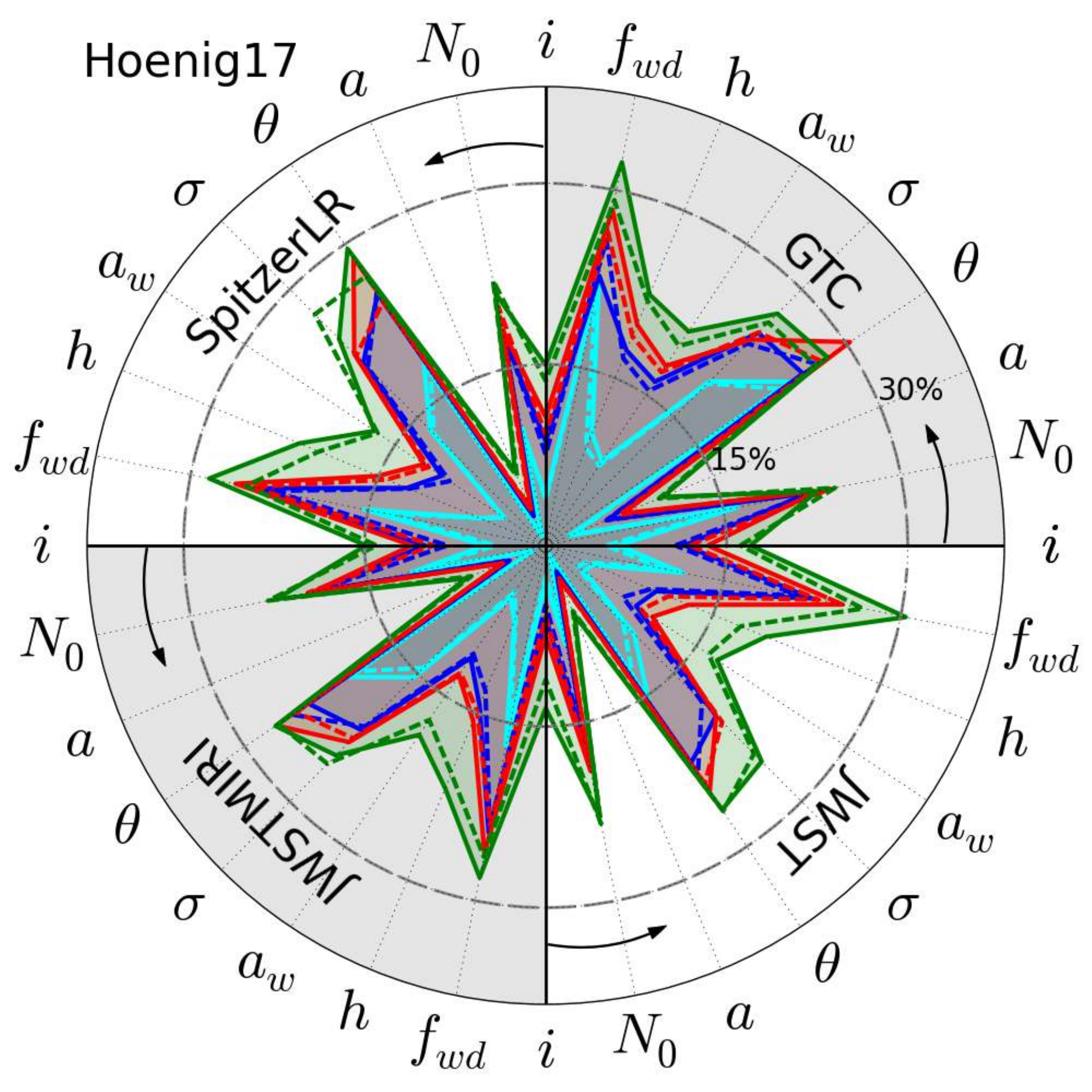}
\caption{Average percentage of error on the parameter estimate per model (each circle) and per instrumental setup (each quarter of the circle). Cyan, dark blue, red, and green lines (and same color shaded areas) link the percentage of error using a simulated spectrum with a $\rm{12\,\mu m}$ flux of $\rm{10\, Jy}$, $\rm{300\, mJy}$,  $\rm{100\, mJy}$, and $\rm{30\, mJy}$, respectively. The arrows shown in counterclock direction indicate the first parameter within the instrument setup. Continuous and dotted lines show the results for 1,000 synthetic spectra with random parameters and that of $\rm{3^N}$ synthetic spectra covering 3 fixed values per parameter (being N the number of parameters). The long-dashed circles highlight the confidence error within 15\% and 30\% of the parameter range.}
\label{fig:ParError}
\end{flushleft}
\end{figure*}

\subsection{Spectral shapes}\label{sec:spectralshape}

We produced a set of 1,000 spectra per model using random parameters. The synthetic SEDs with fixed parameters will be used to ensure a good coverage of the parameter space. For each of them, we used the {\sc fakeit} task to produce synthetic spectra for GTC/CanariCam, \emph{Spitzer}/IRS, \emph{JWST}/MIRI, and \emph{JWST}/(MIRI + NIRSpec) instruments that can be fitted within XSPEC. We scaled the normalization parameter to four 12\,$\rm{\mu m}$ fluxes ($\rm{\sim}$30 mJy, 100 mJy, 300 mJy, and 10\,Jy) to simulate different S/N ratio (equivalent to S/N$\rm{\sim}$ 3-10, 20-30, 40-60, and 100-150, respectively). Therefore, we produced 12,000 SEDs per model using random realizations (1,000 SEDs per model, four instruments, and three S/N levels) and 9,234 SEDs using fixed steps for the six models studied (times the 12 instrumental setups, see Fig.\,\ref{fig:flowchart_syntheticspectra}). These SEDs are used to study the accuracy of the parameter determination in Section \ref{sec:Modelconstrain}. 

We compute the following spectral parameters to compare the spectral shapes provided by these models, following previous works on the infrared spectra of AGN \citep[e.g.][]{Hernan-Caballero15,Garcia-Gonzalez17,Hoenig17}, although optimized to the spectral range provided by \emph{Spitzer}/IRS. Note that this analysis has been focused on the synthetic spectra provided by \emph{Spitzer}/IRS for consistency with the analysis performed in Paper II.

\begin{itemize} 
\item Spectral slopes: We computed three spectral slopes of the form $\rm{\alpha = - log(F_{\nu}(\lambda_{2})/F_{\nu}(\lambda_{1}))}$ /$\rm{log(\lambda_{2}/\lambda_{1})}$ (where $\rm{\lambda_2 > \lambda_1}$). Note that, under this nomenclature, negative (positive) values mean that the flux increases (decreases) with wavelength. We called $\rm{\alpha_{NIR}}$, $\rm{\alpha_{MIR}}$, and $\rm{\alpha_{FIR}}$ to the slopes evaluated at [$\rm{\lambda_1,\lambda_2}$] equal to [5.5,7.5], [7.5,14], and [25,30]\,$\rm{\mu m}$, respectively. 

\item Silicate features strength: We also computed the silicate feature strength using the formula $\rm{Si_{\lambda}=- ln(F_{\nu}(\lambda)/F_{\nu}(continuum))}$, for the two Silicate features located at $\rm{\sim 9.7\mu m}$ and $\rm{\sim 18\mu m}$. Silicate features in emission (absorption) show negative (positive) $\rm{Si_{\lambda}}$. We anchor the continuum at the $\rm{\sim 9.7\mu m}$ ($\rm{\sim 18\mu m}$) silicate feature using a linear fit to the continuum at 7-7.5 and 14-15\,${\mu m}$ (14-15 and 25-26\,${\mu m}$). 
\end{itemize}

\begin{table}
\scriptsize
\begin{center}
\begin{tabular}{ l c c c c c}
\hline \hline
 			& $\rm{\alpha_{NIR}}$ & $\rm{\alpha_{MIR}}$ & $\rm{\alpha_{FIR}}$ & $\rm{Si_{9.7\mu m}}$ & $\rm{Si_{18\mu m}}$ \\ \hline
F06 		& 		$[-3.5,-0.5]$	&		$[-3.0,-1.0]$		&	$[-4.0, 1.0]$		&		$[-1.5, 4.0]$		&	$[-0.5, 0.6]$		\\
N08	& 		$[-7.0,-1.0]$	&		$[-7.5,-1.0]$		&	$[-5.0, 0.0]$		&		$[-1.5, 1.5]$		&	$[-0.6, 0.4]$		\\
H10	& 		$[-7.0,0.0]$	&		$[-4.0, 0.0]$		&	$[-1.0, 3.0]$		&		$[-1.0, 1.0]$		&	$[-0.6, 0.2]$		\\
S15	& 		$[-8.0,-1.0]$	&		$[-8.0,-2.0]$		&	$[-6.0, 0.0]$		&		$[-1.0, 5.0]$		&	$[-0.4, 0.4]$		\\
S16	& 		$[-3.5,-1.0]$	&		$[-3.0,-1.0]$		&	$[-2.5, 0.5]$		&		$[-1.5, 2.5]$		&	$[-0.4, 0.3]$		\\
H17	& 		$[-8.0,0.0]$	&		$[-4.0, 0.0]$		&	$[-1.0, 2.0]$		&		$[-0.8,0.8]$		&	$[-0.6, 0.0]$		\\
\hline \hline
\end{tabular}
\caption{Range of the spectral parameters per model. F06: [Fritz06]; N08: [Nenkova08]; H10: [Hoenig10]; S15: [Sieben15]; S16: [Stalev16]; and H17: [Hoenig17].}
\label{tab:shape}
\end{center}
\end{table}

Figs.\,\ref{fig:genfit1}-\ref{fig:genfit3} show $\rm{\alpha_{MIR}}$ versus $\rm{\alpha_{NIR}}$, the $\rm{\alpha_{FIR}}$ versus $\rm{\alpha_{MIR}}$, and $\rm{Si_{18\mu m}}$ versus $\rm{Si_{9.7\mu m}}$, respectively. Table \ref{tab:shape} shows the range of values for each parameter and model. Synthetic spectra produced with fixed steps (small dots) always show spectral parameters well within those obtained using the synthetic spectra produced with random parameters (large circles). This ensures that 1,000 random realizations are enough to cover the spectral shapes covered by the SEDs. 

Figs.\,\ref{fig:genfit1} and \ref{fig:genfit2} show that all the models overlap in a range of $\rm{\alpha_{NIR}}$ and $\rm{\alpha_{MIR}}$ with $\rm{\sim}$[-3,-1.0], and in a range of $\rm{\alpha_{FIR}}$ with $\rm{\sim}$[-1,0], except for [Sieben15], which lack of $\rm{\alpha_{MIR}}$ in the range [-2,-1] (see also Table\,\ref{tab:shape}). [Nenkova08] ([Sieben15]) shows values of these slopes as low as  $\rm{\alpha_{NIR}\sim-7}$ ($\rm{\alpha_{NIR}\sim-8}$), $\rm{\alpha_{MIR}\sim-7.5}$ ($\rm{\alpha_{MIR}\sim-8}$), and $\rm{\alpha_{FIR}\sim-5}$ ($\rm{\alpha_{FIR}\sim-6}$). [Hoenig10] and [Hoenig17], also extend $\rm{\alpha_{NIR}}$ and $\rm{\alpha_{MIR}}$ to low values with $\rm{\alpha_{NIR}\sim-8}$ and  $\rm{\alpha_{MIR}\sim-4}$ but also to larger values up to $\rm{\alpha_{NIR}\sim 0}$ and  $\rm{\alpha_{MIR}\sim 0}$. Some of these characteristics might explain the need of non-realistic stellar and ISM components by the data when using some of the models (see Paper II).

The strength of the silicate features also differ depending on the model used (see Fig.\,\ref{fig:genfit3}). All of them overlap in range of $\rm{Si_{9.7\mu m}=[-1,1]}$ and $\rm{Si_{18\mu m}=[-0.4,0]}$, i.e. producing relatively weak emission features. However, [Hoenig10] and [Hoenig17] almost do not include SEDs with 18$\rm{\mu m}$ absorption features while [Fritz06] contains SEDs with 18$\rm{\mu m}$ silicate features with strengths as high as  $\rm{Si_{18\mu m}=0.6}$. Similarly, [Fritz06] and [Sieben15] also contain SEDs with absorption 9.7$\rm{\mu m}$ silicate features with strengths as high as $\rm{Si_{18\mu m}=4-5}$. These plots and Table \ref{tab:shape} could be used to study the adequacy of the models to AGN mid-infrared spectra (see also Paper II).

\begin{figure*}[!ht]
\begin{center}
\includegraphics[width=1.0\columnwidth]{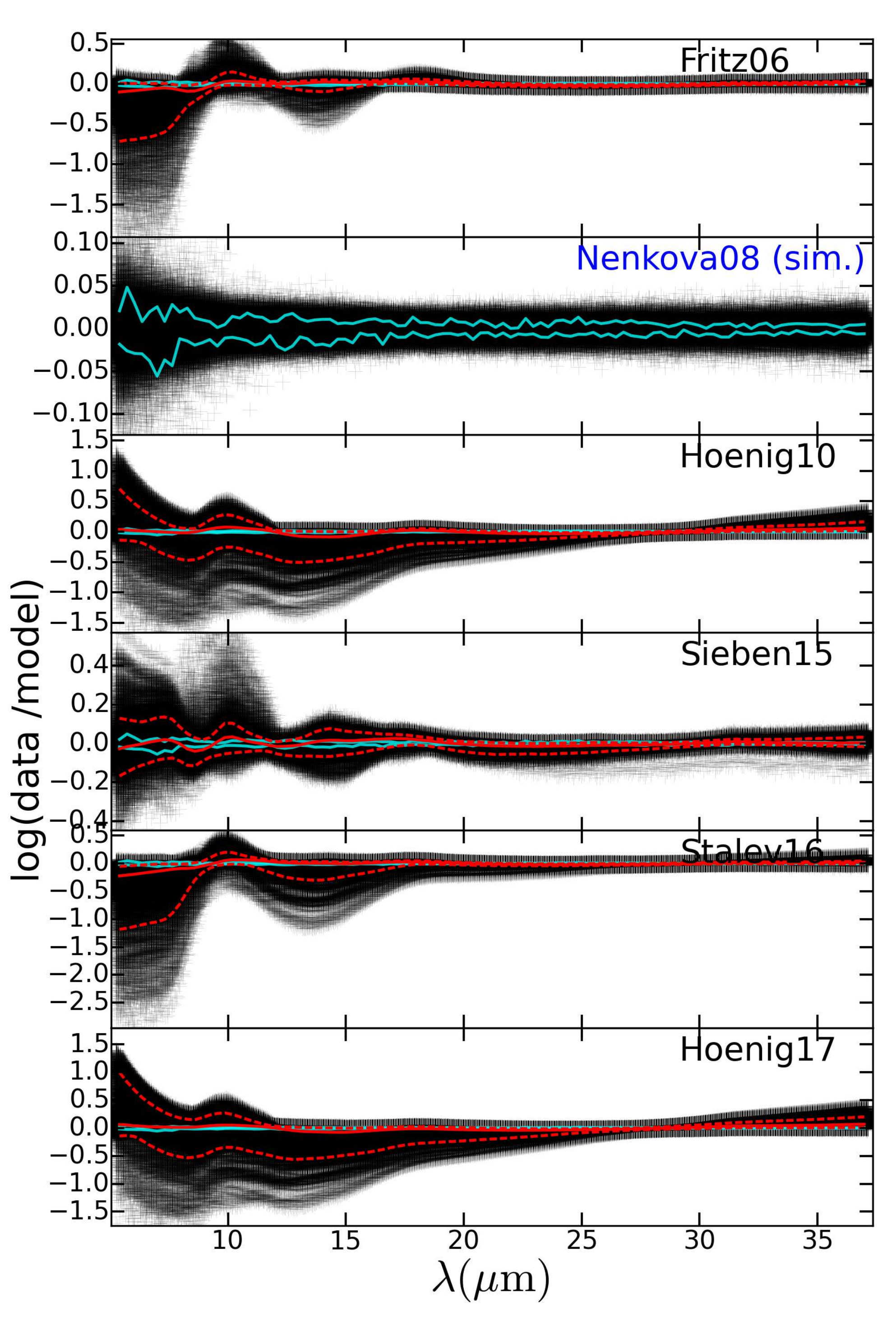}
\includegraphics[width=1.0\columnwidth]{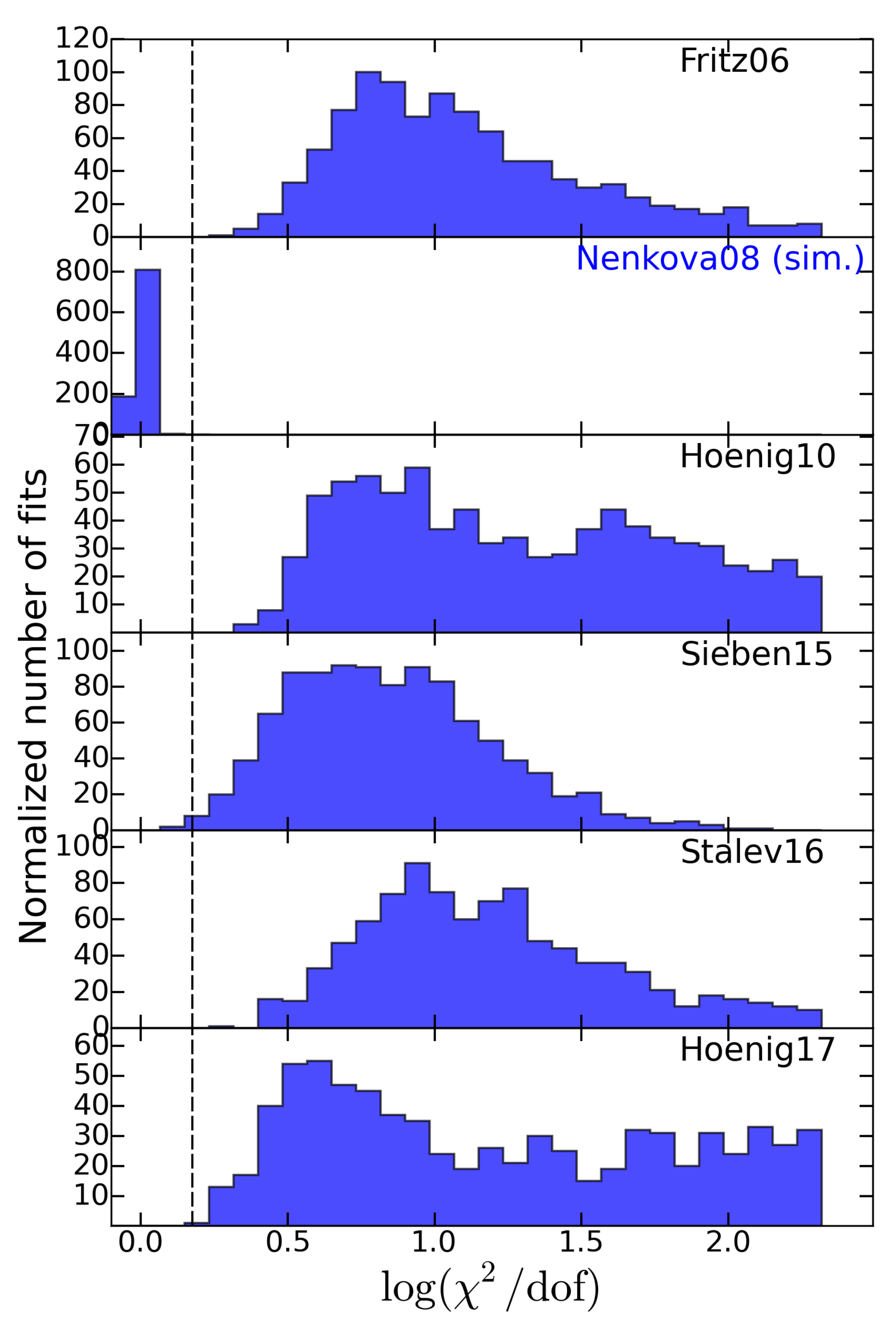}
\caption{Residuals as the ratio between data and model (left) and distribution of $\rm{\chi^2/dof}$ (right) obtained when fitting the 1,000 synthetic spectra produced with \citet{Nenkova08B} SEDs to each of the six models. Panels, from top to bottom, show the results of fitting the synthetic spectra to [Fritz06], [Nenkova08], [Hoenig10], [Sieben15], [Stalev16], and [Hoenig17]. The self-fit (i.e. fitting the synthetic spectra to the same model used to produce them) is highlighted with blue letters. Cyan continuous lines show the 15\% and 85\% of the residuals when the synthetic spectra are self-fitted to the same model (i.e. [Nenkova08]). Red continuous and dashed lines show the median, 15\%, and 85\% of the residuals when the synthetic spectra are fitted with other models. The vertical-dashed line in the right panel shows the locus of $\rm{\chi^2/dof=1.5}$. The same figures for the synthetic spectra obtained with the six models are included in Figs. \ref{appendix:FitResiduals1} and \ref{appendix:FitResiduals2}. }
\label{fig:FitResiduals}
\end{center}
\end{figure*}

\subsection{Parameter determination}\label{sec:Modelconstrain}

We automatically fitted the synthetic SEDs used in the previous section with the same model used to produce them. This allows us to study how accurate is the parameter determination for these instruments. Note that, in order to fit [Sieben15] using the \emph{JWST}/(MIRI + NIRSpec) combination of instruments, we needed to rule out the spectral range below $\rm{4\,\mu m}$ because their SEDs show a chaotic behavior in the near-infrared domain which produces a failure in the chi-squared procedure. 

Fig.\,\ref{fig:Par2Par} shows the results obtained for [Nenkova08] with $\rm{f(12\mu m)=300\, mJy}$. Each panel shows the estimated versus simulated values for one of the parameters of the model. The results using random parameters for the four instrumental setups are included in each panel using different colors and symbols in the bottom panels. The top panels show the histogram of the resulting values obtained for the synthetic spectra produced using fixed parameter steps (the simulated parameters are marked with vertical dashed lines). The best estimates of the parameters are obtained for $Y$, $q$, and $\tau_{v}$. This is clearly seen in the small dispersion around the simulated value using both random or fixed parameter values. \citet{Ramos-Almeida14} found that $Y$ is sensitive to wavelengths above $\rm{\sim20\,\mu m}$ while $q$ and $\tau_{v}$ can be restricted through the silicate features. However, $i$ and $\sigma$ are more sensitive to the near-to-mid infrared slope. This might explain why $Y$, $q$, and $\tau_{v}$ are better recovered than $i$ and $\sigma$. 

The best way to analyze the parameter determination is to produce the posterior distribution function (PDF) per parameter. However, producing PDFs requires $\rm{\sim 5}$min per parameter and spectral fit. Since 250,000 fits requires longer than 2 year computer time, we performed this calculation only for the $\emph{Spitzer}$ /IRS data in Paper II. The results obtained in Paper II are consistent with the analysis performed here. In the synthetic spectra we estimate the accuracy on the parameter determination using as the error the difference between the computed and simulated values compared to the parameter space range. We consider that a parameter is well determined if this error is within 15\% of the parameter space on average. Fig.\,\ref{fig:ParError} shows the average percentage error compared to the parameter space range per model and instrumental setup. The 15\% of the parameter space  is highlighted as the first dashed circle. The results for different S/N simulated spectra are linked with cyan continuous, blue long- and red short-dashed lines. Except for [Hoenig17], all the parameters are recovered well within 15\% of error for a source with $\rm{f(12\mu m)=100\, mJy}$. However, sources with lower fluxes (green-filled area in Fig.\,\ref{fig:ParError}) shows errors larger than 15\% of the parameter space. Note that the results are irrespective of the use of synthetic spectra using random initial parameters (continuous lines) or fixed parameter values (dashed lines). We can still recover four ($i$, $h$, $a$, and $a_w$) out of the eight parameters for [Hoenig17] using any of the instrumental setups. The large errors obtained for [Hoenig17] are associated to the large number of parameters involved in this model.

It is also worth mentioning that, when the angular width of the torus is a free parameter on the models (i.e. [Fritz06], [Nenkova08], [Hoenig10], [Stalev16], and [Hoenig17]), it is the less constrained parameter, except for [Stalev16]. Several authors have pointed out that the viewing angle of the torus is very difficult to constrain without near-infrared data, at least for [Nenkova08] \citep[][]{Ramos-Almeida11,Alonso-Herrero11,Ramos-Almeida14}. We found that the viewing angle is well constrained for all the models analyzed, even using GTC/CanariCam data, although this can be improved with \emph{JWST}/(MIRI+NIRSpec) spectra. This could be explained by the fact that we use here full coverage mid-infrared spectra while they used photometry and N-band spectroscopy. Moreover, the inclusion of host galaxy dilution might be key to understand why near-infrared data are needed (see Section \ref{sec:CircumModelconstrain}). [Sieben15] is particularly well determined with less than 10\% error for all the instrumental setups, except for the faintest sources with $\rm{f(12\mu m) \sim 30\,mJy}$.

In a broad comparison among instruments, an increment on the errors below 5\% is obtained for GTC/CanariCam data compared to other instruments. The difficulty in the determination of parameters using GTC/Canaricam data is related to the limited wavelength range. Furthermore, \emph{Spitzer}/IRS spectra are better suited to get a good determination of the models than \emph{JWST}/MIRI and only a marginal improvement is obtained when using \emph{JWST}/(MIRI+NIRSpec). This is due to the larger wavelength range of the \emph{Spitzer}/IRS spectra (5-38\,$\rm{\mu m}$) compared to \emph{JWST}/MIRI (5-30\,$\rm{\mu m}$). However, we emphasize here that \emph{Spitzer}/IRS spectra include a larger portion of the galaxy compared to \emph{JWST}/MIRI or ground-based GTC/CanariCam, imposing a further limitation on the determination of the torus parameters (see Section \ref{sec:CircumModelconstrain}). 

\begin{figure*}[!ht]
\begin{flushleft}
\includegraphics[width=0.68\columnwidth]{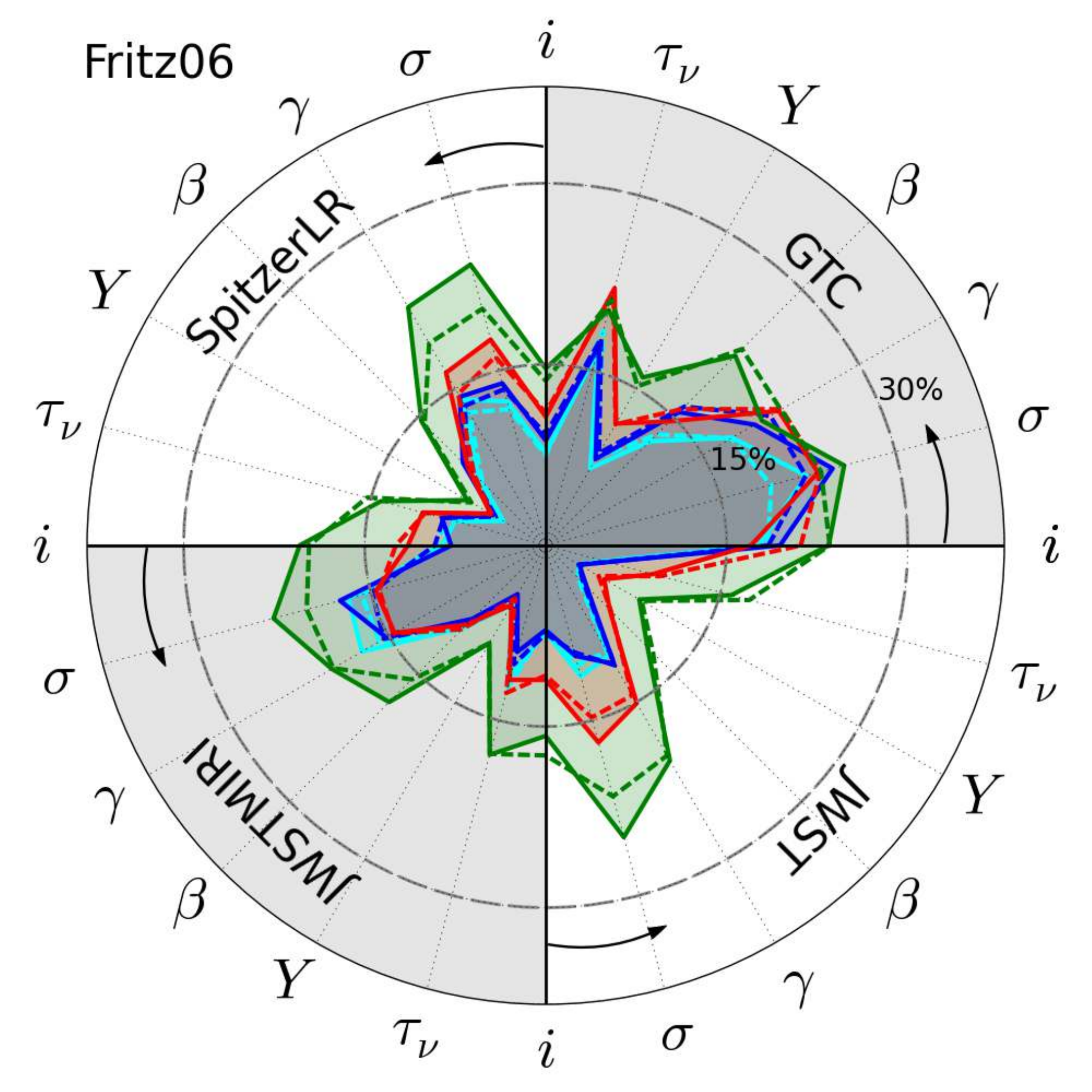}
\includegraphics[width=0.68\columnwidth]{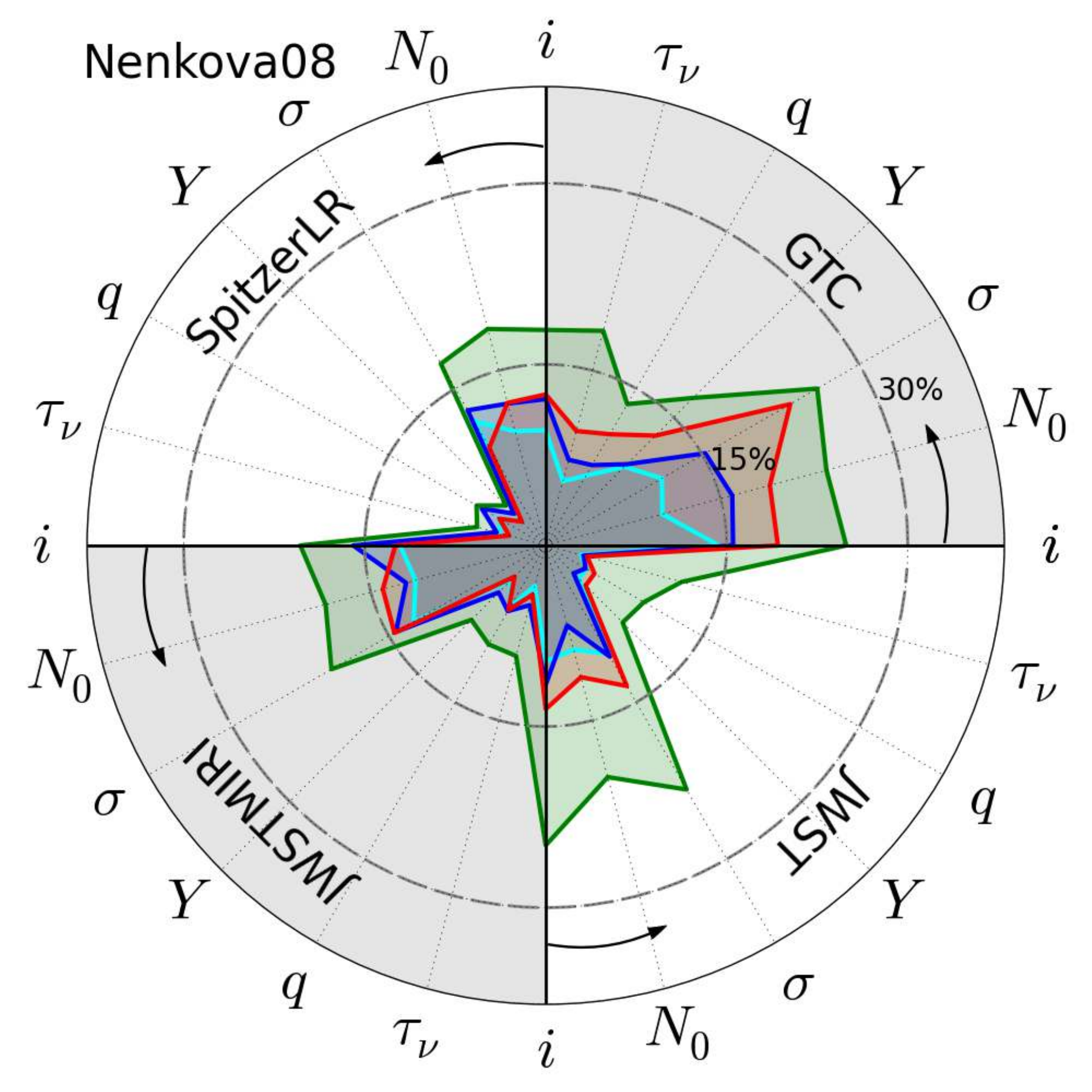}
\includegraphics[width=0.68\columnwidth]{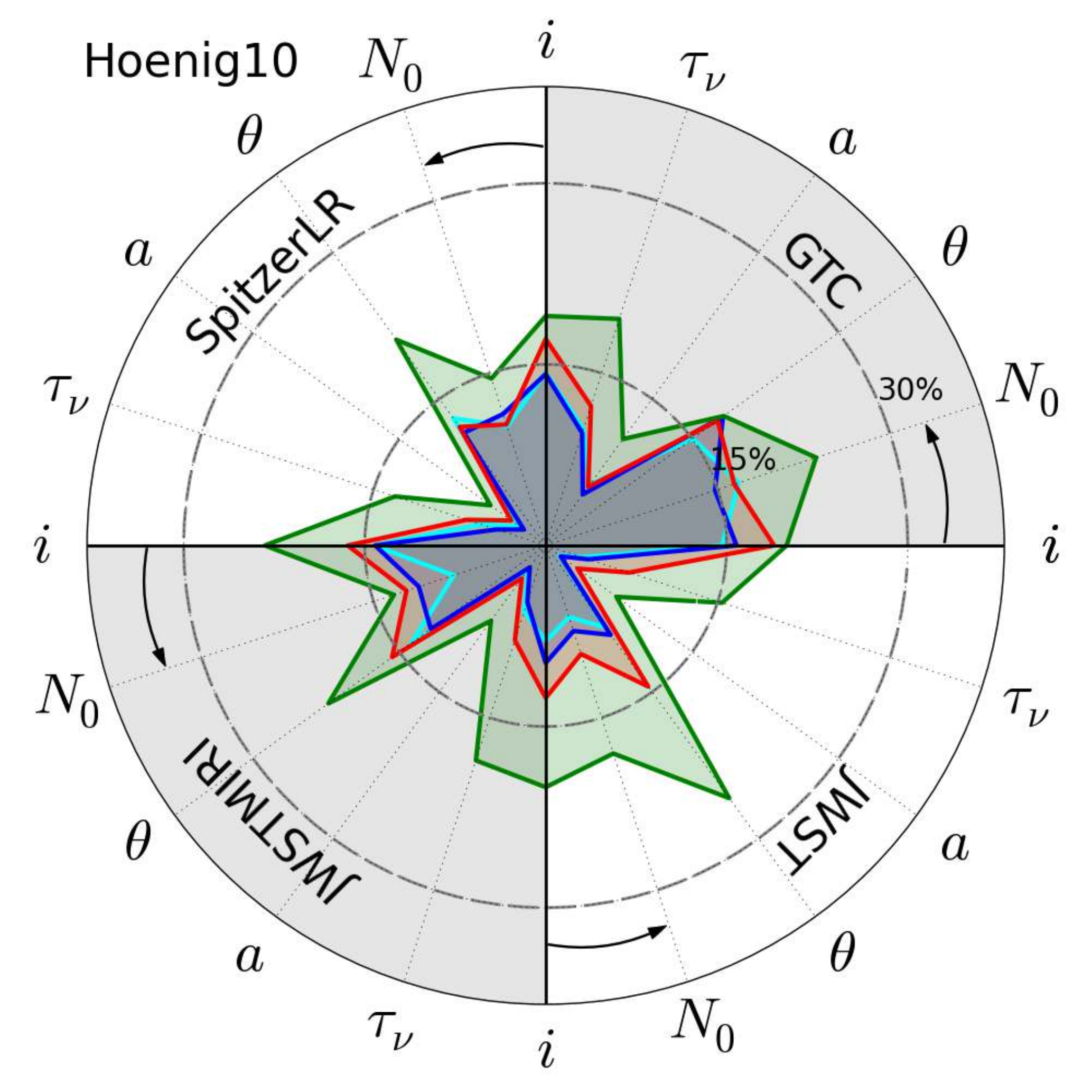}
\includegraphics[width=0.68\columnwidth]{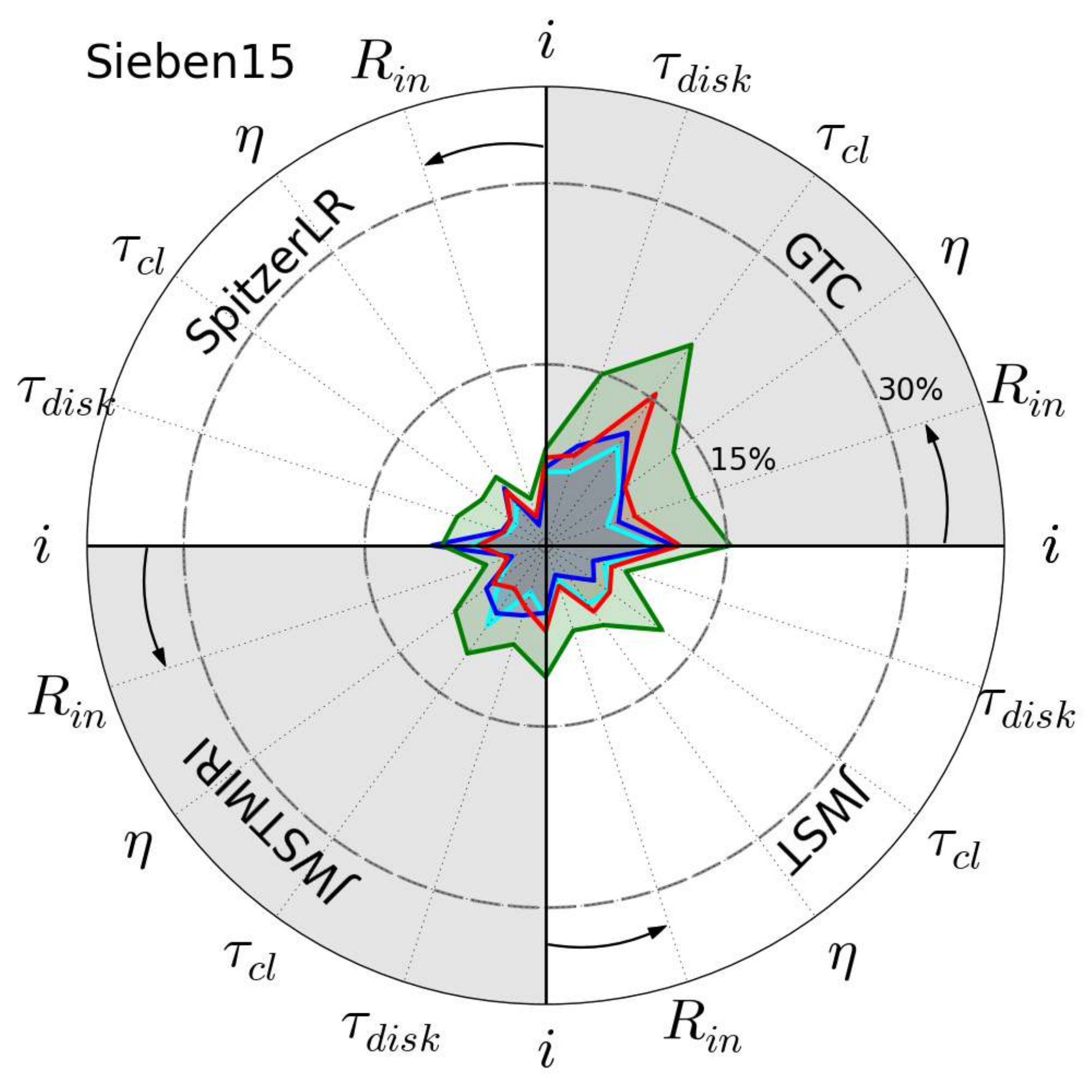}
\includegraphics[width=0.68\columnwidth]{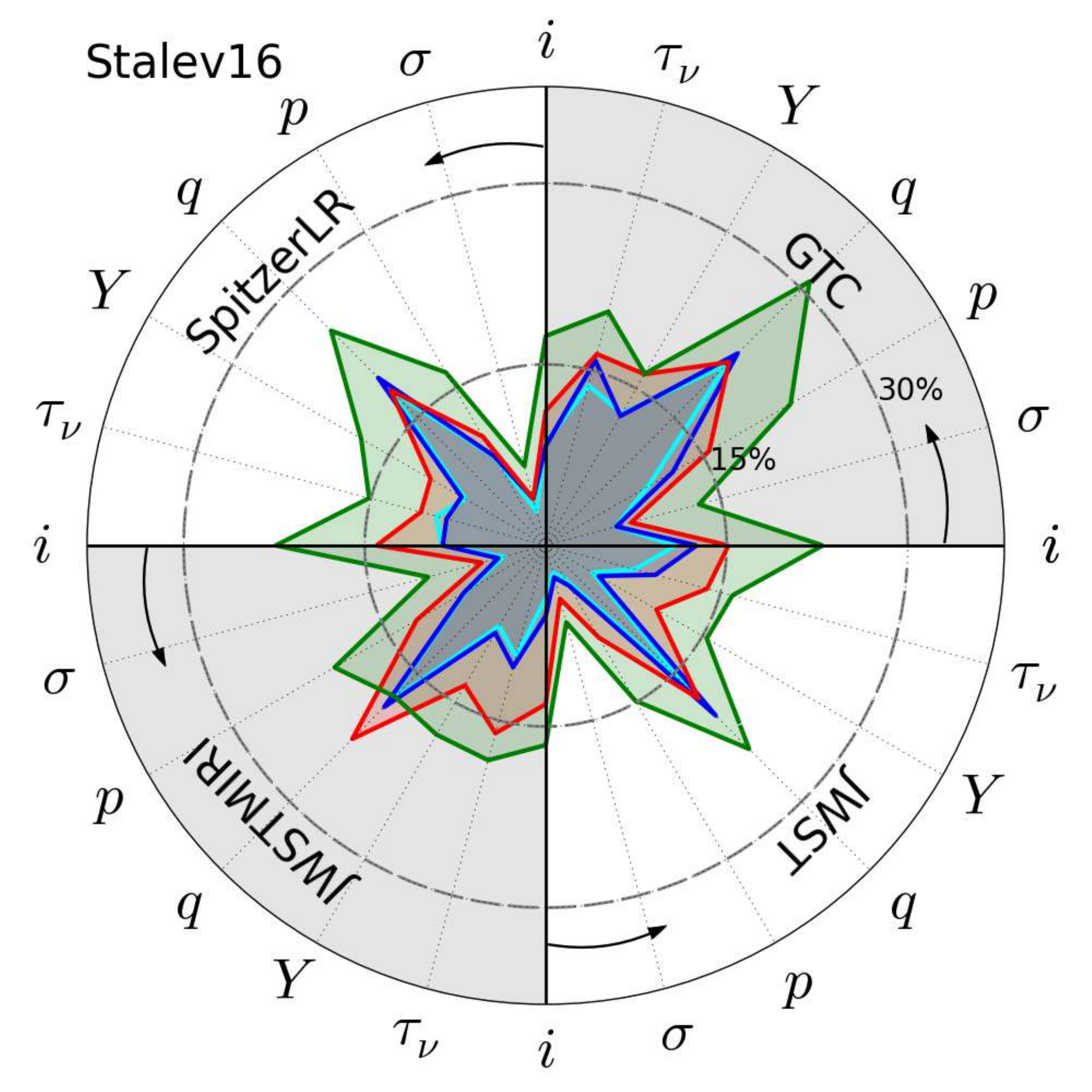}
\includegraphics[width=0.68\columnwidth]{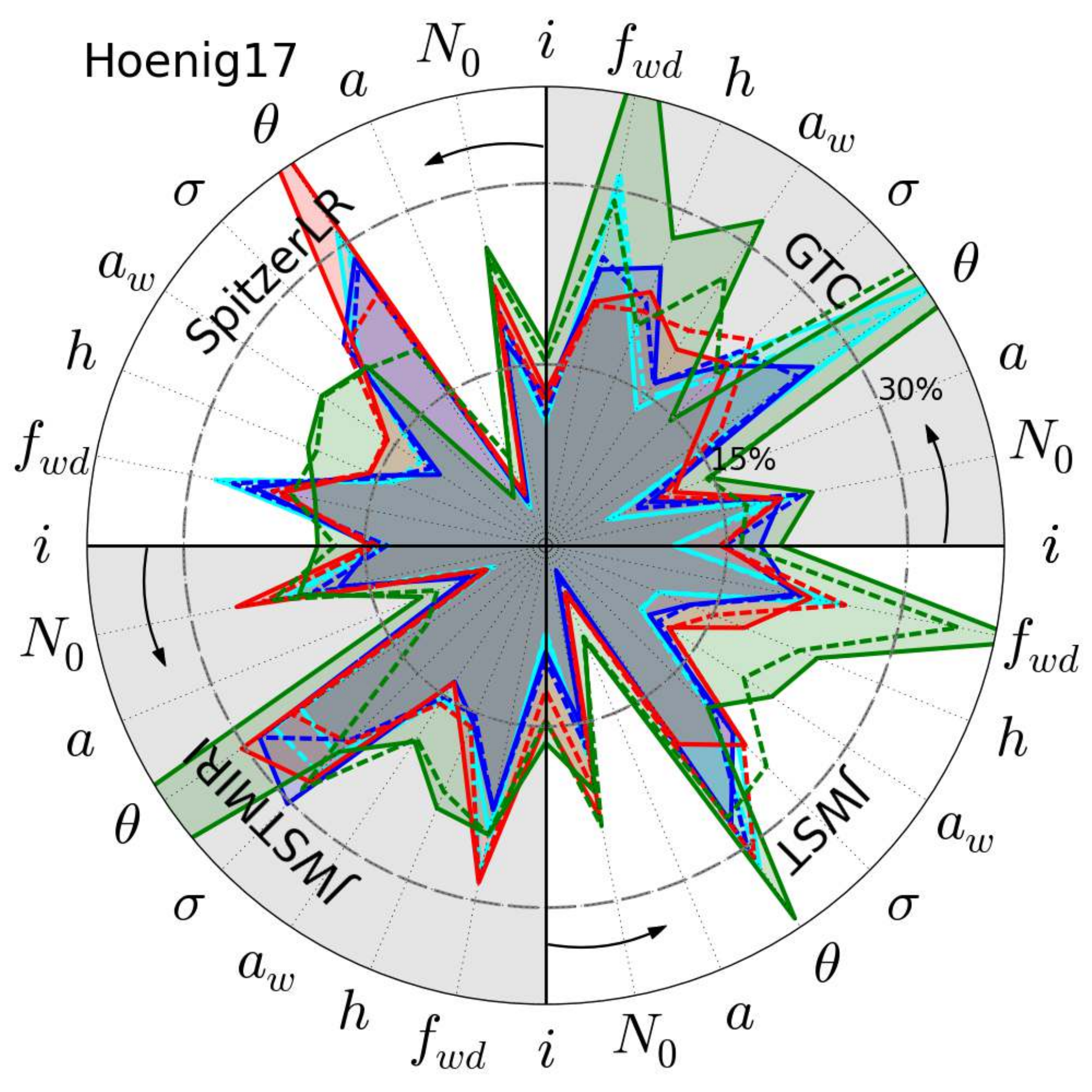}
\end{flushleft}
\begin{center}
\caption{ Average percentage of error on the parameter estimate per model (each circle) and per instrumental setup (each quarter of a circle) when stellar contribution is included in the \emph{Spitzer}/IRS simulated spectra. The arrows shown in counterclock direction indicate the first parameter within the instrument setup. All the spectra are set to a continuum flux of $\rm{f(12 \mu m) = 300}$ mJy. The cyan, dark blue, red, and green lines link results using a stellar contribution of 10\%, 50\%, and 100\% of the flux of the AGN component at 5\,$\rm{\mu m}$ and 100\% of the flux of the AGN component at 10\,$\rm{\mu m}$, respectively. Continuum and short-dashed lines show the results when using 100 and 200 simulated spectra, respectively. Note that the results for the 200 simulated spectra have been computed only for [Fritz06] and [Hoenig17] to ratify that the number of iterations do not have an impact on the results. The long-dashed circles highlight the confidence error within 15\% and 30\% of the parameter range.}
\label{fig:ParErrorDilution1}
\end{center}
\end{figure*}

\begin{figure*}[!ht]
\begin{flushleft}
\includegraphics[width=0.68\columnwidth]{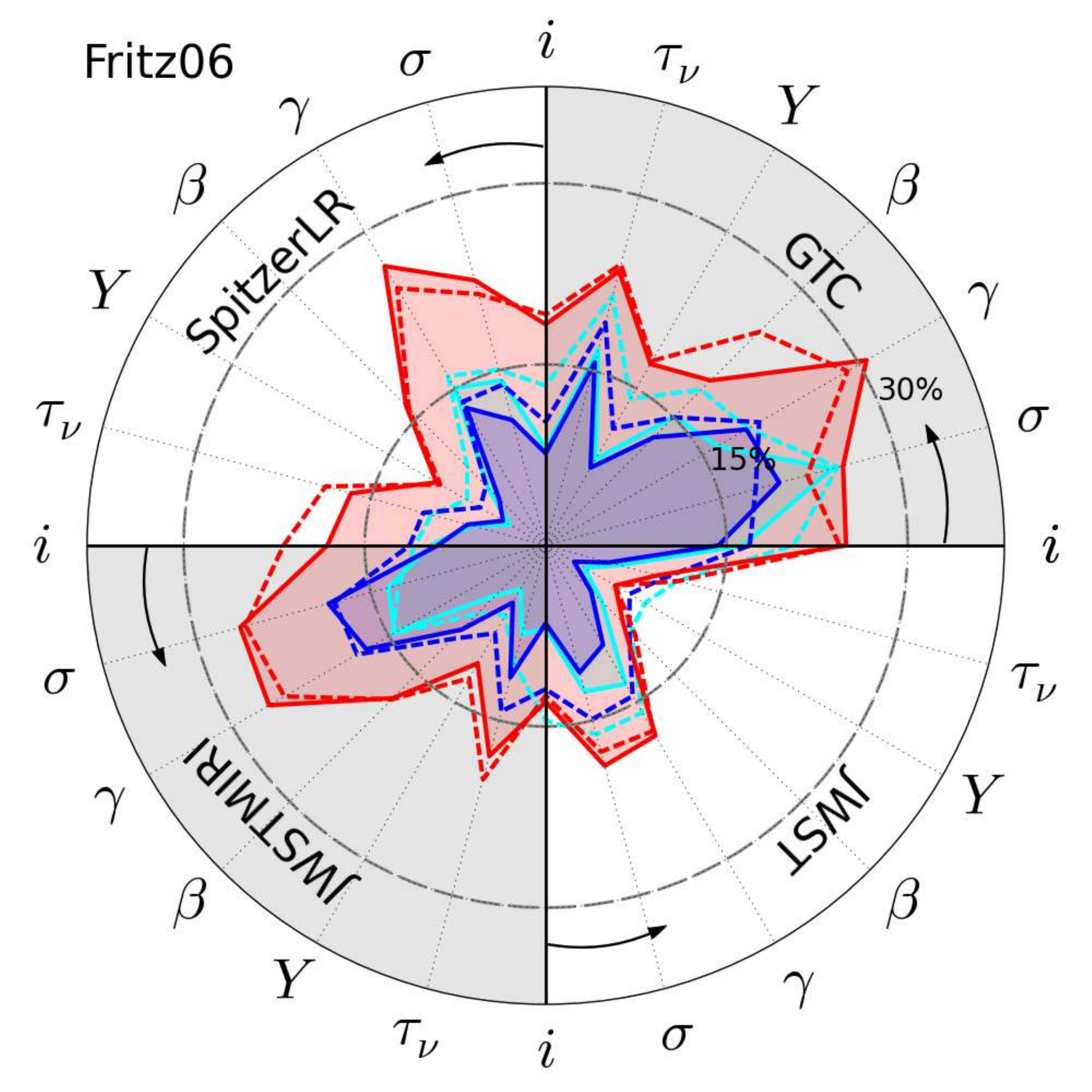}
\includegraphics[width=0.68\columnwidth]{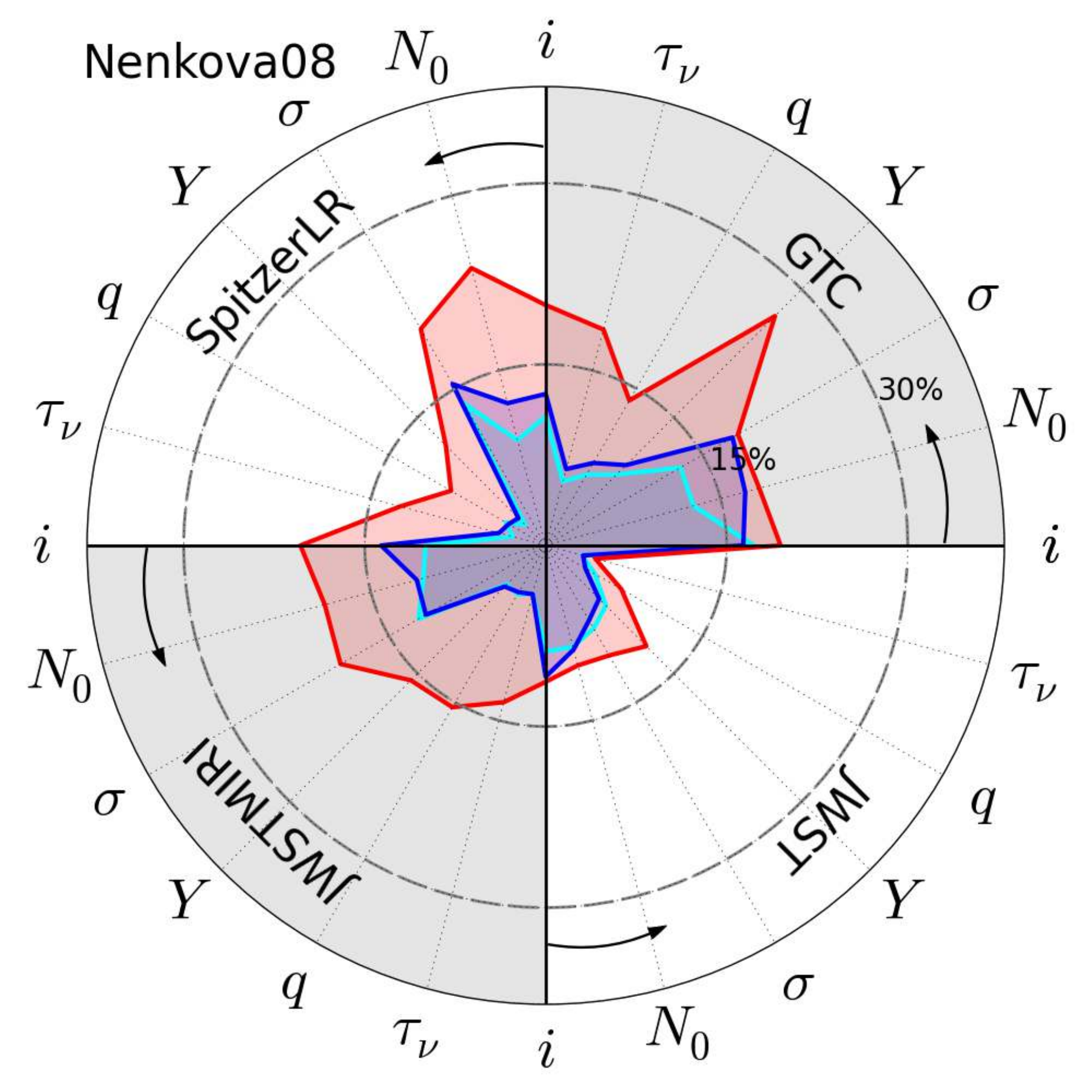}
\includegraphics[width=0.68\columnwidth]{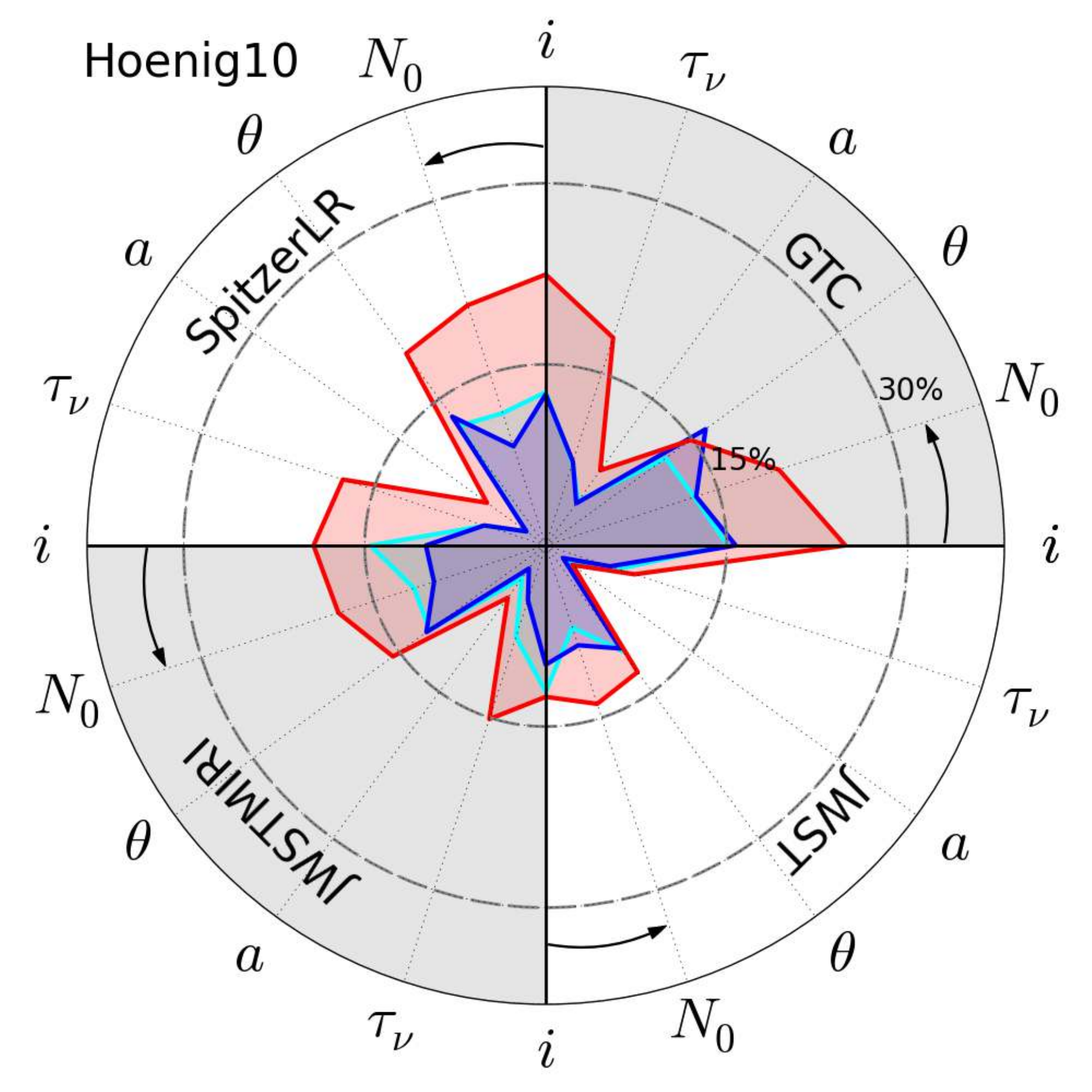}
\includegraphics[width=0.68\columnwidth]{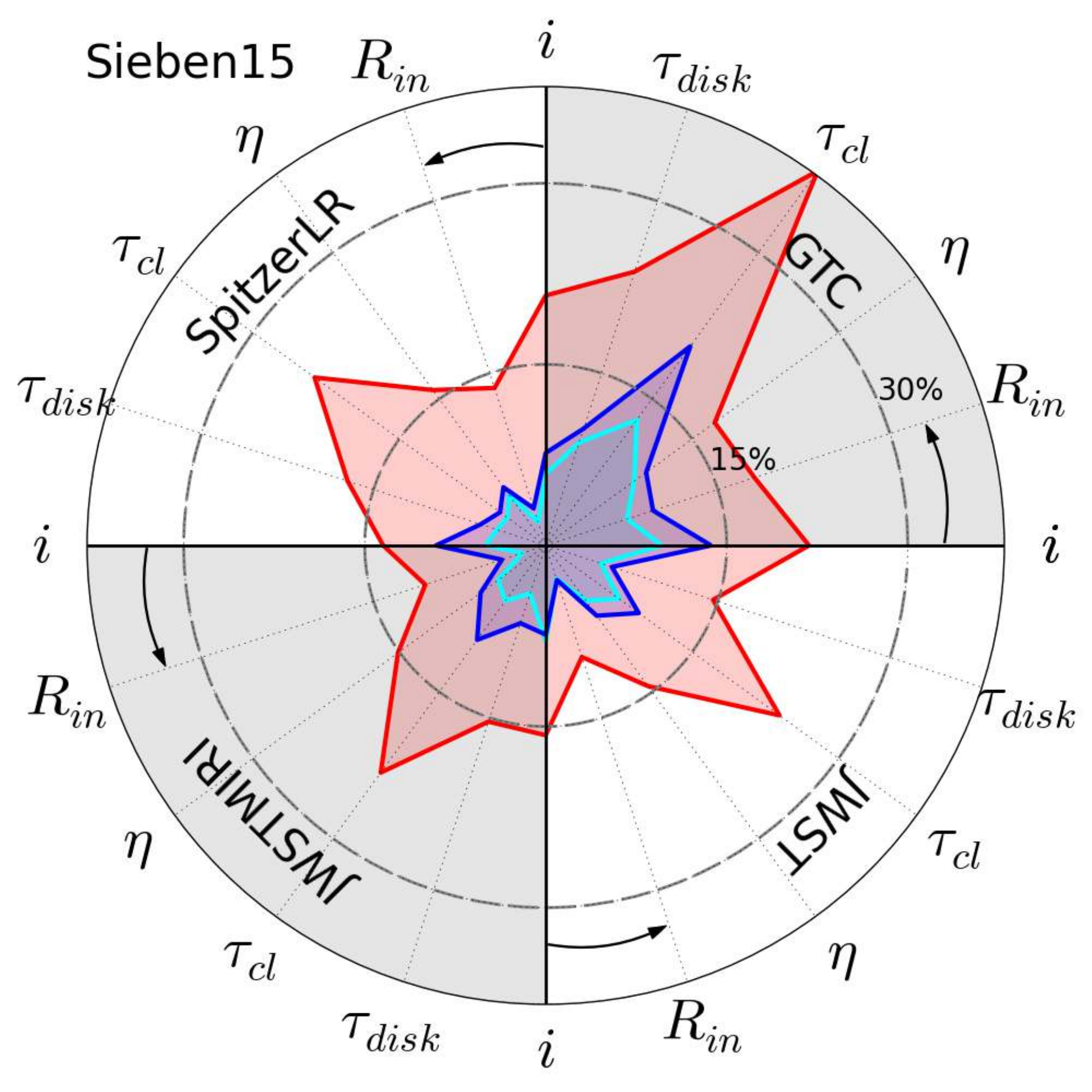}
\includegraphics[width=0.68\columnwidth]{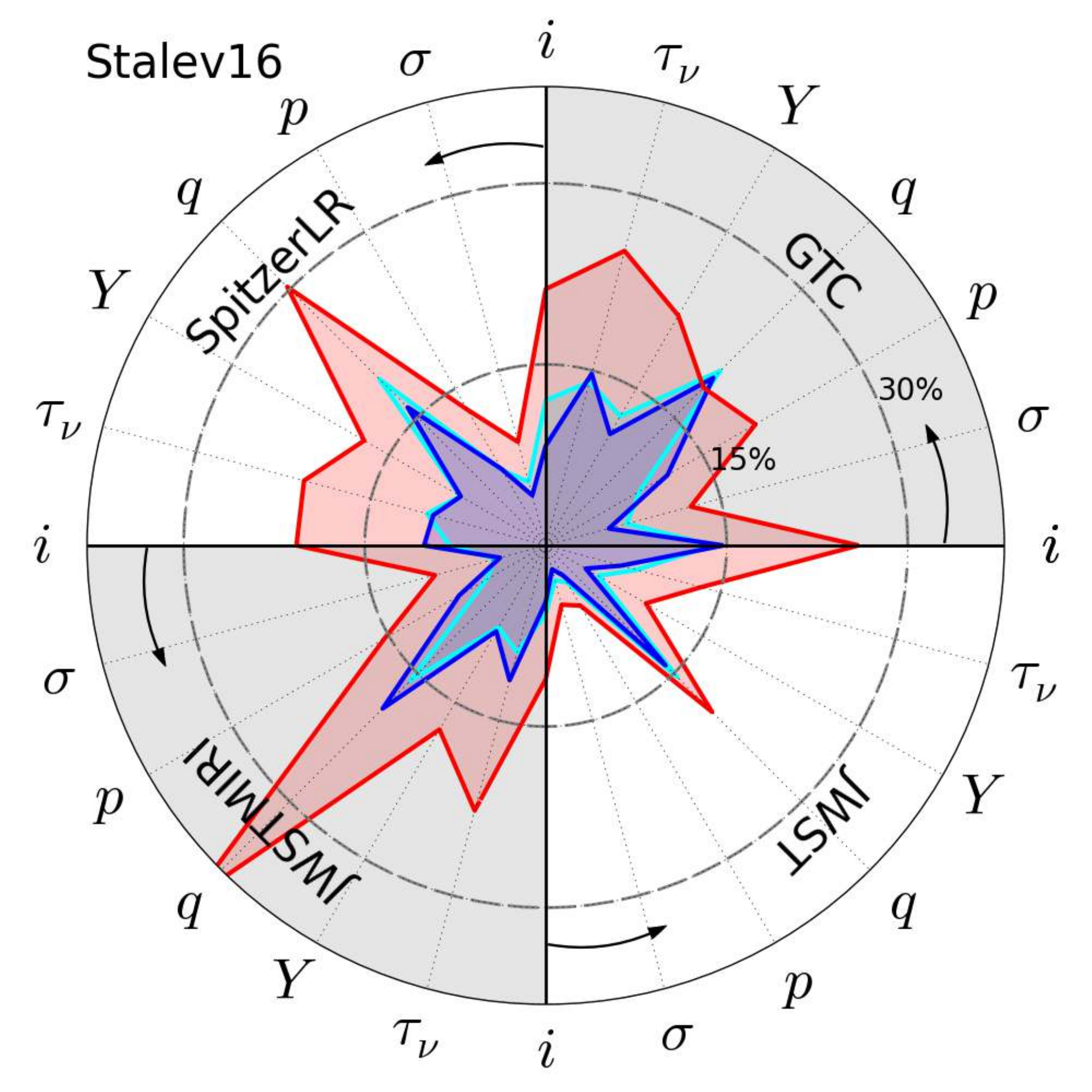}
\includegraphics[width=0.68\columnwidth]{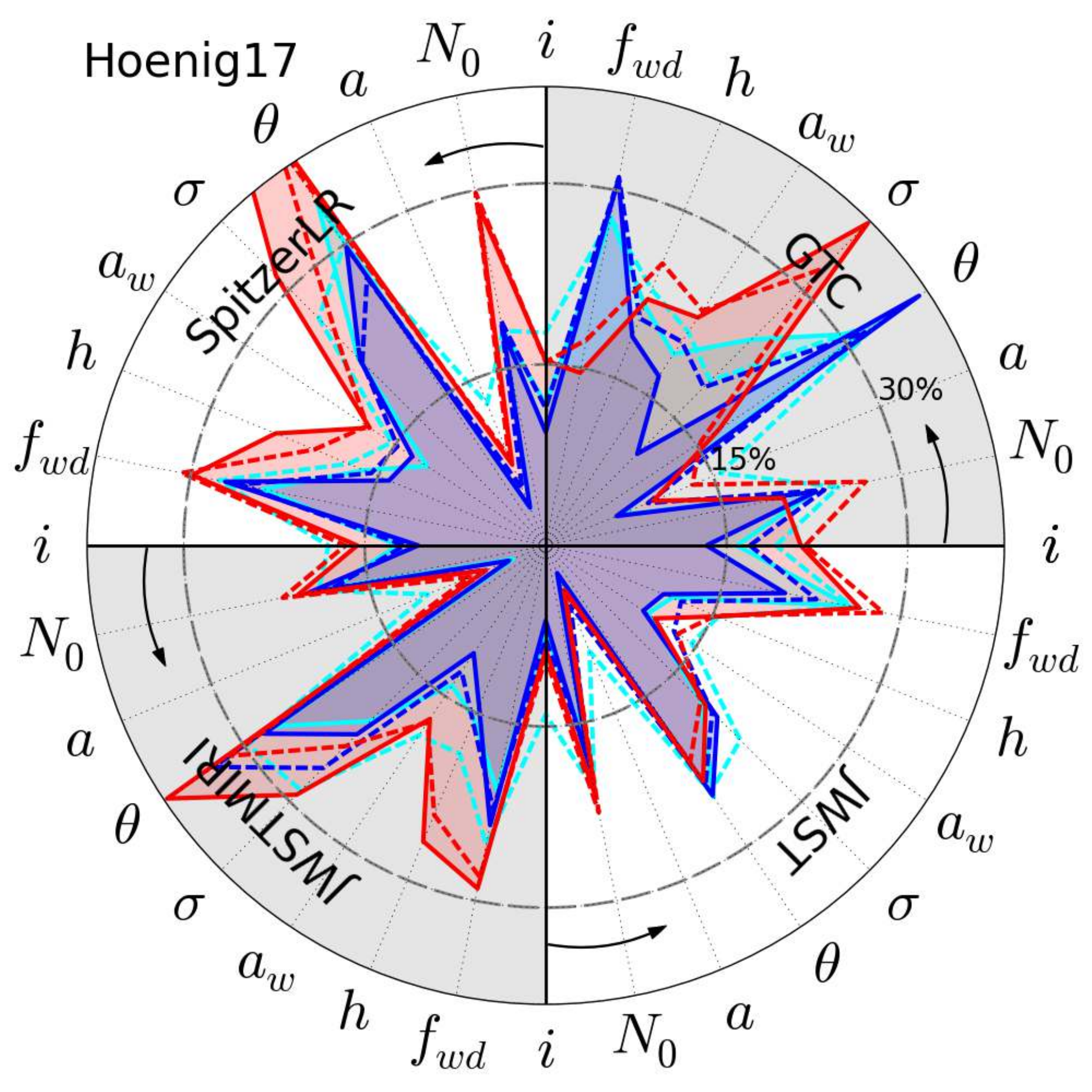}
\end{flushleft}
\begin{center}
\caption{ Average percentage of error on the parameter estimate per model (each circle) and per instrumental setup (each quarter of a circle) when ISM contribution is included in the \emph{Spitzer}/IRS simulated spectra. The arrows shown in counterclock direction indicate the first parameter within the instrument setup. All the spectra are set to a continuum flux of $\rm{f(12 \mu m) = 300}$ mJy. The cyan, dark blue, and red lines link results using an ISM contribution of 10\%, 50\%, and 100\% of the flux of the AGN component at 30\,$\rm{\mu m}$, respectively. Continuum and short-dashed lines show the results when using 100 and 200 simulated spectra, respectively. Note that the results for the 200 simulated spectra have been computed only for [Fritz06] and [Hoenig17] to ratify that the number of iterations do not have an impact on the results. The long-dashed circles highlight the confidence error within 15\% and 30\% of the parameter range. Note that the results for ISM dilution using \emph{JWST}/(MIRI+NIRSpec) are not taken into account along the text due to an improper coverage of this component at near-infrared wavelengths (see text).}
\label{fig:ParErrorDilution2}
\end{center}
\end{figure*}

\begin{figure*}[!ht]
\begin{flushleft}
\includegraphics[width=0.68\columnwidth]{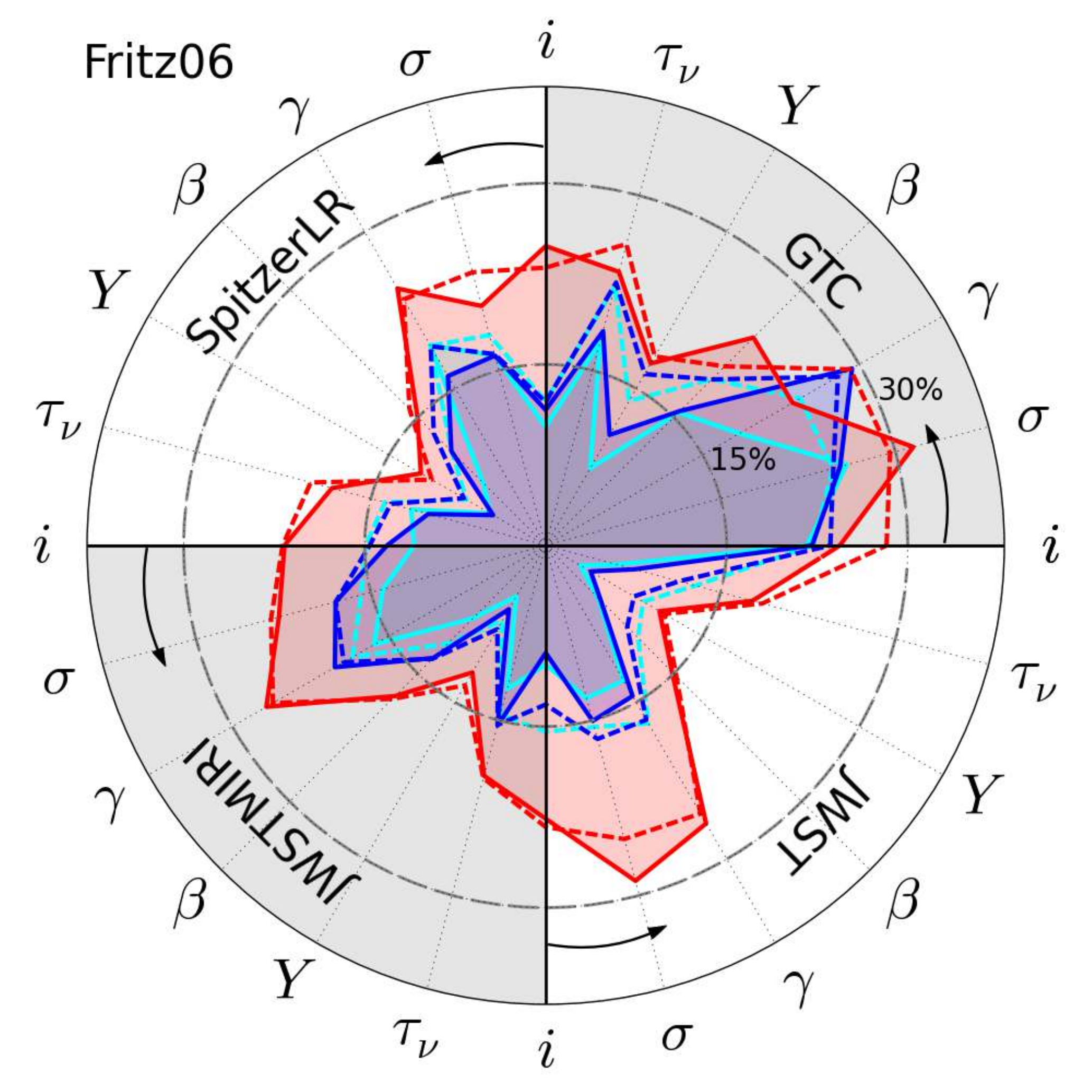}
\includegraphics[width=0.68\columnwidth]{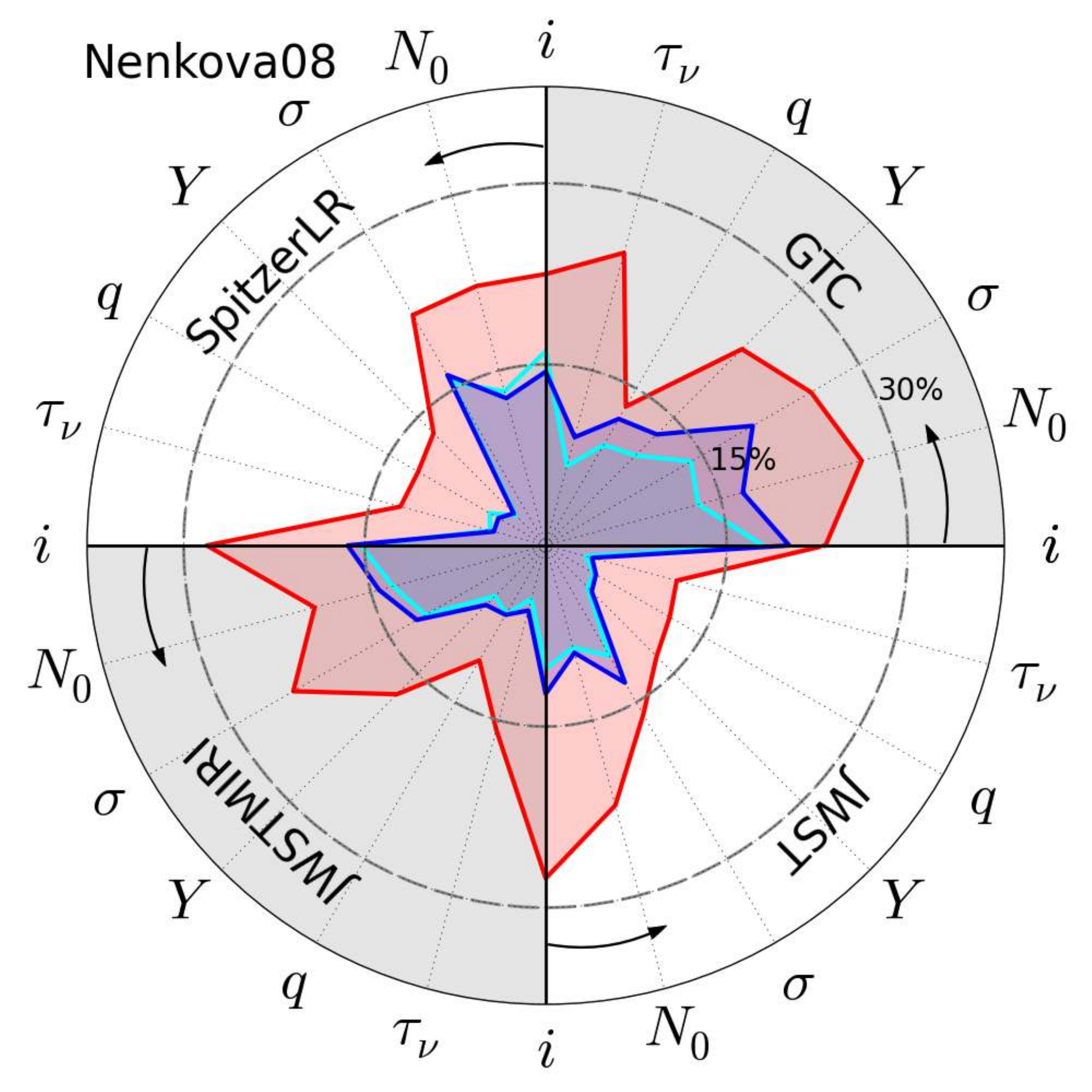}
\includegraphics[width=0.68\columnwidth]{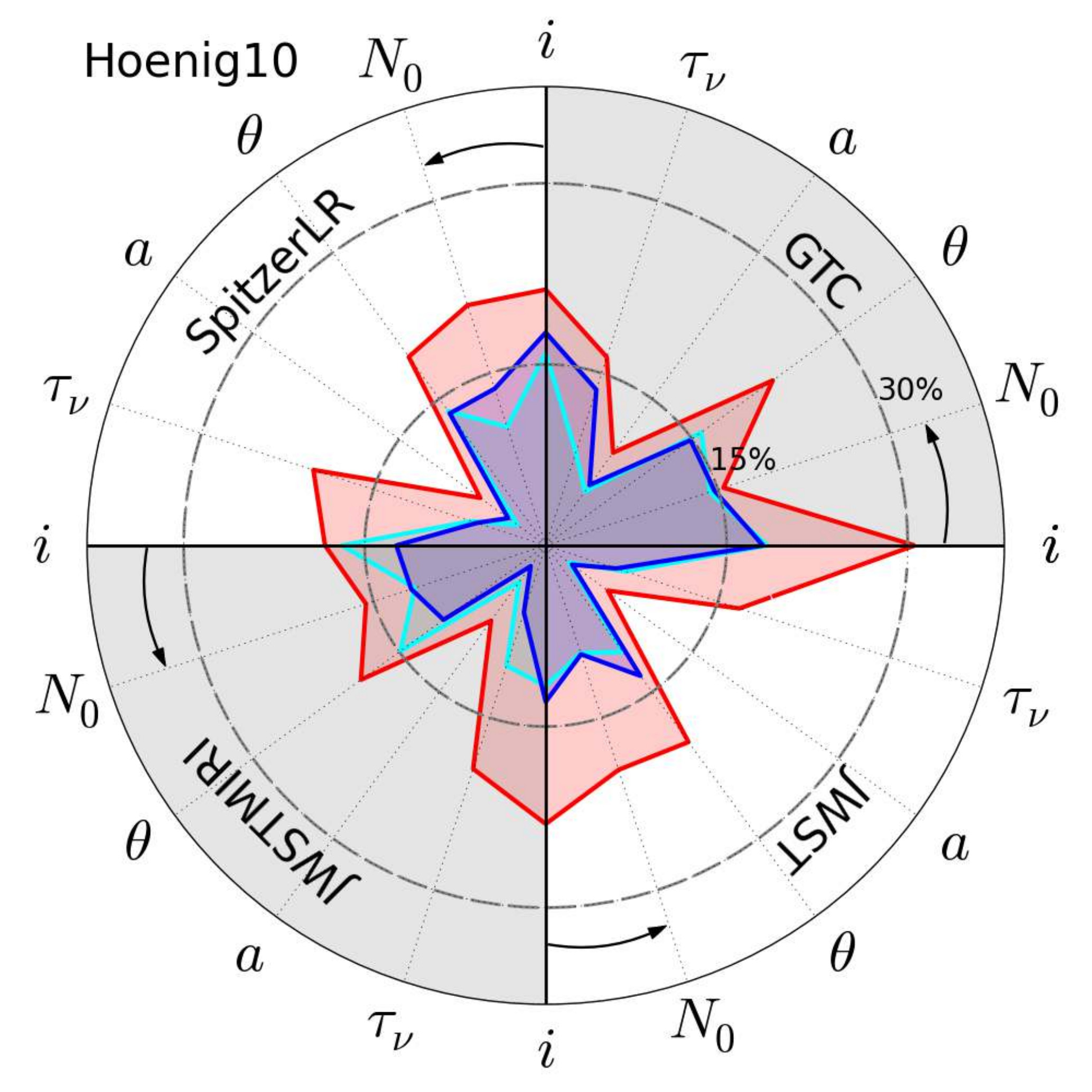}
\includegraphics[width=0.68\columnwidth]{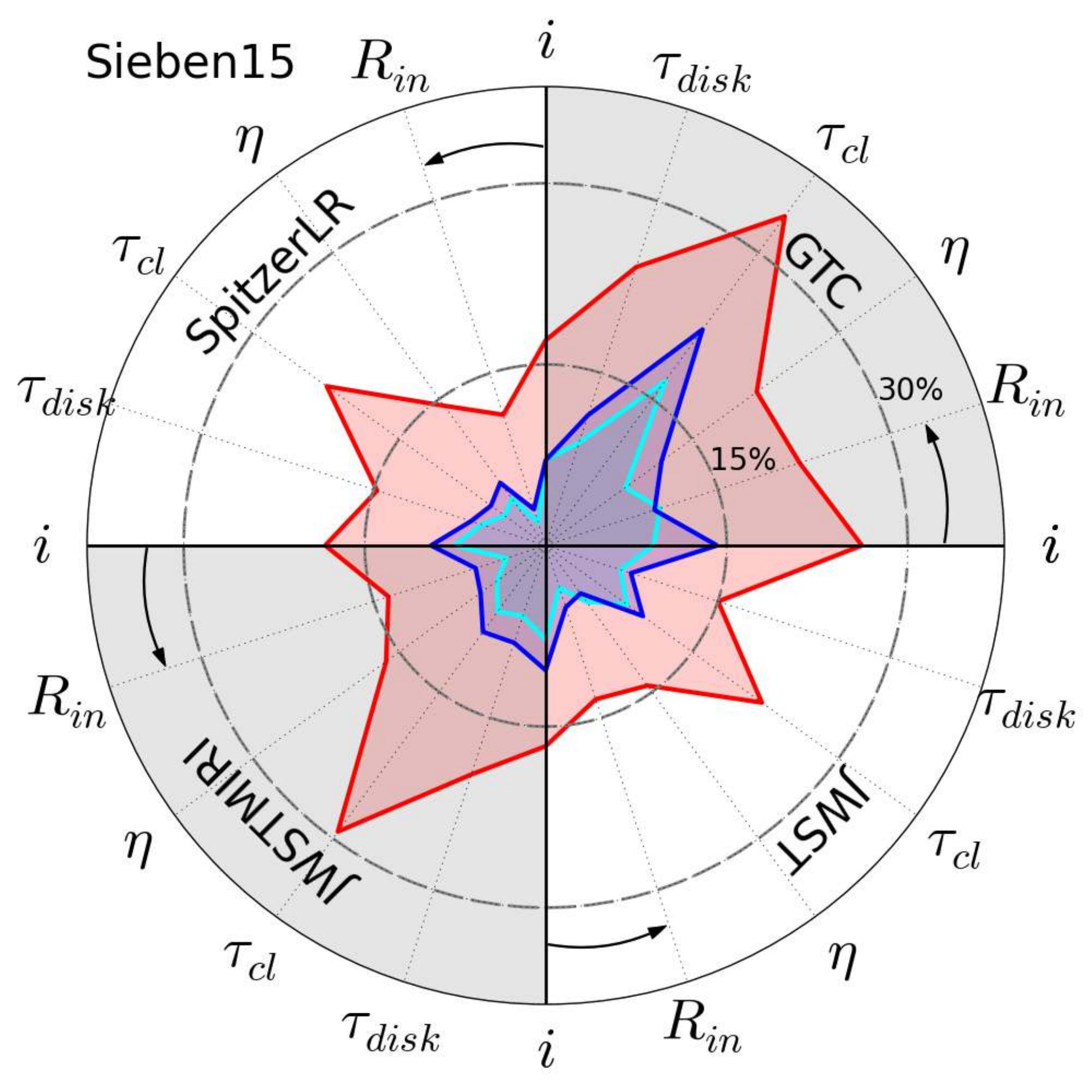}
\includegraphics[width=0.68\columnwidth]{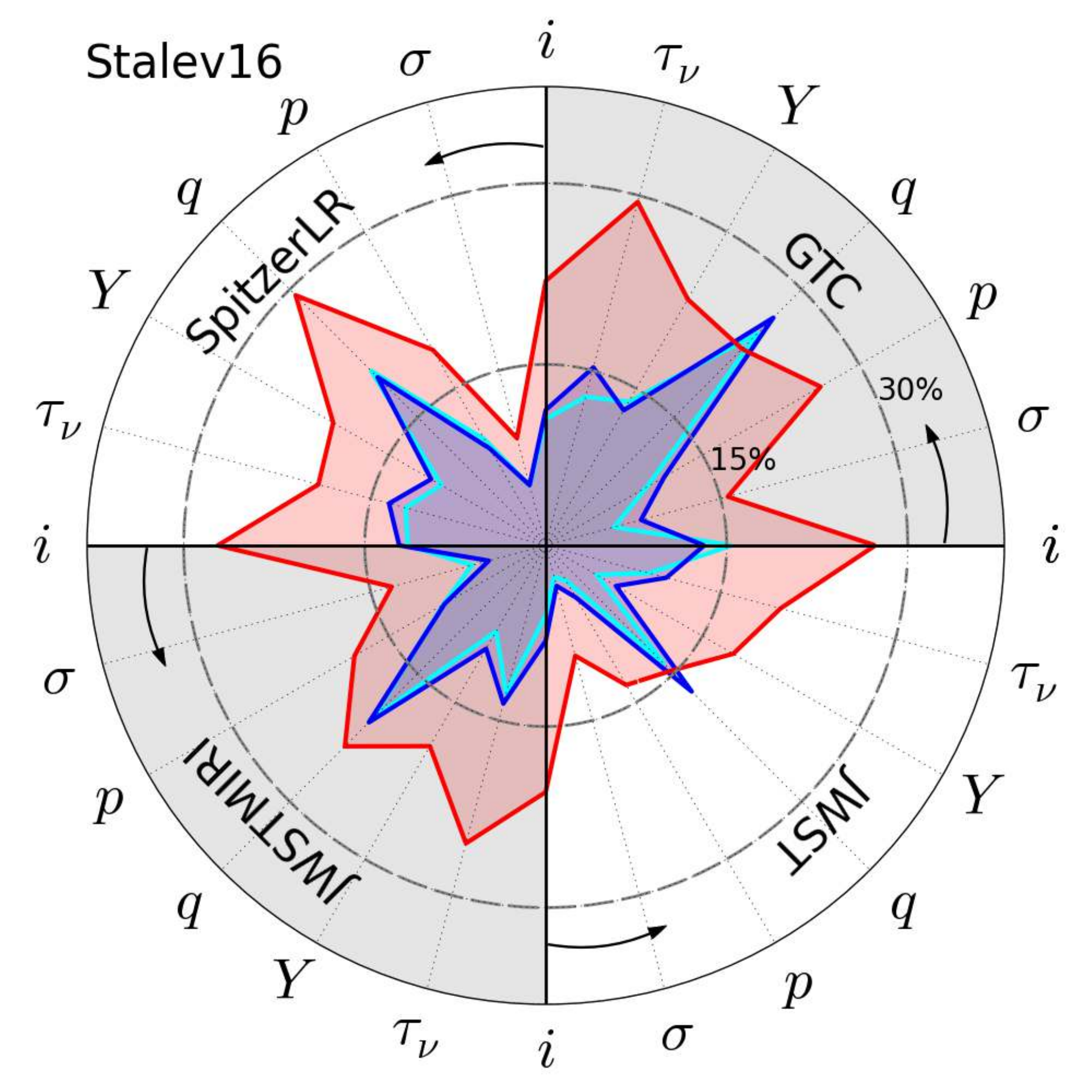}
\includegraphics[width=0.68\columnwidth]{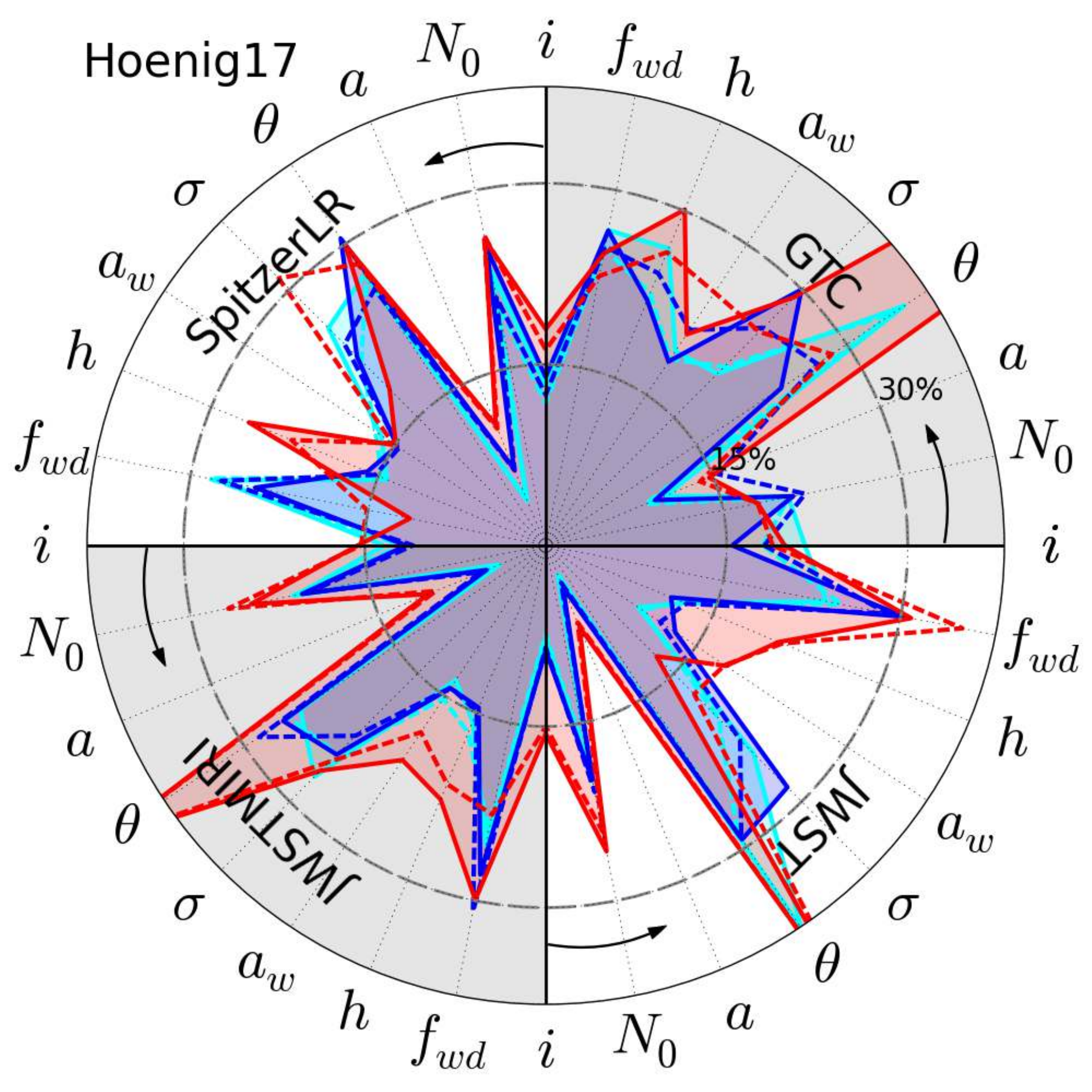}
\end{flushleft}
\begin{center}
\caption{ Average percentage of error on the parameter estimate per model (each circle) and per instrumental setup (each quarter of a circle) when the  combination of ISM+stellar contributions are included in the \emph{Spitzer}/IRS simulated spectra. The arrows shown in counterclock direction indicate the first parameter within the instrument setup. All the spectra are set to a continuum flux of $\rm{f(12 \mu m) = 300}$ mJy. The cyan, dark blue, and red lines link results using an ISM and stellar contributions of 10 and 10\%, 50 and 50\%, and 100 and 100\% of the flux of the AGN component at 5 and 30\,$\rm{\mu m}$, respectively. Continuum and short-dashed lines show the results when using 100 and 200 simulated spectra, respectively. Note that the results for the 200 simulated spectra have been computed only for [Fritz06] and [Hoenig17] to ratify that the number of iterations do not have an impact on the results. The long-dashed circles highlight the confidence error within 15\% and 30\% of the parameter range.}
\label{fig:ParErrorDilution3}
\end{center}
\end{figure*}

\subsection{Model discrimination}\label{sec:ModelvsModel}

We investigate in this section if the spectrum from a particular model can be distinguished from the others. We used for this purpose the 6,000 synthetic \emph{Spitzer}/IRS spectra (1,000 spectra per model) obtained with a $\rm{12\,\mu m}$ flux of 300 mJy (same as those used in Section \ref{sec:Modelconstrain}, see Fig.\,\ref{fig:flowchart_syntheticspectra}).

Each synthetic spectrum is fitted with the six models, including the one used to produce the synthetic spectrum. Fig.\,\ref{fig:FitResiduals} shows the residuals (left) and distribution of $\rm{\chi^2/dof}$ (right) obtained when fitting the 1,000 synthetic spectra produced with [Nenkova08]. The distributions of reduced $\rm{\chi^2}$ show that the only model able to obtain a good fit for these synthetic spectra is its own model. The main differences are: (1) the 9.7 and 18 \,$\rm{\mu m}$ silicate features, (2) deficit (for [Fritz06]), excess (for [Hoenig10] and [Hoenig17]), or both (for [Sieben15]) on the flux below 7\,$\rm{\mu m}$, and (3) steep slopes above $\rm{\sim}$25\,$\rm{\mu m}$ for [Hoenig10] and [Hoenig17]. Figs.\,\ref{appendix:FitResiduals1} and \ref{appendix:FitResiduals2} show the results for all the simulated spectra (one panel per model). In general, all the models can be distinguished based on the $\rm{\chi^2}$ statistics and residuals except for the comparisons [Hoenig10] versus [Hoenig17] and [Fritz06] versus [Stalev16]. This is somewhat expected because the CAT3DWIND model reported by \citet{Hoenig17} is based in the CAT3D model reported by \citet{Hoenig10B} and the two media torus model produced by \citet{Stalevski16} use similar geometry and dust composition than that used by \citet{Fritz06}. Nevertheless, it is worth to notice that [Stalev16] synthetic spectra are more easily recovered with [Fritz06] model than the other way around. This is due to the largest parameter space covered by [Fritz06] compared to [Stalev16]. Note that [Stalev16] has enough spectral coverage to reproduce real data (see Paper II). In general, the main differences are the same; i.e. a different slope below $\rm{\sim}$7\,$\rm{\mu m}$, different shape for the $\rm{10 \ and \ 18 \mu m}$ silicate features, and a steeper slope above $\rm{\sim}$25\,$\rm{\mu m}$ for [Hoenig10] and [Hoenig17] compared to [Fritz06], [Nenkova08], [Sieben15], and [Stalev16].

We also repeat our analysis for a source with $\rm{f(12 \mu m)\sim}$10 Jy and 100 mJy. In the former, the models are so different that we could not converge to a good solution. In the case of a source with $\rm{f(12 \mu m)\sim}$100 mJy, our results are similar to those obtained for a source with $\rm{f(12 \mu m)\sim}$300 mJy. Thus, except for the combinations [Hoenig10] versus  [Hoenig17] and [Fritz06] versus [Stalev16], models can be distinguished based solely in \emph{Spitzer}/IRS spectra. Note, however, that we also repeated the analysis with $\rm{f(12 \mu m)\sim}$30 mJy source. At this low S/N limit models are indistinguishable irrespective of the instrumental setup or model used.  

\subsection{Host galaxy dilution}\label{sec:CircumModelconstrain}

In previous sections, we showed that mid-infrared spectra are useful to discern which is the best model and to constrain the parameters of the models for relatively low-flux AGN ($\rm{f(12 \mu m)>}$100 mJy). However, a question remains open: the effect of the dilution by external contributors (i.e. host galaxy) on the estimated parameters. In order to analyze that, we have included 100 synthetic spectra with stellar, ISM components, and a combination of the two per AGN model used. Note that we also produced 100 additional synthetic spectra for [Fritz06] and [Hoenig17] to ensure that the relatively low number of SEDs does not affect our result. The ISM component is taken from \citet{Smith07}, which are averaged Starburst templates in the $\rm{\sim}$5-160\,$\rm{\mu m}$ wavelength range for different 6.2, 7.7, 11.3, and 17\,$\rm{\mu m}$ PAH feature strengths (see their Fig.\,13). Note that the results for ISM dilution using \emph{JWST}/(MIRI+NIRSpec) are not accurate because these templates lack of spectral information at near-infrared wavelengths. The stellar component corresponds to a stellar population of $\rm{10^{10}}$ years and solar metallicity from the stellar libraries provided by \citet{Bruzual03}. We set the parameters of the ISM component to the first ISM template for simplicity. The stellar component contributes mostly to the shortest wavelengths ($\rm{<10\,\mu m}$) and the ISM component contributes to the entire mid-infrared wavelength range with the highest contribution at longer wavelengths ($\rm{>20\,\mu m}$). 

These libraries have been converted into XSPEC following the same procedure explained in Section \ref{sec:XspecModel}. We produced four combinations for each synthetic spectrum using a stellar contribution scaled to 10\%, 50\%, and 100\% of the torus flux at 5\,$\rm{\mu m}$ and 100\% of the torus flux at 10\,$\rm{\mu m}$. Note, however, that we include 100\% contribution of stellar component at mid-infrared wavelength for completeness although it is well documented in the literature that this percentage is never reached at mid-infrared \citep[stellar contribution always below 50-60\% for Sy2 and almost negligible for Sy1, see][]{Rodriguez-Espinosa87,Dultzin88,Dultzin94,Dultzin96}. Similarly, we produced three combinations for each synthetic spectrum using an ISM contribution scaled to 10\%, 50\%, and 100\% of the torus flux at 30\,$\rm{\mu m}$ (see Fig.\,\ref{fig:flowchart_syntheticspectra}). Finally, we also produced 100 synthetic spectra using a combination of both ISM and stellar components as follows: (1) 10\% of stellar contribution at 5$\mu m$ + 10\% of ISM contribution at 30$\rm{\mu m}$; (2) 50\% of stellar contribution at 5$\mu m$ + 50\% of ISM contribution at 30$\rm{\mu m}$; and (3) 100\% of stellar contribution at 5$\mu m$ + 100\% of ISM contribution at 30$\rm{\mu m}$. The dusty torus spectra are simulated with $\rm{f(12 \mu m)\sim}$300 mJy (i.e. the intermediate S/N$\rm{\sim 40-60}$ studied in Sections\,\ref{sec:Modelconstrain} and \ref{sec:ModelvsModel}). We then repeated the analysis performed in Section \ref{sec:Modelconstrain} to study how good is the determination of the parameters. 

Note that the simulated fractional contributions of host galaxy (both stellar and ISM) are well recovered with less than 5\% error. Figs.\,\ref{fig:ParErrorDilution1}-\ref{fig:ParErrorDilution3} show the average percentage error (compared to the parameter range) per AGN dust model and instrumental setup, after including dilution by the stellar, ISM and stellar + ISM component into the synthetic spectra, respectively. Our results are robust since tests increasing the number of synthetic spectra do not change them. Better results are obtained when using \emph{Spitzer}/IRS and \emph{JWST}/(MIRI + NIRSpec) compared to GTC and \emph{JWST}/MIRI. However, this is due to the lack of near-infrared coverage of the ISM templates. However, is worth to mention that the superb spatial resolution obtained with GTC/CanariCam allows to isolate the nuclear component from host contributors much better than \emph{Spitzer}/IRS (see below). Indeed a large portion of the \emph{Swift}/BAT sample with \emph{Spitzer}/IRS spectra is highly contaminated by circumnuclear contributors (see Paper II). 

We were able to constrain all the parameters (roughly within 15\% error) with up to 50\% of stellar contribution at 5\,$\rm{\mu m}$ or 50\% of ISM contribution at 30\,$\rm{\mu m}$ for all the models (except [Hoenig17] model and GTC/CanariCam instrumental setup, see Figs.\,\ref{fig:ParErrorDilution1} and \ref{fig:ParErrorDilution2}). [Nenkova08] and [Hoenig10] can further constrain the parameters with less than 15\% error for 100\% of the stellar component at 5\,$\rm{\mu m}$. [Sieben15] is able to constrain the parameters with less than 15\% uncertainty even for 100\% of the stellar component at 10\,$\rm{\mu m}$. [Hoenig17], due to a large number of parameters, shows the largest percentage errors even for 10\% of stellar component at 5\,$\rm{\mu m}$ or 10\% of ISM component at 30\,$\rm{\mu m}$. The parameters are still within 15\% uncertainty only for [Nenkova08], [Hoenig10] and [Sieben15] (excluding in this case GTC/CanariCam) when the stellar and ISM contributions are combined up to 10-50\% (see Fig.\,\ref{fig:ParErrorDilution3}). The parameters for [Fritz06] and [Stalev16] could also be estimated including up to 50\% of combined stellar and ISM contributions except for $\rm{\sigma}$ and $\rm{\gamma}$ for [Fritz06] and $p$ for [Stalev16] (with all the instrumental setups except GTC/CanariCam).

The impact of galaxy dilution on the parameter estimate is stronger than the instrumental setup. In general, 100\% of stellar component at 10\,$\rm{\mu m}$ doubles the error on the resulting parameters while 100\% of stellar component at 5\,$\rm{\mu m}$ introduces a 50\% additional error on the resulting parameters. Similarly, 100\% ISM component at 30\,$\rm{\mu m}$ doubles the error obtained in the resulting parameters. The combination of 100\% of the stellar + ISM components includes twice the error found with 100\% of the stellar component and similar error than that found when including 100\% of ISM component. Thus, infrared high spatial resolution spectra are key to study the AGN dust by restricting the host galaxy contribution to the lowest level. \emph{JWST}/MIRI or ground-based GTC/CanariCam spectra are very useful to estimate the torus parameters because they better isolate the AGN from the host galaxy. This is clearly seen in Fig.\,\ref{fig:ParErrorDilution2} if we compare, for instance, the error on the parameter estimate of GTC/CanariCam or \emph{JWST} spectra and 50\% of ISM at 30\,$\rm{\mu m}$ (data points linked with continuum lines) with \emph{Spitzer}/IRS spectra and 100\% of ISM at 30\,$\rm{\mu m}$ (data points linked with long-dashed lines). Thus, high spatial resolution GTC/CanariCam spectra still play an important role on the parameter estimate until \emph{JWST} is able to give both, spectral coverage and spatial resolution.  

\section{Discussion} \label{sec:discussion}

Radiative transfer models have proven successful in reproducing the infrared SED of AGN \citep[e.g.][]{Fritz06,Ramos-Almeida09,Alonso-Herrero11,Hoenig17}. These models differ on the dust chemical composition, distribution, morphology or dynamics. We attempt to discuss for the first time how a large variety of these models could be distinguished based on SED fitting. We have found that four out of the six models could be distinguished based on the slope below $\rm{\sim}$7\,$\rm{\mu m}$, the slope above $\rm{\sim}$25\,$\rm{\mu m}$, and the silicate features. Indeed, one of the main controversial aspects of the AGN model unification regarding the dust emission is associated to the silicate features. Most type-1 Seyferts exhibit a rather weak emission feature \citep{Hoenig10A} and AGN in general lack deep 9.7 \,$\rm{\mu m}$ silicate absorption features \citep{Hao05,Hao07}. These weak features are naturally produced when large graphite grains are dominating the dust composition \citep{Hoenig10B}. The peak wavelength is above 10.2\,$\rm{\mu m}$ in about 65\% of the silicate emission features, whereas the shift of the 9.7\,$\rm{\mu m}$ absorption feature is small \citep[see also][]{Nikutta09,Hatziminaoglou15,Hoenig10A}. \citet{Nenkova02} proposed that these findings might be explained by a clumpy distribution of dusty clouds because after the inclusion of effects as directly illuminated clumps and clouds illuminated by others, these feature are never deep \citep[see also][]{Levenson06,Spoon07,Sirocky08}. Nevertheless, different sublimation temperatures of the silicate and graphite grains can also produce these weak features using smooth models \citep{Fritz06,Hoenig10A,Schartmann08}. Indeed, we also found that the smooth toroidal model [Fritz06] and the smooth + clumpy toroidal model [Stalev16] produces a deep absorption feature at 9.7$\rm{\mu m}$ compared to the other models (see Fig.\,\ref{fig:genfit3}). 

The first question that can be tackled with our analysis is whether we can distinguish between smooth and clumpy torus models. \citet{Dullemond05} compared their own radiative transfer models of smooth and clumpy tori, finding that, despite the distinct nature of the models, it was not possible to distinguish between them based on the SEDs \citep[similar results are found by][]{Stalevski12}. \citet{Schartmann08} also compared smooth and clumpy torus models developed by their own group both based on the radiative transfer code MC3D. They found that the main difference is that a clumpy medium allows the central source to heat the clouds at large radii. The only work comparing models not based on similar prescriptions for the dust composition is presented by \citet{Feltre12}. They compared the smooth model [Fritz06] and the clumpy model [Nenkova08B], finding that even with matched parameters they do not produce similar SEDs. Due to the different chemical composition in [Fritz06] compared to [Nenkova08], the behavior of the silicate features is quite distinctive \citep{Feltre12}. They can also be distinguished throughout the near-infrared slopes, because the clumpy torus model [Nenkova08] are sensitive to a much wider range of near-infrared slopes than the smooth torus model [Fritz06] (see Fig.\,\ref{fig:genfit1}). Interestingly, this narrow range of slopes is also found for the two-phases model [Stalev16]. \citet{Feltre12} suggested that the difference in the near-infrared slopes might be due to the different slope of the primary emission from the accretion disk. However, all the models except [Fritz06] and [Stalev16] show a wider range for the near-infrared slopes. Note, however, that these steep near-infrared slopes found for [Nenkova08], [Hoenig10], [Sieben15], and [Hoenig17] are not required for the AGN \emph{Spitzer}/IRS spectra analyzed in Paper II, that might indicate an over-sampling of some of the parameter spaces.

We can also attempt to answer if models with similar cloud distributions produce similar SEDs. Using identical cloud distributions, our synthetic spectral analysis shows that the two clumpy models [Nenkova08] and [Hoenig10] can be distinguished by looking to their silicate feature residuals which are mainly attributed to different cloud compositions \citep[by ][for $\rm{[Nenkova08]}$ and $\rm{[Hoenig10]}$, respectively]{Ossenkopf92,Draine84}. This can also be seen in Fig.\,\ref{fig:genfit3}. [Nenkova08] can produce 18$\rm{\mu m}$ silicate features in absorption (i.e. positive values) while [Hoenig10] can only produce these features in emission. After exploring the resulting spectral shapes for different ranges on the parameters, we found that these differences cannot be attributed to unmatched parameter space for these SED library. Indeed, \citet{Feltre12} showed that, even though clumpy and smooth dust models produce different SEDs, most of the differences arise from the model assumptions and not from the dust distribution. This might also explain why [Hoenig10] and [Hoenig17], despite their major differences in overall components and morphology of the distribution of dust, are undistinguishable when it comes to SED fitting, as shown here. This also happens for the parent models [Fritz06] and [Stalev16]. Therefore, it seems that the dust composition needs to be better explored because it might have a major impact on the shape of the silicate features.  

The clumpy torus model [Hoenig10] and its disk+wind version [Hoenig17], naturally produces flat near- and mid-infrared slopes (see Fig.\,\ref{fig:genfit1}), lack of steep far-infrared slopes (see Fig.\,\ref{fig:genfit2}), and a narrower range of silicate feature strengths (see Fig.\,\ref{fig:genfit3}) compared to other models. These main differences are fully consistent with previous results. \citet{Hoenig10A} pointed out that the silicate features and the near-infrared slopes are some of the main differences among models \citep[see also][]{Feltre12,Stalevski12,Ramos-Almeida14,Hoenig17}. The smooth torus model [Fritz06] and its two phase version [Stalev16] can also be distinguished from the others (see above). However, parent models do not always produce the same spectral features. \citet{Garcia-Gonzalez17} found that [Hoenig10] produces stronger near-infrared emission and bluer mid-infrared spectral slopes than its previous version.

Finally, another important result of the present analysis is that the parameters for each model can be determined (within less than 15\% of the parameter space). This is true for any model except [Hoenig17] for a source with a 12$\mu m$ flux above $\rm{\sim 100}$mJy and for any instrumental configuration except for GTC/CanariCam (for which sources with 12$\mu m$ flux larger than $\rm{\sim 300}$mJy are required). However, the isolation of the AGN mid-infrared emission is key; more than 50\% of stellar contribution at 5$\rm{\mu m}$, 50\% of ISM contribution at 30$\rm{\mu m}$, or a combination of 10\% of stellar and 10\% of ISM, can double the error on the parameters. This is the main reason to require high resolution infrared observations, as those provided by the future \emph{JWST}.

\section{Summary} \label{sec:summary}

We have investigated a set of six SED libraries with the aim at reproducing the dust infrared emission of AGN. We produced synthetic spectra for GTC/CanariCam, \emph{Spitzer}/IRS, \emph{JWST}/NIRSpec, and \emph{JWST}/MIRI instrumental setups and four sensitivities (equivalent to $\rm{30 mJy <F_{12\mu m} <10\,Jy}$ or $\rm{3<S/N<150}$). We fitted them with the set of models and using a general parametrization which includes three slopes and the strength of the silicate features. The main results are:

\begin{enumerate}
    
    \item Each model can be distinguished from the others based on the chi-squared statistics. The exceptions are the comparison between [Hoenig10] versus [Hoenig17] and [Fritz06] versus [Stalev16]. This is probably due to the fact that [Hoenig17] ([Stalev16]) is based on [Hoenig10] ([Fritz06]), and therefore, they used similar prescriptions for the dust composition. 
    
    \item In general, residuals show that the main differences among the models are the slopes of the spectra below $\rm{\sim}$7\,$\rm{\mu m}$ and above $\rm{\sim}$25\,$\rm{\mu m}$, and the shape of the 10 and 18\,$\rm{\mu m}$ silicate features. We also show that these models can be distinguished based on the continuum slopes and the silicate feature strengths.

    \item The parameters can be well determined within 15\% error (compared to their parameter space) for all the models using either \emph{Spitzer}/IRS, \emph{JWST}/MIRI, or \emph{JWST}/(MIRI+NIRSpec). This is true except for [Hoenig17] that has a large number of free parameters. In this case the inclusion of \emph{JWST}/(MIRI+NIRSpec) data and the selection of the brightest AGN at mid-infrared are needed to improve the resulting parameter determination. We do not see any significant improvement by including near-infrared to the mid-infrared data for other models. However, this might be relevant for host galaxy decontamination. Slightly worse results are obtained when using GTC N- and Q-bands. 
        
    \item Dilution plays an important role on the parameter determination. Either more than 50\% of stellar contribution compared to the AGN contribution at 5\,$\rm{\mu m}$, more than 50\% of ISM compared to the AGN contribution at 30\,$\rm{\mu m}$, or a combination of 50\% of each component prevents the parameter determination. 
    
\end{enumerate}

Infrared dust models (either in the form of a torus, disk or wind) could be distinguished and their parameter constrained using the spectral fit to the 5-30\,$\rm{\mu m}$ wavelength range. However, high spatial resolution data are required to isolate this AGN emission from host galaxy dilution. For that, current ground-based mid-infrared (e.g. GTC/CanariCam) and future \emph{JWST} data are required. In Paper II, we fit these models to a sample of 110 AGN selected from the \emph{Swift}/BAT survey with available \emph{Spitzer}/IRS spectra. 

\acknowledgments
We thank to the anonymous referee for his/her comments and suggestions which have improved significantly the results of this research. This research is mainly funded by the UNAM PAPIIT project IA103118 (PI OG-M). MM-P acknowledges support by KASI postdoctoral fellowships. IM and JM acknowledge financial support from the research project AYA2016-76682-C3-1-P (AEI/FEDER, UE). JMR-E acknowledge support from the Spanish Ministry of Science under grant AYA2015-70498-C2-1, and AYA2017-84061-P.  IG-B acknowledges financial support from the Spanish Ministry of Science and Innovation (MICINN) through projects PN AYA2015-64346-C2-1-P and AYA2016-76682-C3-2-P. I.M. and J.M. acknowledge financial support from the State Agency for Research of the Spanish MCIU through the ``Center of Excellence Severo
Ochoa" award for the Instituto de Astrof\'isica de Andaluc\'ia (SEV-2017-0709). D. E.-A. acknowledges support from a CONACYT scholarship. D-D acknowledges PAPIIT UNAM support from grant IN113719. This research has made use of dedicated servers maintained by Jaime Perea (HyperCat at IAA-CSIC), Alfonso Ginori Gonz\'alez, Gilberto Zavala, and Miguel Espejel (Galaxias, Posgrado04, and Arambolas at IRyA-UNAM) and Daniel D\'iaz-Gonz\'alez (IRyAGN1 and IRyAGN2). All of them are gratefully acknowledged.

\appendix

\section{Synthetic spectra complementary figures}\label{appendix:Par2Par}

\begin{figure*}[!ht]
\begin{center}
\includegraphics[width=0.3\columnwidth]{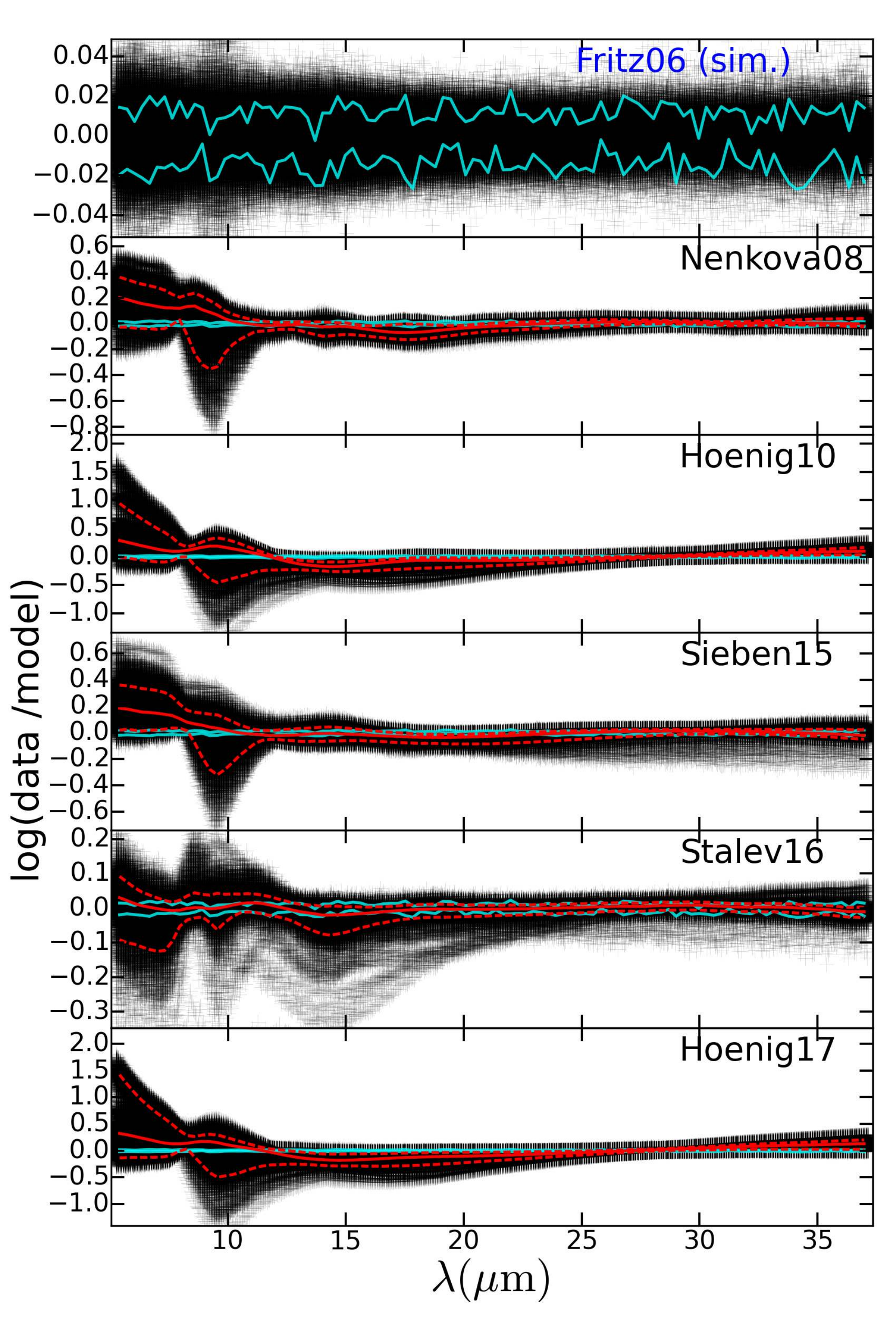}
\includegraphics[width=0.3\columnwidth]{RESULTSNGC3516_Nenkova08_SpitzerLR_res_of.pdf}
\includegraphics[width=0.3\columnwidth]{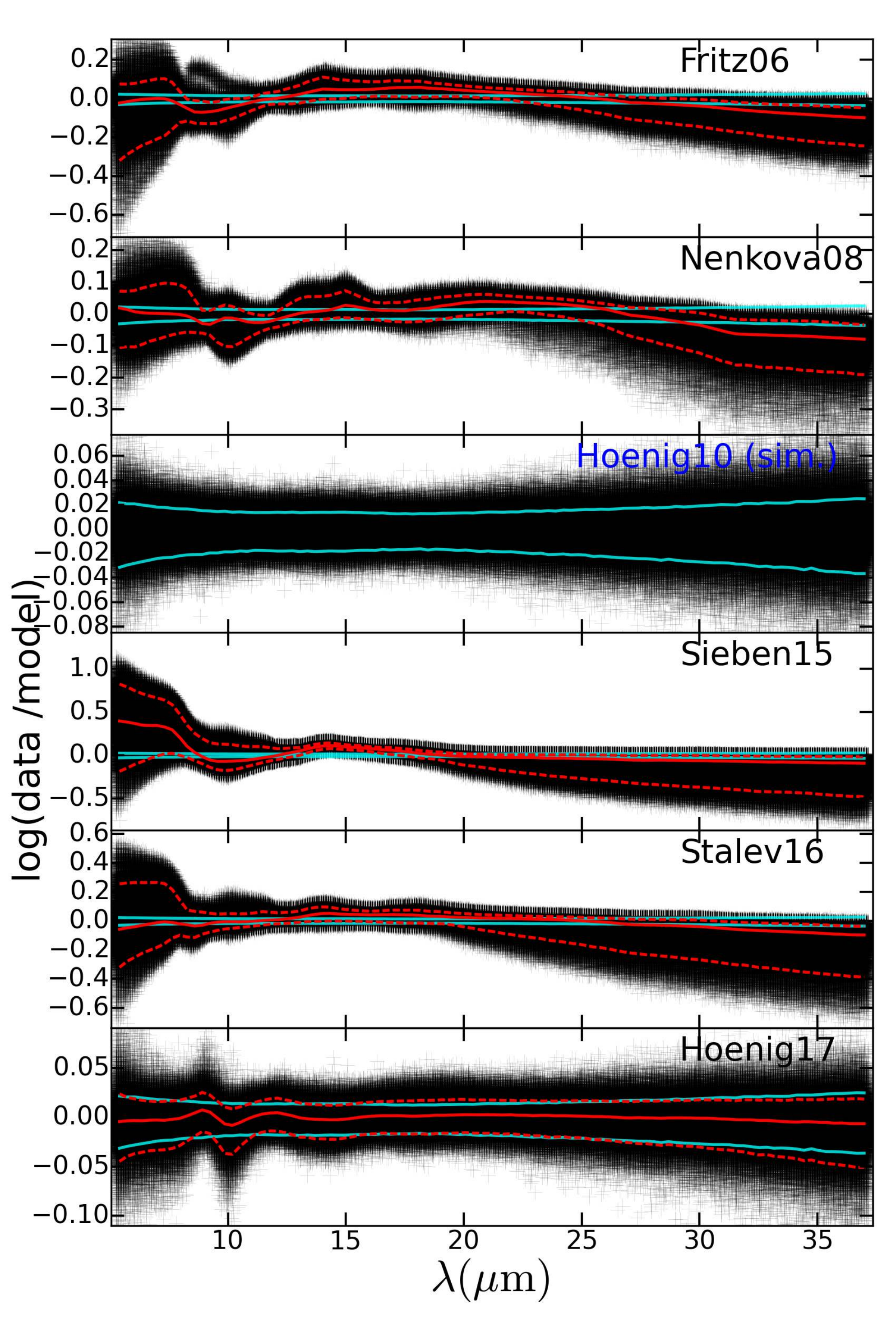}
\includegraphics[width=0.3\columnwidth]{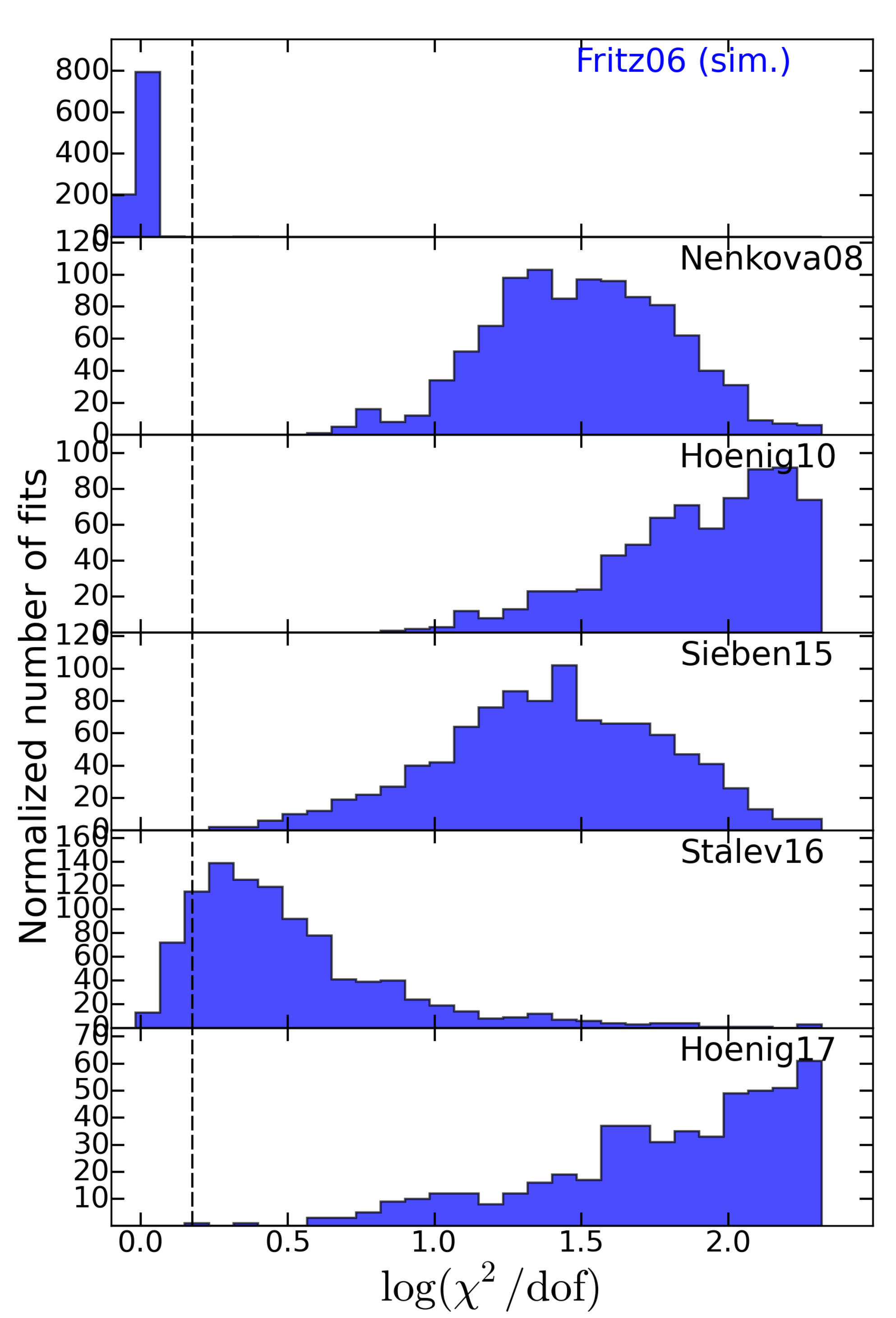}
\includegraphics[width=0.3\columnwidth]{RESULTSNGC3516_Nenkova08_SpitzerLR_Chi_of.pdf}
\includegraphics[width=0.3\columnwidth]{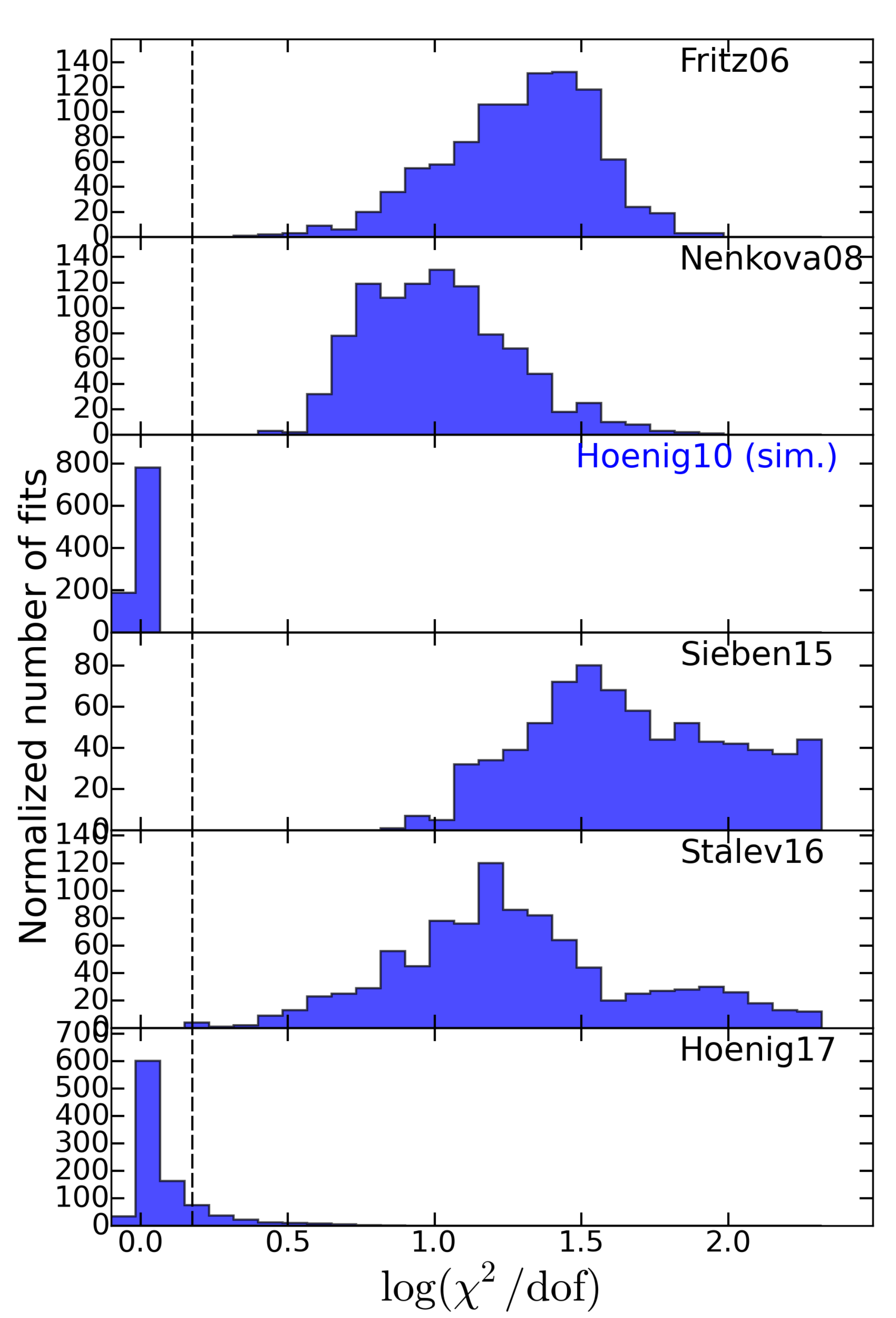}
\caption{Residuals (as the ratio between data and model, top) and distribution of $\rm{\chi^2/dof}$ (bottom) obtained when fitting the 1,000 synthetic spectra produced using the models reported by \citet{Fritz06} (left), \citet{Nenkova08B} (middle), \citet{Hoenig10B} (right). Panels, from top to bottom in each plot, show the results of fitting the synthetic spectra to \citet{Fritz06}, \citet{Nenkova08B},  \citet{Hoenig10B}, \citet{Siebenmorgen15}, \citet{Stalevski16}, and \citet{Hoenig17}. The self-fit is highlighted with blue letters in the second panel. Cyan continuous lines show the 15\% and 85\% of the residuals when the synthetic spectra are fitted with the same model as the one simulated \citep[i.e.][]{Nenkova08B}. Red continuous and dashed lines show the median, 15\%, and 85\% of the residuals when the synthetic spectra are fitted with other models. The vertical-dashed line in the right panel shows the locus of $\rm{\chi^2/dof=1.5}$.}
\label{appendix:FitResiduals1}
\end{center}
\end{figure*}

\begin{figure*}[!ht]
\begin{center}
\includegraphics[width=0.3\columnwidth]{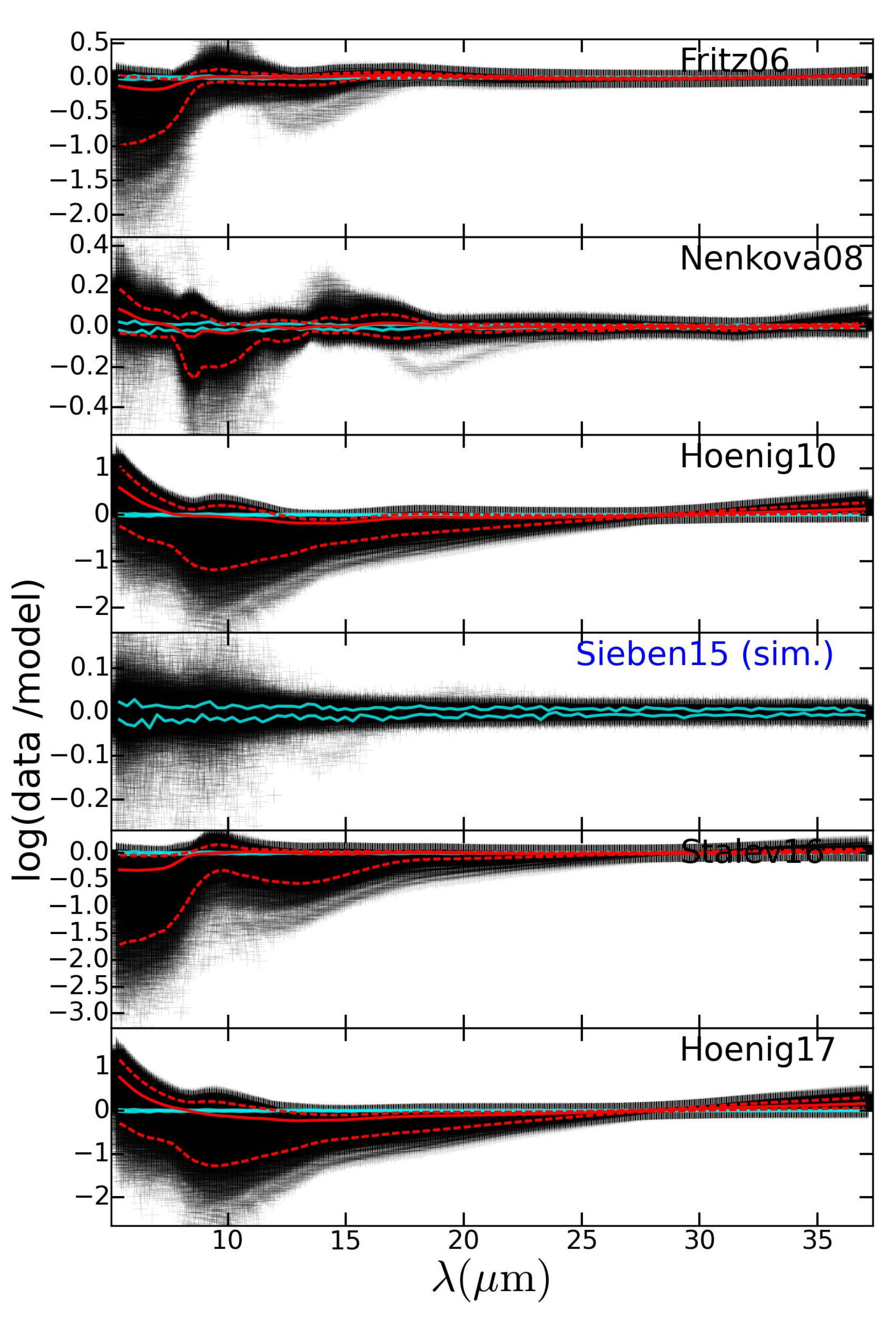}
\includegraphics[width=0.3\columnwidth]{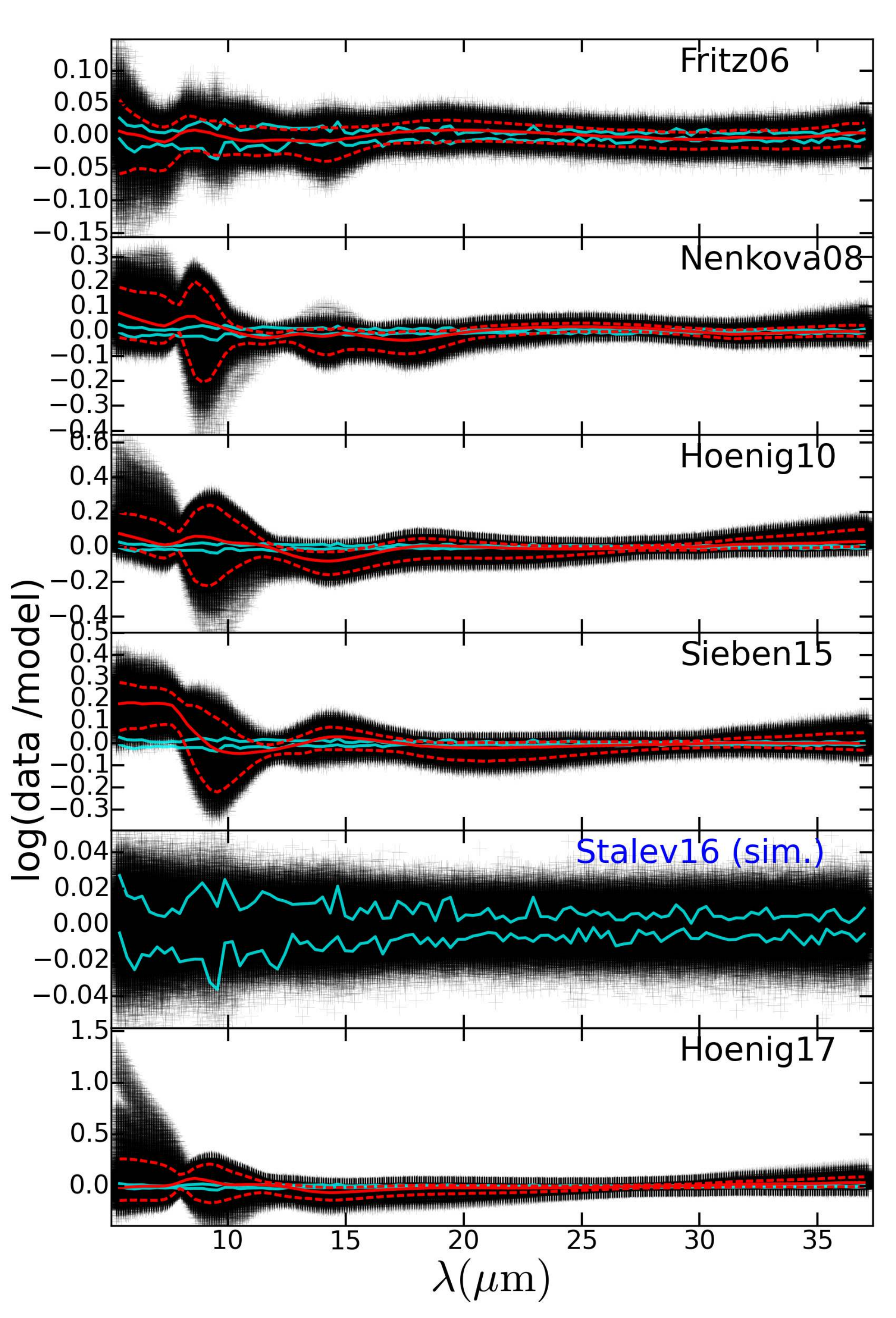} 
\includegraphics[width=0.3\columnwidth]{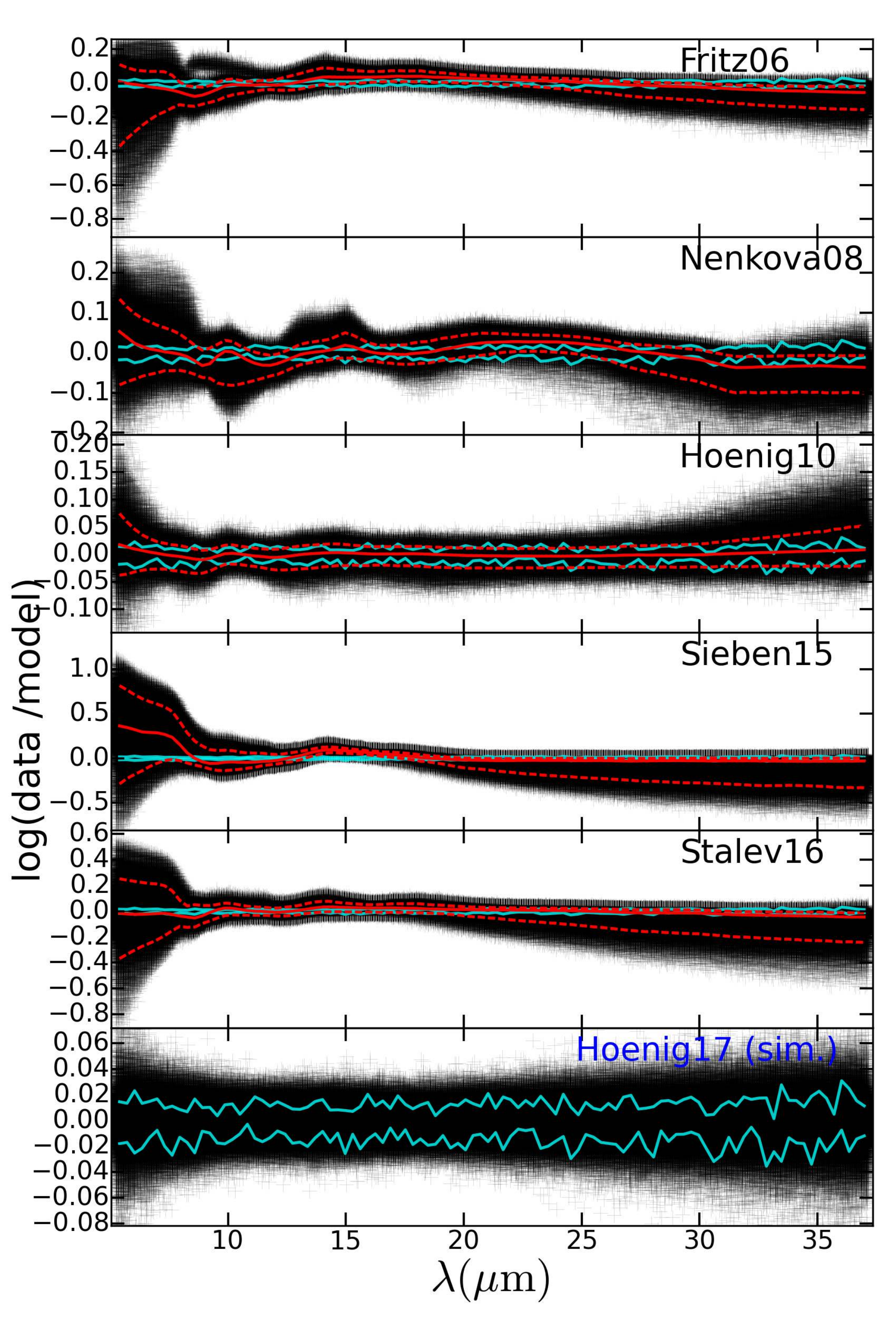} 
\includegraphics[width=0.3\columnwidth]{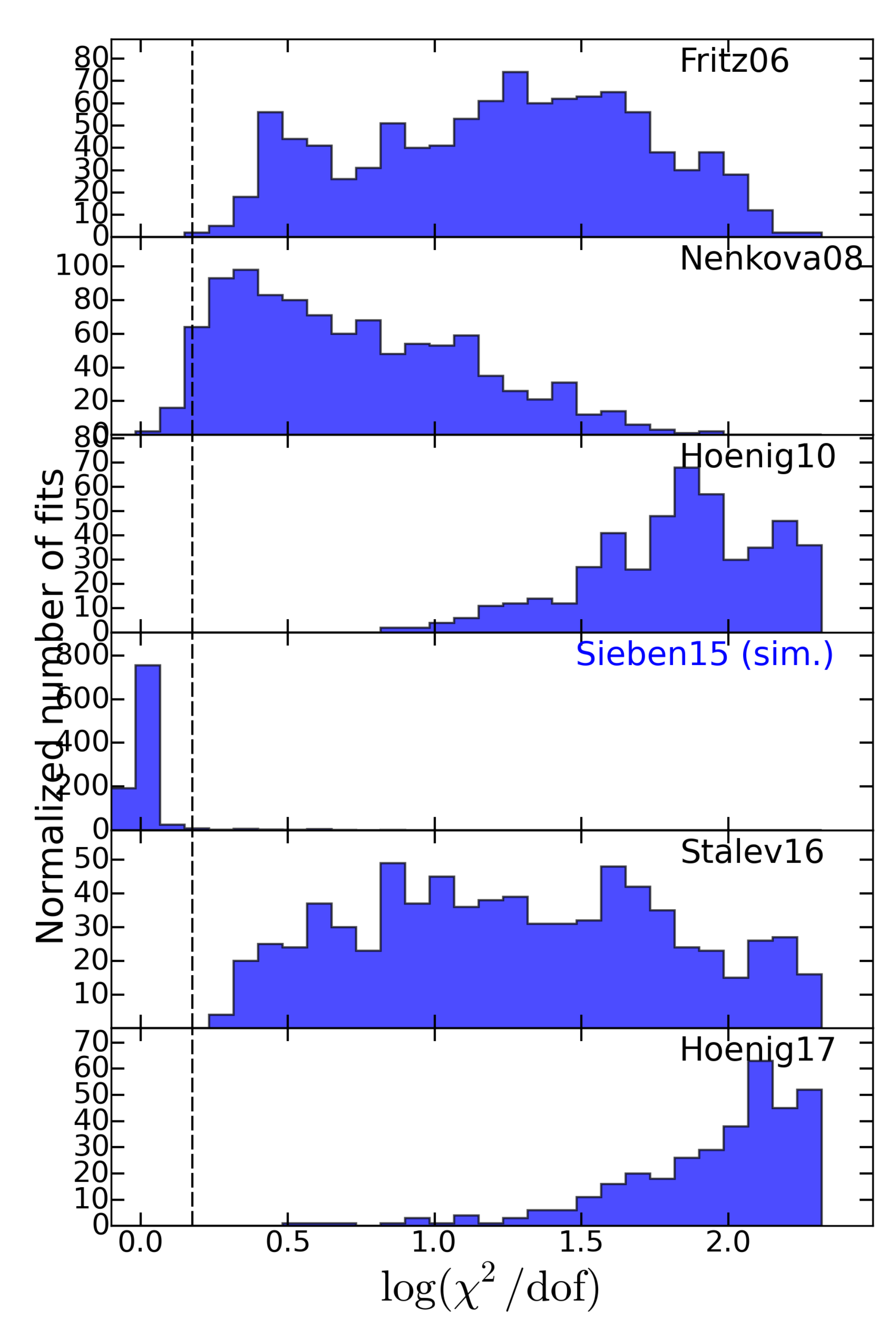}
\includegraphics[width=0.3\columnwidth]{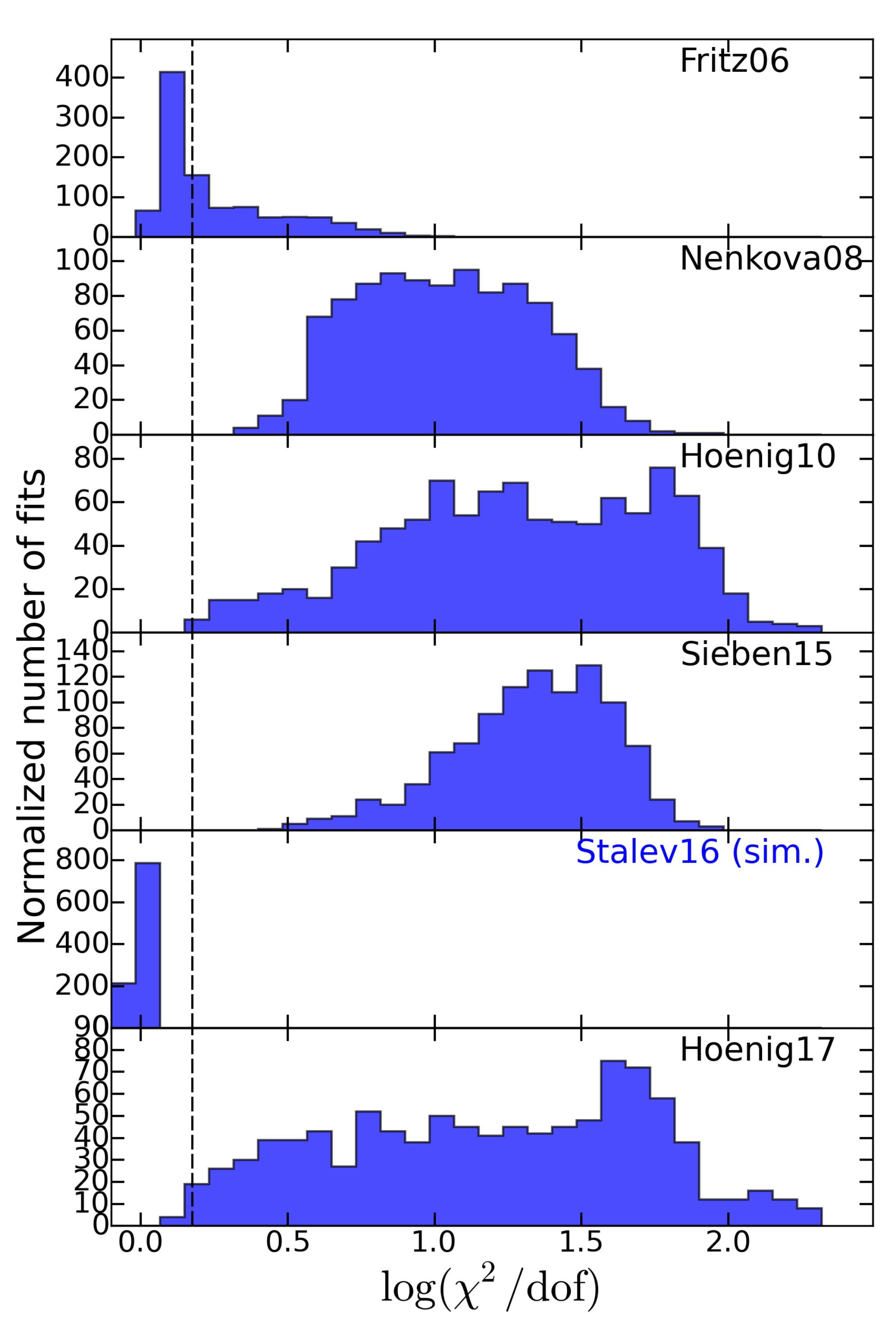}
\includegraphics[width=0.3\columnwidth]{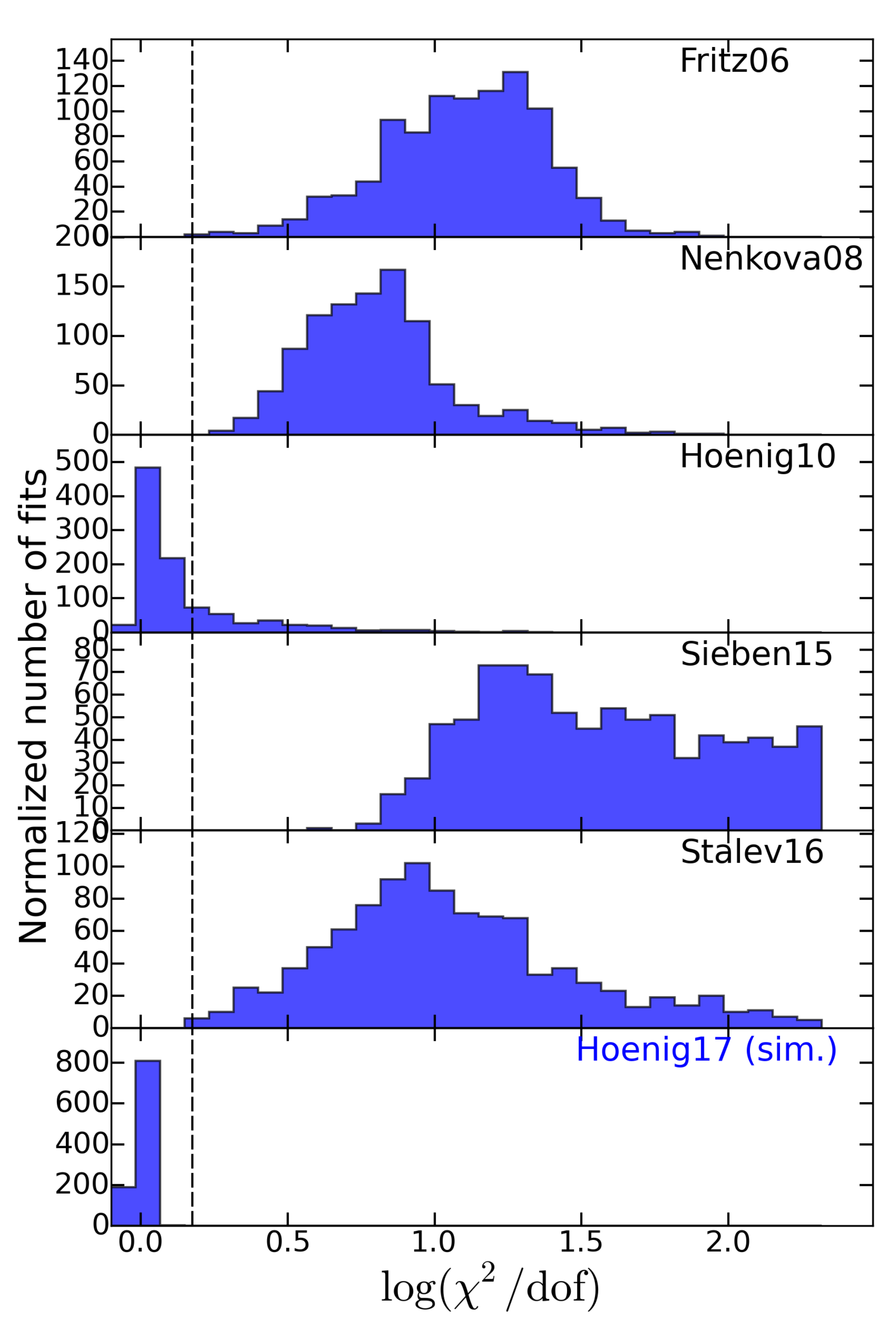}
\caption{Same as Fig.\,\ref{appendix:FitResiduals1} for the synthetic spectra produced with the model described by \citet{Siebenmorgen15} (left), \citet{Stalevski16} (middle), and \citet{Hoenig17} (right).}
\label{appendix:FitResiduals2}
\end{center}
\end{figure*}

\end{document}